\documentclass[linenumber]{aastex631}
\usepackage{CJK}

\shortauthors{Wei et al.}
\usepackage{graphicx}
\usepackage{amsmath}
\usepackage{txfonts}
\usepackage{hyperref}
\usepackage{threeparttable}
\usepackage{multirow}
\usepackage{threeparttable}
\usepackage{natbib}
\usepackage{enumitem}
\defcitealias{TM99}{TM99}
\makeatletter

\newcommand{\Rmnum}[1]{\expandafter\@slowromancap\romannumeral #1@}
\makeatother

\begin{document}

\begin{CJK*}{UTF8}{gbsn}

\title{Constraining the Supernova Remnant Environment of FRB 190520B with Dispersion Measure and Scattering Timescale}

\author{Jia-Peng Wei}
\affiliation{Tsung-Dao Lee Institute, Shanghai Jiao Tong University, Shanghai 201210, P. R. China}

\author{Chen Deng}
\affiliation{School of Astronomy and Space Science, Nanjing University, Nanjing 210023, China}

\author{Gwenael Giacinti}
\affiliation{Universit\'e Paris Cit\'e, CNRS, AstroParticule et Cosmologie, F-75013 Paris, France}

\author{Ze-Cheng Zou}
\affiliation{School of Astronomy and Space Science, Nanjing University, Nanjing 210023, China}

\author{Chen-Ran Hu}
\affiliation{School of Astronomy and Space Science, Nanjing University, Nanjing 210023, China}

\author{Yong-Feng Huang}
\affiliation{School of Astronomy and Space Science, Nanjing
University, Nanjing 210023, China} \affiliation{Key Laboratory of
Modern Astronomy and Astrophysics (Nanjing University), Ministry
of Education, China}

\author{Jin-Jun Geng}
\affiliation{Purple Mountain Observatory, Chinese Academy of Sciences, Nanjing 210023, China}

\correspondingauthor{Gwenael Giacinti}
\email{ggiacinti@apc.in2p3.fr}

\correspondingauthor{Yong-Feng Huang}
\email{hyf@nju.edu.cn}

\begin{abstract}

FRB 190520B is a repeating fast radio burst source whose large
dispersion measure (DM) and temporal broadening suggest a dense
and evolving local environment. In this work, we test the
possibility that FRB 190520B originates from the core-collapse of
a massive star so that its central engine is embedded in a
supernova remnant (SNR) expanding into a wind environment, whose
evolution is described by the self-similar solution. We use the
observed DM and scattering timescale of FRB 190520B to constrain
the physical parameters of its surrounding SNR and host-galaxy DM.
Twenty typical cases are considered, arising from four ejecta
profiles and five scattering prescriptions. It is found that only
6 cases are retained and provide acceptable fits. All retained
cases have a shallow ejecta profile and a young source age of
$t_0=79.8$--$169.8~{\rm yr}$. The ejecta mass is inferred to be
large for all six cases, while the kinetic energy and mass-loss
rate span a wide range. The secular DM evolution is reproduced
better than the detailed scattering evolution. The up-drift
behavior of the scattering residual suggests an additional
component or more complicated structures inside the SNR. All
retained cases are self-consistent within the adopted scattering
theory and the circum-burst medium becomes transparent for GHz
bursts before the inferred source ages.

\end{abstract}

\keywords{Interstellar medium (847) --- Supernova remnants (1667) --- Radio bursts (1339) --- Radio transient sources (2008) --- Compact radiation sources (289)}

\section{Introduction}
\label{sec:intro}

Fast radio bursts (FRBs) are energetic radio transients with millisecond durations, originating from the deep sky (\citealt{Lorimer2007}).
Their extremely high brightness temperatures strongly suggest a coherent emission mechanism (\citealt{Zhang2023}).
The observed dispersion measures (DMs) of FRBs are generally larger than the Galactic contribution estimated by the NE2001 model, providing compelling evidence for their extragalactic origin (\citealt{CordesLazio2003}; \citealt{Thornton2013}; \citealt{Spitler2014,Spitler2016}; \citealt{Niu2022}).
Recently, the Canadian Hydrogen Intensity Mapping Experiment (CHIME) reported more than 3600 newly detected FRBs in CHIME/FRB Catalog 2, bringing the total number of known FRB sources to over 4500 (\citealt{CHIMEcatalog2}).
Observationally, FRB sources are commonly divided into two categories: repeaters and apparent non-repeaters (\citealt{Thornton2013}; \citealt{Spitler2016}).
There are several differences between the two populations in observations such as the time-frequency down-drifting behavior and the association with persistent radio sources (PRSs) for repeaters (\citealt{Chatterjee2017}; \citealt{Ricci2021}; \citealt{Niu2022}).

\begin{figure}
    \centering
    \includegraphics[width=0.45\linewidth]{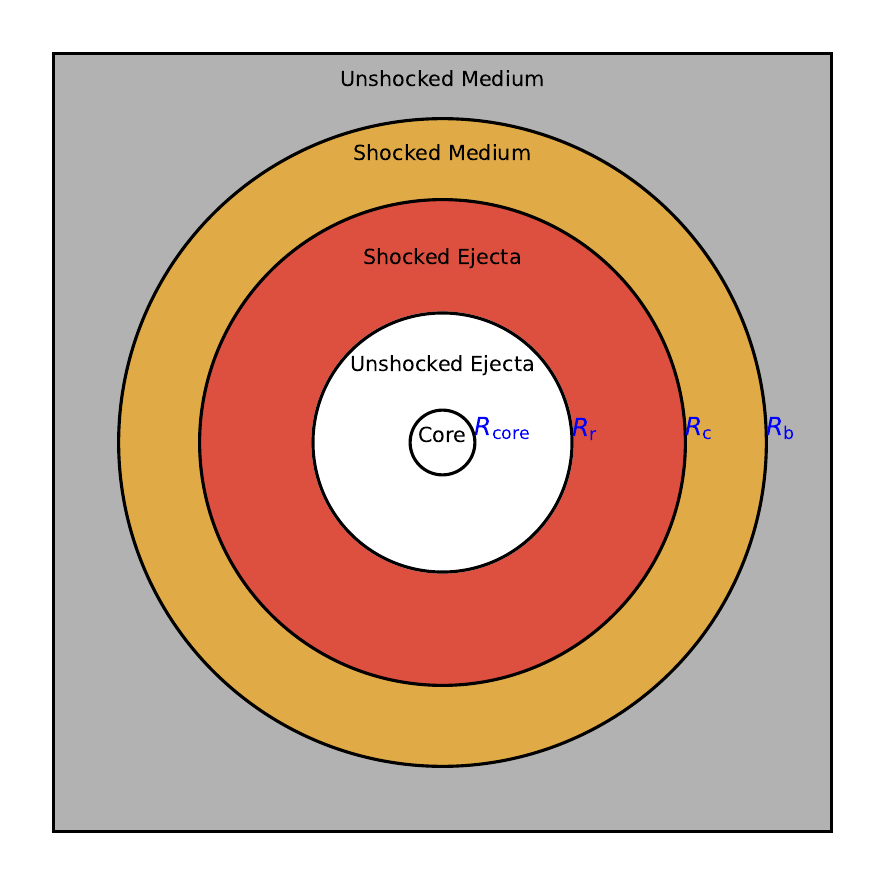}
    \caption{Schematic illustration of a SNR in the self-similar solution.}
    \label{figsketch}
\end{figure}

Despite rapid observational progresses, the origin of FRBs remains
under debate. Researchers have proposed many theoretical models,
trying to explain the mechanisms responsible for FRBs
(\citealt{GengHuang2015}; \citealt{DaiZG2016};
\citealt{ZhangB2017}; \citealt{Lyubarsky2020};
\citealt{Beloborodov2020}). Convincing support for magnetars as
FRB progenitors came from the discovery of FRB 200428, which was
associated with the Galactic magnetar SGR 1935+2154
(\citealt{Bochenek2020}; \citealt{CHIME2020magnetar}). Magnetars
can be produced through at least two distinct evolutionary
channels: the core-collapse of massive stars and compact binary
mergers (\citealt{Yoon2007}; \citealt{Giacomazzo2013};
\citealt{Zhong2020}; \citealt{Jacco2020}). In both formation
channels, an explosive event launches rapidly expanding ejecta and
initiates the ejecta-dominated (ED) phase of remnant evolution.
Because the ejecta expands supersonically into the ambient medium,
a forward shock is formed and propagates outward, while a reverse
shock is driven back into the ejecta by the pressure of the
shocked ambient material (\citealt{TM99}, hereafter TM99). The
shocked ejecta and shocked ambient medium are separated by a
contact discontinuity (CD), across which the density and
temperature generally change sharply (\citetalias{TM99}). The
hydrodynamical evolution of such systems has been extensively
studied in the context of supernova remnants (SNRs). Self-similar
solutions for SNR evolution were developed by
\citet{Chevalier1982} and \citet{Nadezhin1985}, in which the
ejecta density profile is described by a uniform-density core and
a power-law envelope. The structure of a SNR in the self-similar
solution is shown in Figure \ref{figsketch}, where $R_{\rm core}$,
$R_{\rm r}$, $R_{\rm c}$, and $R_{\rm b}$ are the radii of the
core, reverse shock, CD shell, and forward shock, respectively.

FRB 190520B is a particularly intriguing source because of its
exceptionally large DM of $\sim 10^{3}\ {\rm pc\ cm^{-3}}$ and its
extreme, time-variable rotation measure (RM), which has exhibited
two sign reversals (\citealt{Niu2022,NiuCH2026}; \citealt{anna2023}; \citealt{Katz2022}). The source is localized to a star-forming
region within a dwarf galaxy at a redshift of $\sim 0.241$ and is
associated with a PRS, showing remarkable similarities to FRB
121102 (\citealt{Chatterjee2017, Tendulkar2017, Niu2022}). It is
well known that the radio emissions are dispersed and scattered as
they pass through a plasma medium, resulting in a
frequency-dependent delay in arrival time and temporal pulse
broadening. Before reaching the Earth, FRB signals may traverse
several media, including the interstellar medium (ISM) of the
Milky Way, the intergalactic medium (IGM), the ISM of the host
galaxy, and the local environment of the source.

In this work, we model the local environment of FRB 190520B as a
SNR described by the self-similar solution, with a core-collapse
massive-star progenitor, and consider several prescriptions for
temporal broadening induced by plasma density fluctuations
(\citealt{Rickett1977, Rickett1990}; \citealt{Xu2016}). Combining
with the observed DMs and scattering timescales, we constrain the
properties of the local environment and source system, including
the ejecta mass, ejecta kinetic energy, and the mass-loss rate of
the progenitor, as well as the source age and the DM contribution
from its host galaxy. Compared with using DM alone, the joint
consideration of DM and scattering timescale can provide tighter
and more informative constraints on the SNR model. For the compact
binary merger scenario, the evolutions of DM and RM have been
investigated by \citet{ZhaoZY2021}. We assume that the observed DM
includes contributions from all media along the line of sight,
whereas the observed scattering timescale is dominated by the
local SNR environment, with contributions from other regions being
negligible (\citealt{Luan2014}; \citealt{Katz2016}), which will be
discussed later.

This paper is organized as follows. In Section \ref{SNR}, we
describe the self-similar SNR model and its dynamical evolution.
Section \ref{scatter} introduces the theoretical framework of
temporal broadening induced by plasma density irregularities. In
Section \ref{MCMC}, we present the Bayesian inference and the
Markov Chain Monte Carlo (MCMC) fitting procedure. The numerical
results are given in Section \ref{result}. Finally, discussions
and summary are provided in Sections \ref{discussion} and
\ref{Summary}.

\section{The Self-Similar Solution of SNR}
\label{SNR}

The evolution of a SNR is commonly divided into four phases
(\citealt{Jacco2020}): (i) the ED phase, during which the swept-up
ambient mass is negligible compared with the ejecta mass and the
ejecta expands approximately homogenously; (ii) the Sedov--Taylor
(ST) phase, when the swept-up mass becomes comparable to and even
larger than the ejecta mass; (iii) the radiative or snowplow
phase; and (iv) the merging phase, in which the remnant gradually
dissipates into the ambient medium. In this work, we focus
exclusively on the ED phase, although our approximate solutions
are also valid for the ST phase.

\subsection{Characteristic scales}
\label{sub2.1}

Following \citetalias{TM99}, the hydrodynamical evolution can be
conveniently described by using dimensionless variables
constructed from the ejecta mass, kinetic energy, and the
normalization of the ambient density. The ambient medium is
assumed to follow a density profile $\rho_{\rm a}=\eta_{\rm s}
r^{-s}$, where $\eta_{\rm s}$ is a constant and $s$ is the
power-law index (\citetalias{TM99}). To simplify the subsequent
expressions, we define three characteristic scales from these
independent parameters, i.e. the characteristic mass, length, and
time as given by
\begin{equation}\label{func1}
    M_{\rm ch} = M_{\rm ej},
\end{equation}
\begin{equation}\label{func2}
    R_{\rm ch}=M_{\rm ej}^{1/(3-s)}\eta_{\rm s}^{-1/(3-s)},
\end{equation}
\begin{equation}\label{func3}
    t_{\rm ch} = E_{\rm k}^{-1/2}M_{\rm ej}^{(5-s)/[2(3-s)]}\eta_{\rm s}^{-1/(3-s)}.
\end{equation}
In Section \ref{sub2.3} below, a physical quantity $X$ normalized
by these characteristic scales is denoted as $X^*$, i.e.,
$X^*=X/M_{\rm ch}^{x_1}R_{\rm ch}^{x_2}t_{\rm ch}^{x_3}$, where
$x_1$, $x_2$, and $x_3$ are constants determined by dimensional
analysis.

Since we focus on the core-collapse scenario, the progenitor is
naturally assumed to be embedded in a stellar wind environment.
For a wind environment, we have $s=2$ and $\eta_{\rm
s}=\dot{M}_{\rm w}/4\pi v_{\rm w}$, where $\dot{M}_{\rm w}$ is the
mass-loss rate of the progenitor and $v_{\rm w}$ is the wind
velocity. The characteristic length and time are then given by
\begin{equation}\label{func4}
    R_{\rm ch} = 12.9~{\rm pc}~\left(\frac{M_{\rm ej}}{M_\odot} \right) \left( \frac{10^{-5}~M_\odot~\rm{yr}^{-1}}{\dot{M}_{\rm w}} \right) \left(\frac{v_{\rm{w}}}{10~\rm{km}~\rm{s}^{-1}} \right),
\end{equation}
\begin{equation}\label{func5}
    t_{\rm{ch}} = 1772~{\rm yr}~\left( \frac{10^{51}~\rm{erg}}{E_{\rm{k}}} \right)^{1/2} \left( \frac{M_{\rm{ej}}}{M_\odot} \right)^{3/2} \left( \frac{10^{-5}~M_\odot~\rm{yr}^{-1}}{\dot{M}_{\rm w}} \right) \left( \frac{v_{\rm{w}}}{10 \, \rm{km} \, \rm{s}^{-1}} \right).
\end{equation}

\subsection{Initial conditions}
\label{sub2.2}

Following \citetalias{TM99}, the density profiles of the ejecta
$\rho_{\rm ej}$ and the ambient medium $\rho_{\rm a}$ can be
expressed as
\begin{equation}\label{func6}
\rho (r,t)=
    \begin{cases}
    \rho_{\rm ej}(r,t)=\frac{M_{\rm ej}}{R_{\rm ej}^3}f(\frac{r}{R_{\rm ej}}),&                      r\leq R_{\rm ej},\\
    \rho_{\rm a}=\eta_{\rm s}r^{-s},&r>R_{\rm ej},
    \end{cases}
\end{equation}
where $f(\frac{r}{R_{\rm ej}})$ is the structure function of
ejecta, and $R_{\rm ej}$ is the outer layer radius of the ejecta.
The structure function is assumed to follow a broken power-law
profile (\citetalias{TM99}; \citealt{ZhaoZY2021})
\begin{equation}\label{func7}
    f(w)=
    \begin{cases}
        f_0,&0\leq w\leq w_{\rm core},\\
        f_{\rm n}w^{-n}, & w_{\rm core}\leq w \leq 1,
    \end{cases}
\end{equation}
where $w=r/R_{\rm ej}$, $w_{\rm core}=R_{\rm core}/R_{\rm ej}$,
$f_{\rm n}=f_0w_{\rm core}^n$ , and $n$ is the power-law index of
the envelope of ejecta density. For $s=2$, when $n<3$, the ejecta
profile contains no flat core so that $w_{\rm core}=0$. In this
case the first branch of Equation~(\ref{func7}) vanishes, and the
ejecta profile reduces to a single power law, \(f(w)=f_{\rm
n}w^{-n}\) for \(0\leq w\leq1\); when $n>3$, a finite core is
present and we adopt a fiducial value $w_{\rm core}=0.1$ following
\citet{Tang2017}. The ejecta mass is conserved during the ED
phase.
 Using mass conservation, the two normalization factors, $f_{\rm n}$ and $f_0$, can be written as (\citetalias{TM99})
\begin{subequations}\label{func8}
\begin{align}
     f_{\rm n} =&\frac{3-n}{4\pi}, &n<3,\label{func8a}\\
     f_0 = & \frac{3}{4\pi w_{\rm{core}}^n} \left[ \frac{1 - (n/3)}{1 - (n/3) w_{\rm{core}}^{3 - n}} \right], &n>3.\label{func8b}
\end{align}
\end{subequations}

\subsection{Approximate solutions of SNR}
\label{sub2.3}

During the ED phase, two classes of approximate solutions can be
obtained, depending on the power-law index $n$ of the ejecta
envelope. If the envelope is shallow ($n<5$), the solution is
referred to as the free-expansion (FE) solution. If the envelope
is steep ($n>5$), the corresponding solution is the self-similar
driven wave (SSDW) solution. In this work, we consider the ejecta
density profiles with $n = 2, 4, 6,$ and $8$, which span both the
shallow-envelope ($n<5$, FE) and steep-envelope ($n>5$, SSDW)
regimes. We again emphasize that we focus on the core-collapse
scenario, corresponding to $s=2$. This means that the values of
some parameters involved in the approximate solutions, which will
be presented later, are only valid for the scenario of $s=2$.

For the FE solution, following \citet{Tang2017}, the evolution of
the reverse shock, CD shell, and forward shock is characterized by
\begin{equation}\label{func9}
    R^*_{\rm r}  = q_{\rm r} R_{\rm c}^*,
\end{equation}
\begin{equation}\label{func10}
    R_{\rm{c}}^{*} = \left[ \left( \lambda_{\rm{c}} t^{*} \right)^{-a} + \left( c t^{*b} \right)^{-a} \right]^{-1/a},
\end{equation}
\begin{equation}\label{func11}
    R_{\rm{b}}^{*}(t^{*}) = \left[ (\lambda_{\rm{b}} t^{*})^{-2\alpha} + (\xi t^{*2})^{-2\alpha/(5-s)} \right]^{-1/(2\alpha)},
\end{equation}
where $q_{\rm r}=q_{\rm b}/l_{\rm ED}$, with $q_{\rm b}=1.19$ and
\begin{equation}\label{func12}
    l_{\rm ED}=1+\frac{8}{n^2}+\frac{0.4}{4-s}.
\end{equation}
The parameters of $\lambda_{\rm c}$, $\lambda_{\rm b}$, and $\xi$
are given by \citet{Tang2017} as,
\begin{equation}\label{func13}
    \lambda_{\rm c}=
    \begin{cases}
        [2(\frac{5-n}{3-n})]^{1/2}, &n<3,\\
        [2w_{\rm core}^{-2}(\frac{5-n}{3-n})(\frac{w_{\rm core}^{n-3}-n/3}{w_{\rm core}^{n-5}-n/5})]^{1/2}, &n>3,
    \end{cases}
\end{equation}
with $\lambda_{\rm b}=q_{\rm b}\lambda_{\rm c}$ and $\xi=3/2\pi$.
The parameters $a$, $b$, $c$, and $\alpha$ depend on the ejecta
power-law index $n$ and the ambient density profile $s$, and their
numerical values are tabulated in Tables 4 and 6 of
\citet{Tang2017}. We further introduce the transition time $t_{\rm
tran}^*$, which marks the end of the ED phase and the onset of the
ST phase. In the FE solution, the transition time can be expressed
as (\citealt{Tang2017})
\begin{equation}\label{func14}
    t_{\rm tran}^*=\left(\frac{\xi}{\lambda_{\rm b}^{5-s}} \right)^{1/(3-s)}.
\end{equation}

For the SSDW solution, the CD shell and forward shock evolve as (\citealt{Tang2017})
\begin{equation}\label{func15}
    R_{\rm{c}}^{*} = \left[ \left( \zeta_{\mathrm{c}} \, t^{* (n-3)/(n-s)} \right)^{-a} + \left( c \, t^{* b} \right)^{-a} \right]^{-1 /a},
\end{equation}
\begin{equation}\label{func16}
    R_{\rm{b}}^{*}(t^{*}) = \left[ \left( \zeta_{\rm{b}} \, t^{* (n-3)/(n-s)} \right)^{-\alpha} + \left( \xi \, t^{* 2} \right)^{-\alpha/(5-s)} \right]^{-1 /\alpha},
\end{equation}
where $\zeta_c$ and $\zeta_b$ are also tabulated in Tables 4 and 6
of \citet{Tang2017}. The evolution of the reverse shock in this
case is more complicated and is divided into two regimes by the
time $t^*_{\rm core}$, when the reverse shock reaches the flat
core (\citealt{Micelotta2016}). This timescale is given by
(\citealt{Hwang2012}; \citealt{Micelotta2016})
\begin{equation}\label{func17}
    t_{\rm{core}}^{*} = \left[ \frac{l_{\rm{ED}}^{s-2}}{\phi_{\rm{ED}}} \frac{3}{4\pi} \frac{(3-s)^{2}}{n(n-3)} \right]^{1/(3-s)} \frac{1}{v_{\rm{core}}^{*}},
\end{equation}
where
\begin{equation}\label{func18}
    \phi_{\rm ED} = (0.65-\exp(-n/4))\sqrt{1-\frac{s}{8}}.
\end{equation}
Before the reverse shock hits the core (i.e. $t^*<t^*_{\rm core}$), the reverse shock radius is (\citealt{Micelotta2016})
\begin{equation}\label{func19}
    R_{\rm{r}}^{*} =
    \left\{v_{\rm{core}}^{*~(n-3)} \frac{(3-s)^{2}}{n(n-3)} \frac{3}{4\pi}
    \frac{l_{\rm{ED}}^{\,n-2}}{\phi_{\rm{ED}}}\right\}^{\frac{1}{n-s}}
    \frac{t^{*\,\frac{n-3}{n-s}}}{l_{\rm ED}}.
\end{equation}

When $t^* > t^*_{\rm core}$, the self-similar description of the
ejecta structure breaks down because the reverse shock has
penetrated into the flat core. Following \citet{Laming2003} and
\citet{Micelotta2016}, we approximate the forward shock radius and
velocity entering the post-core reverse shock solution by their
values at $t^* = t^*_{\rm core}$. Under this prescription, the
reverse-shock radius for $t^* > t^*_{\rm core}$ can be written as
\begin{equation}\label{func20}
    R_{\rm r}^{*}(t^*) =
    \left[
    \frac{ R_{\rm b}^{*} ( t^*_{\rm core} ) }{ l_{\rm ED} \, t^*_{\rm core} }
    - \frac{3-s}{n-3}
    \frac{ v_{\rm b}^{*} ( t^*_{\rm core} ) }{ l_{\rm ED} }
    \ln \left( \frac{ t^* }{ t^*_{\rm core} } \right)
    \right] t^* .
\end{equation}
For the SSDW solution, the transition time is (\citealt{Tang2017})
\begin{equation}\label{func21}
    t_{\rm tran}^*=\left(\frac{\xi}{\zeta_{\rm b}^{5-s}} \right)^{(n-s)/[(n-5)(3-s)]}.
\end{equation}
In the following, we use these solutions to model the time
evolution of the DM contributed by different regions of the SNR.

\subsection{The evolution of DM}
\label{sub2.4}

The observed DM consists of several components:
\begin{equation}\label{func22}
    DM_{\rm obs}=DM_{\rm MW}+DM_{\rm MW,halo}+DM_{\rm IGM}+\frac{DM_{\rm host}+DM_{\rm local}}{1+z_0},
\end{equation}
where $z_0$ is the redshift of FRB source, $DM_{\rm MW}$, $DM_{\rm
MW,halo}$, $DM_{\rm IGM}$, $DM_{\rm host}$, and $DM_{\rm local}$
denote the contributions from the ISM of the Milky Way, the
Galactic halo, the IGM, the ISM of host galaxy, and the local SNR
environment, respectively. For FRB 190520B, following
\cite{Niu2022}, we adopt $DM_{\rm MW}=113\pm17~\rm{pc~cm^{-3}}$
and $DM_{\rm IGM}=195^{+110}_{-70}~\rm{pc~cm^{-3}}$. Since
$DM_{\rm MW,halo}$ is expected to lie in the range $30$--$80~{\rm
pc~cm^{-3}}$, we adopt the midpoint value, $55~{\rm pc~cm^{-3}}$,
as the fiducial contribution and take half of the interval width
as a symmetric uncertainty. This yields $DM_{\rm MW,halo} = 55 \pm
25~{\rm pc~cm^{-3}}$. In this work, $DM_{\rm host}$ is treated as
a free parameter. The local contribution $DM_{\rm local}$ is
further divided into four components from the SNR: the unshocked
core ($r<R_{\rm core}$), the unshocked ejecta envelope ($R_{\rm
core}<r<R_{\rm r}$), the shocked ejecta ($R_{\rm r}<r<R_{\rm c}$),
and the shocked ambient medium ($R_{\rm c}<r<R_{\rm b}$), as shown
in Figure \ref{figsketch}.

When $n<3$, there is no flat core and the unshocked ejecta extends
to the center of SNR. To avoid the singularity of DM at the SNR
center, we introduce a cutoff radius, $R_{\rm cut}=0.01R_{\rm r}$.
Therefore, the DM contribution from unshocked ejecta in this case
is
\begin{equation}\label{fun23}
    DM_{\rm unsh,ej} = \frac{M_{\rm ej}}{\mu m_{\rm p}}\eta f_{\rm n} \frac{R_{\rm r}^{1-n}-R_{\rm cut}^{1-n}}{(1-n)R_{\rm c}^{3-n}},
\end{equation}
where $\eta$ is the ionization fraction, $\mu$ is the mean atomic weight, and $m_{\rm p}$ is the proton mass.
In this work, we take $\mu\sim1$ and $\eta=0.03$ (\citealt{Chevalier2017}; \citealt{ZhaoZY2021}).
We also adopt $R_{\rm ej}\sim R_{\rm c}$ in the ED phase, which is a reasonable approximation in the self-similar framework (\citealt{Chevalier1982}).
When $n>3$, the DM contributions from the unshocked core and the unshocked ejecta are
\begin{equation}\label{func24}
    DM_{\rm unsh,core} = \frac{M_{\rm ej}}{\mu m_{\rm p}R^3_{\rm c}}\eta f_0 R_{\rm core},
\end{equation}
\begin{equation}\label{func25}
    DM_{\rm unsh,ej} = \frac{M_{\rm ej}}{\mu m_{\rm p}}\eta f_0 w_{\rm core}^n \frac{R_{\rm r}^{1-n}-R_{\rm core}^{1-n}}{(1-n)R_{\rm c}^{3-n}}.
\end{equation}

The downstream density of a strong shock is compressed by 4 times (\citealt{ZB2018}).
Accordingly, for the forward shock, the electron density in the shocked ambient medium is
\begin{equation}\label{func26}
    n_{\rm e,a} = 4n_0= \frac{4\eta_{\rm s}}{\mu m_{\rm p}}r^{-s},
\end{equation}
where $n_0$ is the density of the unshocked ambient medium.
Under the thin-shell approximation, the electron density in the shocked ejecta is (\citealt{Chevalier1982})
\begin{equation}\label{func27}
    n_{\rm e,ej} = \frac{(n-3)(n-4)}{(3-s)(4-s)}n_{\rm e,a}.
\end{equation}
The corresponding DM contributions from the shocked ejecta and the shocked ambient medium are then
\begin{equation}\label{func28}
    DM_{\rm sh,ej} = \frac{4\eta_{\rm s}}{\mu m_{\rm p}}\frac{(n-3)(n-4)}{(3-s)(4-s)}\frac{R_{\rm c}^{1-s}-R_{\rm r}^{1-s}}{1-s},
\end{equation}
\begin{equation}\label{func29}
    DM_{\rm sh,a} = \frac{4\eta_{\rm s}}{\mu m_{\rm p}}\frac{R_{\rm b}^{1-s}-R_{\rm c}^{1-s}}{1-s}.
\end{equation}

\section{Temporal Broadening of Radio Pulses in Turbulent Plasma}
\label{scatter}

We assume that the three-dimensional power spectrum of electron density irregularity
in the turbulent plasma takes the form (\citealt{Rickett1977,Rickett1990})
\begin{equation}\label{func30}
    P_{\rm 3N}(k)=C_N^2(r)(k^2+(\frac{2\pi}{L})^2)^{-\beta/2}\exp{(-k^2/(\frac{2\pi}{l_0})^2)},
\end{equation}
where $k$ is the spatial wavenumber and $\beta$ is the spectral index.
Here, $L$ and $l_0$ are the outer and inner scales of turbulence, corresponding to the energy injection and dissipation scales, respectively (\citealt{Xu2016}).

Within the inertial range, Equation (\ref{func30}) reduces to
\begin{equation}\label{func31}
    P_{\rm 3N}(k)=C_N^2(r)k^{-\beta}, \quad 2\pi/L\ll k\ll 2\pi/l_0.
\end{equation}
Observations indicate that the spectral index $\beta$ lies between 2 and 4 (\citealt{Lee1975II}; \citealt{Rickett1977}).
From the normalization of the power-law spectrum, $\int P_{3N}(k)d^3\vec{k}=(\delta n_{\rm e})^2$, one obtains (\citealt{Xu2016})
\begin{equation}\label{func32}
    C_N^2 \simeq
    \begin{cases}
    \frac{\beta - 3}{2(2\pi)^{4 - \beta}} (\delta n_{\rm e})^2 L^{3 - \beta}, & \beta > 3, \\
    \frac{3 - \beta}{2(2\pi)^{4 - \beta}} (\delta n_{\rm e})^2 l_0^{3 - \beta}, & \beta < 3,
    \end{cases}
\end{equation}
where $n_{\rm e}$ is the number density of electrons, $\delta n_{\rm e}=n_{\rm e}-\langle n_{\rm e}\rangle$
is the root-mean-square (rms) amplitude of the density fluctuation,  and $<~>$ denotes an ensemble average in the radial direction.
The case of $\beta=3$ is excluded, since in this situation $\delta n_{\rm e}$ becomes scale-independent  (\citealt{Xu2016}).

When the radio waves propagate through a turbulent plasma, their pulse widths are broadened due to the multi-path scattering (\citealt{Rickett1990}; \citealt{Xu2016}).
The wave structure function, $D_{\phi}$, which represents the mean square phase difference between two points separated by $\sigma_{\rm s}$, is given by (\citealt{Rickett1977}; \citealt{Yang2022})
\begin{equation}\label{func33}
    D_{\phi}=
    \begin{cases}
    f_1\pi^2 r_e^2\lambda^2 C_N^2\Delta D l_0^{\beta-4}\sigma_{\rm s}^2, & \sigma_{\rm s} < l_0, \\
    f_2\pi^2 r_e^2\lambda^2 C_N^2\Delta D \sigma_{\rm s}^{\beta-2}, & \sigma_{\rm s} > l_0,
    \end{cases}
\end{equation}
where $r_{\rm e}$ is the classical electron radius, $\lambda$ is the wavelength of the electromagnetic wave, $\Delta D$ is the straight-line path through the turbulent plasma.
The coefficients are $f_1=\Gamma(1-\alpha/2)$, $f_2=[\Gamma(1-\alpha/2)/\Gamma(1+\alpha/2)]\frac{8}{\alpha 2^{\alpha}}$, and $\alpha=\beta-2$.
The approximation of $\int C_N^2dz\approx C_N^2\Delta D$ is adopted.
The diffractive length, $\sigma_{\rm diff}$, is defined by the condition of $D_{\phi}=1$ rad.
It then can be expressed as (\citealt{Xu2016}; \citealt{Yang2022})
\begin{equation}\label{func34}
    \sigma_{\rm diff} =
    \begin{cases}
    (f_1 \pi^2 r_e^2 \lambda^2 C_N^2 \Delta D l_0^{\beta - 4})^{-\frac{1}{2}}, & \sigma_{\rm diff} < l_0, \\
    (f_2 \pi^2 r_e^2 \lambda^2 C_N^2 \Delta D)^{\frac{1}{2-\beta}}, & \sigma_{\rm diff} > l_0.
\end{cases}
\end{equation}
The angular broadening induced by the multi-path scattering is (\citealt{Rickett1990}; \citealt{Xu2016})
\begin{equation}\label{func35}
    \theta_{\rm sc}=\frac{1}{k\sigma_{\rm diff}}.
\end{equation}
The corresponding temporal broadening is given by (\citealt{Lee1975I})
\begin{equation}\label{func36}
    \tau_{\rm sc}=\frac{D\theta_{\rm sc}^2}{2c},
\end{equation}
where $D$ is the distance between the turbulent plasma and the source in the thin-screen case, while for a thick screen it denotes the propagation path length, and $c$ is the speed of light.
We include the cosmological correction and adopt $\Delta D\sim D$ and $DM\approx\delta n_{\rm e}D$ (\citealt{Xu2016}; \citealt{Yang2022}).
To account for clumpy density structures, we introduce the volume filling factor $f=\langle n_{\rm e}\rangle^2/\langle n_{\rm e}^2\rangle$ and replace $\delta n_{\rm e}$ with $\sqrt{f}~\delta n_{\rm e}$ (\citealt{Xu2016}).

For $\beta<3$, $\tau_{\rm sc}$ can be written as
\begin{subequations}\label{func37}
\begin{align}
\tau_{\rm sc} &= \frac{r_{\rm e}^2 \lambda^4 {\rm DM}^2}{16 c (1+z_0)^3}
                 \frac{f_1 (3 - \beta)}{(2\pi)^{4-\beta}}
                 \left( \frac{\delta n_{\rm e}}{n_{\rm e}} \right)^2 f~ l_0^{-1}, &\sigma_{\rm diff}<l_0,
                 \label{func37a}
\\ %
\tau_{\rm sc} &= \frac{r_{\rm e}^{\frac{4}{\beta-2}}\lambda^{\frac{2\beta}{\beta-2}}{\rm DM}^{\frac{\beta}{\beta-2}}}                                {8c\pi^{\frac{2(\beta-4)}{\beta-2}}(1+z_0)^{\frac{\beta+2}{\beta-2}}}
                 \left(\frac{f_2 (3-\beta)}{2(2\pi)^{4-\beta}} \right)^{\frac{2}{\beta - 2}}
                 \left( \frac{\delta n_{\rm e}}{n_{\rm e}} \right)^{\frac{\beta}{\beta - 2}}
                (\delta n_{\rm e})^{\frac{4 - \beta}{\beta - 2}}f^{\frac{2}{\beta-2}}~
                l_0^{\frac{2(3 - \beta)}{\beta - 2}},
                &\sigma_{\rm diff}>l_0.
                \label{func37b}
\end{align}
\end{subequations}
In Equation (\ref{func37a}), $\beta$ affects only the numerical coefficient.
We therefore fix $\frac{f_1 (3 - \beta)}{(2\pi)^{4-\beta}}$ at its maximum value, 0.04.
By contrast, in Equation (\ref{func37b}), $\beta$ is treated as a free parameter.
For $\beta>3$, $\tau_{\rm sc}$ can be written as
\begin{subequations}\label{func38}
\begin{align}
\tau_{\rm sc} &=
\frac{r_{\rm e}^2 \lambda^4 {\rm DM}^2}{16 c (1+z_0)^3}
\frac{f_1 (\beta - 3)}{(2\pi)^{4-\beta}}
\left( \frac{\delta n_{\rm e}}{n_{\rm e}} \right)^2 f~
\left( \frac{l_0}{L} \right)^{\beta - 4} L^{-1},
&\sigma_{\rm diff}<l_0,
\label{func38a} \\
\tau_{\rm sc} &=
\frac{r_{\rm e}^{\frac{4}{\beta-2}}\lambda^{\frac{2\beta}{\beta-2}}
{\rm DM}^{\frac{\beta}{\beta-2}}}
{8c\pi^{\frac{2(\beta-4)}{\beta-2}}
(1+z_0)^{\frac{\beta+2}{\beta-2}}}
\left(\frac{f_2 (\beta-3)}{2(2\pi)^{4-\beta}} \right)^{\frac{2}{\beta - 2}}
\left( \frac{\delta n_{\rm e}}{n_{\rm e}} \right)^{\frac{\beta}{\beta - 2}}
(\delta n_{\rm e})^{\frac{4 - \beta}{\beta - 2}}f^{\frac{2}{\beta-2}}
L^{\frac{2(3 - \beta)}{\beta - 2}},
&\sigma_{\rm diff}>l_0.
\label{func38b}
\end{align}
\end{subequations}
In this work, for $\beta>3$, we adopt the Kolmogorov spectrum, i.e. $\beta=11/3$.

\citet{Rickett1977} suggested that a power-law spectrum with $\beta \geq 4$ is observationally indistinguishable from a Gaussian spectrum.
In addition, a spectral index of $\beta=4$ is expected when the line of sight passes through an abrupt structure in plasmas (\citealt{Armstrong1995}).
Motivated by these considerations, we also include the Gaussian spectrum in this work.
The corresponding temporal broadening is
\begin{equation}\label{func39}
    \tau_{\rm sc}=\frac{r_{\rm e}^2\lambda^4DM^2}{4c\pi^{1/2}(1+z_0)^3}(\frac{\delta n_{\rm e}}{n_{\rm e}})^2f~l_0^{-1}.
\end{equation}
Its derivation follows essentially the same procedure as that for the power-law spectrum.
A detailed derivation is presented in Appendix~\ref{appe}.

For convenience, we assign simplified labels to the different expressions of $\tau_{\rm sc}$:
Equations (\ref{func37a}), (\ref{func37b}), (\ref{func38a}), (\ref{func38b}), and (\ref{func39}) are denoted as models A, B, C, D, and E, respectively.
In this study, we consider ejecta envelopes with power-law indices $n=2, 4, 6,$ and $8$.
To distinguish models with different $n$, we append the value of $n$ to the model label.
For example, model A with $n=2$ is denoted as A2.
The 20 cases explored in this work, spanning all combinations of the five scattering prescriptions (A--E) and four ejecta density profiles ($n=2,4,6,8$), are summarized in Table~\ref{tablecases}.

Before reaching the Earth, FRBs propagate through several media, including its local plasma,
the host galaxy ISM, the IGM, and the Galactic ISM.
\citet{Yang2022} reported that the scintillation bandwidths, $\delta \nu_{\rm sc}$, of order $\sim 1$ MHz observed in FRBs originate from the Milky Way.
Using the uncertainty relation, $2\pi\tau_{\rm sc}\delta\nu_{\rm sc}\sim 1$, we estimate that the temporal broadening contributed by the Milky Way is of order $10^{-4}$ ms, which is far below 1 ms.
We therefore neglect the Galactic contribution to the observed temporal broadening.
\cite{Xu2016} suggested that a spectral index $\beta>3$ is more likely in the IGM.
For such a plasma to contribute significantly to temporal broadening, the outer scale $L$ would need to be smaller than $10^{-2}$ pc.
However, \citet{Luan2014} argued that, for a Kolmogorov spectrum, the outer scale in the IGM is generally larger than $10^5$ pc.
We thus consider the IGM contribution to be negligible.
The host galaxies of FRB 121102A and FRB 190520B share remarkably similar properties, and bursts from FRB 121102A show no significant scattering tails (\citealt{Li2021}; \citealt{Zhang2023}; \citealt{Wei2025}).
It is therefore reasonable, as a working assumption, to neglect the contribution from the host galaxy of FRB 190520B as well.
Accordingly, we attribute the observed temporal broadening primarily to the local environment of FRB 190520B.
In this study, we model its local environment as a SNR comprising four components, as illustrated in Figure \ref{figsketch}.
Strong turbulence is expected to primarily arise from the shocked ejecta and the shocked ambient medium, driven by Rayleigh-Taylor instabilities (\citealt{Shirkey1978}; \citealt{Fraschetti2010}).
We hence assume that the temporal broadening is produced mainly in the shocked ejecta and shocked ambient medium.
The observed scattering timescale is then treated as the convolution of the contributions from these two components.

We adopt the central frequency of the Five-hundred-meter Aperture Spherical radio Telescope (FAST), $\nu=1.25~{\rm GHz}$, as the observing frequency of bursts from FRB 190520B.
The DM contributions from the shocked ejecta and the shocked ambient medium, which are used separately in the calculation of $\tau_{\rm sc}$, are given by Equations (\ref{func28}) and (\ref{func29}), respectively.
According to their definitions, the rms density fluctuation $\delta n_{\rm e}$, the electron number density $n_{\rm e}$, and the volume filling factor $f$ can be derived from Equations (\ref{func26}) and (\ref{func27}) .
Following \cite{Coles1989}, the inner scale $l_0$ for each model is determined by
\begin{equation}\label{func40}
    l_0=3\sqrt{\frac{m_p}{4\pi}}\frac{c}{e}n_e^{-1/2},
\end{equation}
where $e$ is the elementary charge.
The quantities $n_{\rm e}$, $\delta n_{\rm e}$, and $l_0$ are all functions of both time and radius.
Since we are interested in their temporal evolution, at each time slice we radially average the composite terms constructed from these quantities, as shown in scattering formalism.
The outer scale $L$ is obtained from a linear fit of simulation data and is given by (\citealt{Dwarkadas2000})
\begin{equation}\label{func41}
    L=0.037~t~R_{\rm ch}/t_{\rm ch},
\end{equation}
where $R_{\rm ch}$ and $t_{\rm ch}$ are the characteristic length and time defined in Section \ref{sub2.1}.

It is worth noting that ISM-like environments essentially satisfy the three assumptions underlying the standard theory of temporal broadening (\citealt{Rickett1977}):
(i) the deviations of refractive index are much less than unity everywhere;
(ii) the inner scale $l_0$ is much larger than the wavelength of electromagnetic wave, $\lambda$;
and (iii) the outer scale $L$ is much less than the thickness of medium.
In this work, we extend these assumptions to the SNR environment.
As shown in Section \ref{discussion}, our results suggest that, within the parameter space constrained in this work, these assumptions are satisfied in the SNR environment of FRB 190520B, thus validating the results derived from the scattering theory adopted here.

\section{Bayesian Inference}
\label{MCMC}

Based on the theoretical framework of SNR evolution and the temporal broadening models introduced in Sections \ref{SNR} and \ref{scatter}, we now adopt a Bayesian approach to constrain the physical parameters of the SNR of FRB 190520B and its host galaxy, using the observed DMs and scattering timescales.

The DM measurements are compiled from \cite{Niu2022,NiuCH2026} and \cite{anna2023}, yielding a total of 646 data points, while the observed scattering timescales are taken from \citet{Niu2022,NiuCH2026}, comprising 95 measurements.
\citet{NiuCH2026} reported scattering timescales in three frequency bands centered at 1.075, 1.225, and 1.4 GHz.
We take these central frequencies as the observing frequencies of the corresponding bursts and rescale the reported scattering timescales to the frequency of 1.25 GHz using the scaling law $\tau_{\rm sc}\propto \nu^{-4}$.
The rescaled scattering timescales from the three bands are then convolved to obtain the final scattering timescale for each burst reported in \citet{NiuCH2026}.

We consider a wind environment with $s=2$ and ejecta envelopes characterized by power-law indices $n = 2, 4, 6,$ or $8$, which determine the dynamical evolution of density profiles, shell radii, and DM contributions from different SNR regions.
The free parameters in our model are the ejecta mass $M_{\rm ej}$, kinetic energy $E_{\rm k}$, progenitor mass-loss rate $\dot{M}_{\rm w}$, the source age $t_0$, and the DM contribution from host galaxy $DM_{\rm host}$.
The wind velocity is fixed at $v_{\rm w}=100~{\rm km~s^{-1}}$.
For model B, the spectral index $\beta$ is treated as an additional free parameter.
The parameter $t_0$ is defined as the source age, which also means at that time, the first observed burst from FRB 190520B was generated.
Using the arrival time of individual bursts, the generated time of all observed bursts can be determined relative to $t_0$.
Observationally, $DM_{\rm host}$ and $DM_{\rm local}$ are coupled.
By treating $DM_{\rm host}$ as a free parameter, $DM_{\rm local}$ can be inferred from the observed $DM_{\rm obs}$ after subtracting the other contributions.
Its uncertainty is obtained through standard error propagation,
\begin{equation}\label{func42}
    \sigma_{\rm local}^2=(\sigma_{\rm obs}^2+\sigma_{\rm MW}^2+\sigma_{\rm halo}^2+\sigma_{\rm IGM}^2)\times(1+z_0)^2,
\end{equation}
where $\sigma_{\rm MW}$, $\sigma_{\rm halo}$, and $\sigma_{\rm IGM}$ are the uncertainties associated with the Milky Way, Galactic halo, and IGM contributions, respectively (see Section \ref{sub2.4}), and $\sigma_{\rm obs}$ is the uncertainty of the observed DM presented in \citet{Niu2022,NiuCH2026} and \citet{anna2023}.
The factor $(1+z_0)^2$ accounts for the redshift correction when converting the uncertainties of the local DM contribution to the source frame.

We employ \texttt{scipy.optimize.minimize}$\footnote{https://docs.scipy.org/doc/scipy/reference/generated/scipy.optimize.minimize.html}$ at first to obtain reasonable initial
values for the free parameters. With these initial
values, we then perform MCMC fitting to $DM_{\rm local}$ and
scattering timescale data by using \texttt{emcee}$\footnote{https://emcee.readthedocs.io}$. 
In this work, we used the inferred $DM_{\rm local}$ in MCMC procedure, whereas
$DM_{\rm obs}$ is adopted in subsequent discussion. The
logarithmic likelihood functions for $DM_{\rm local}$ and
$\tau_{\rm sc}$ adopted in MCMC are
\begin{subequations}\label{func43}
    \begin{align}
{\rm log}\mathcal{L}_{\rm DM_{\rm local}} &= -\frac{1}{2}\sum_i\left [(\frac{DM_{{\rm local},i}-\hat{DM}_{{\rm local},i}}{\sigma_{DM_{{\rm local},i}}})^2 + {\rm ln}(2\pi\sigma_{DM_{{\rm local},i}}^2) \right],\label{func43a}
\\%
{\rm log}\mathcal{L}_{\rm \tau_{\rm sc}} &= -\frac{1}{2}\sum_i\left [(\frac{\tau_{{\rm sc},i}-\hat{\tau}_{{\rm sc},i}}{\sigma_{\tau_{{\rm sc},i}}})^2 + {\rm ln}(2\pi\sigma_{\tau_{{\rm sc},i}}^2) \right],\label{func43b}
    \end{align}
\end{subequations}
where $DM_{{\rm local},i}$ and $\tau_{{\rm sc},i}$ are the inferred local DM and the observed scattering timescale for the $i$th data point, respectively, and $\sigma_{DM_{{\rm local},i}}$ and $\sigma_{\tau_{{\rm sc},i}}$ are their corresponding uncertainties.
The quantities with hats denote the values predicted by theories.
The total likelihood is taken as the sum of the two logarithmic likelihoods.

In this work, the transition time from the ED phase to the ST phase, $t_{\rm tran}$, is required to be smaller than $10^4$ yr.
The prior ranges for $M_{\rm ej}$, $E_{\rm k}$, and $\dot{M}_{\rm w}$ are set to $1- 20~M_{\odot}$, $10^{50}-10^{52}$ erg, and $10^{-8}-10^{-3}~M_\odot~{\rm yr^{-1}}$, respectively (\citealt{Smartt2009}; \citealt{Smith2014}).
Some previous studies suggested that $t_0$ for FRB 190520B is of the order of a few decades, while \cite{Rahaman2025} argued that the persistent radio source of FRB 190520B has an age of $10-100$ years (\citealt{WangFY2025}; \citealt{Bhattacharya2025}).
We therefore adopt a broader prior range of $10$--$200$ yr for $t_0$.
The prior range for $DM_{\rm host}$ is set to $10-1000~{\rm pc~cm^{-3}}$, estimated by subtracting other contributions from the minimum observed DM in the source frame.
For model B, the additional free parameter $\beta$ is restricted to the range from $2+{\rm e}^{-1}$ to $3$.
The lower limit is set as $2+{\rm e}^{-1}$ to ensure it is sufficiently small and to avoid numerical instabilities in the calculations.
Since we cannot constrain the entire evolution of SNR in the ED phase by observations spanning several years, we use the MCMC method to find the best-fit of free parameters.
For the MCMC sampling, thereby, we run 10,000 steps with 48 walkers.
No burn-in is considered, and thinning factor is set to 1.

\section{Numerical Results}
\label{result}

\begin{figure}
    \centering
    \includegraphics[width=0.33\linewidth]{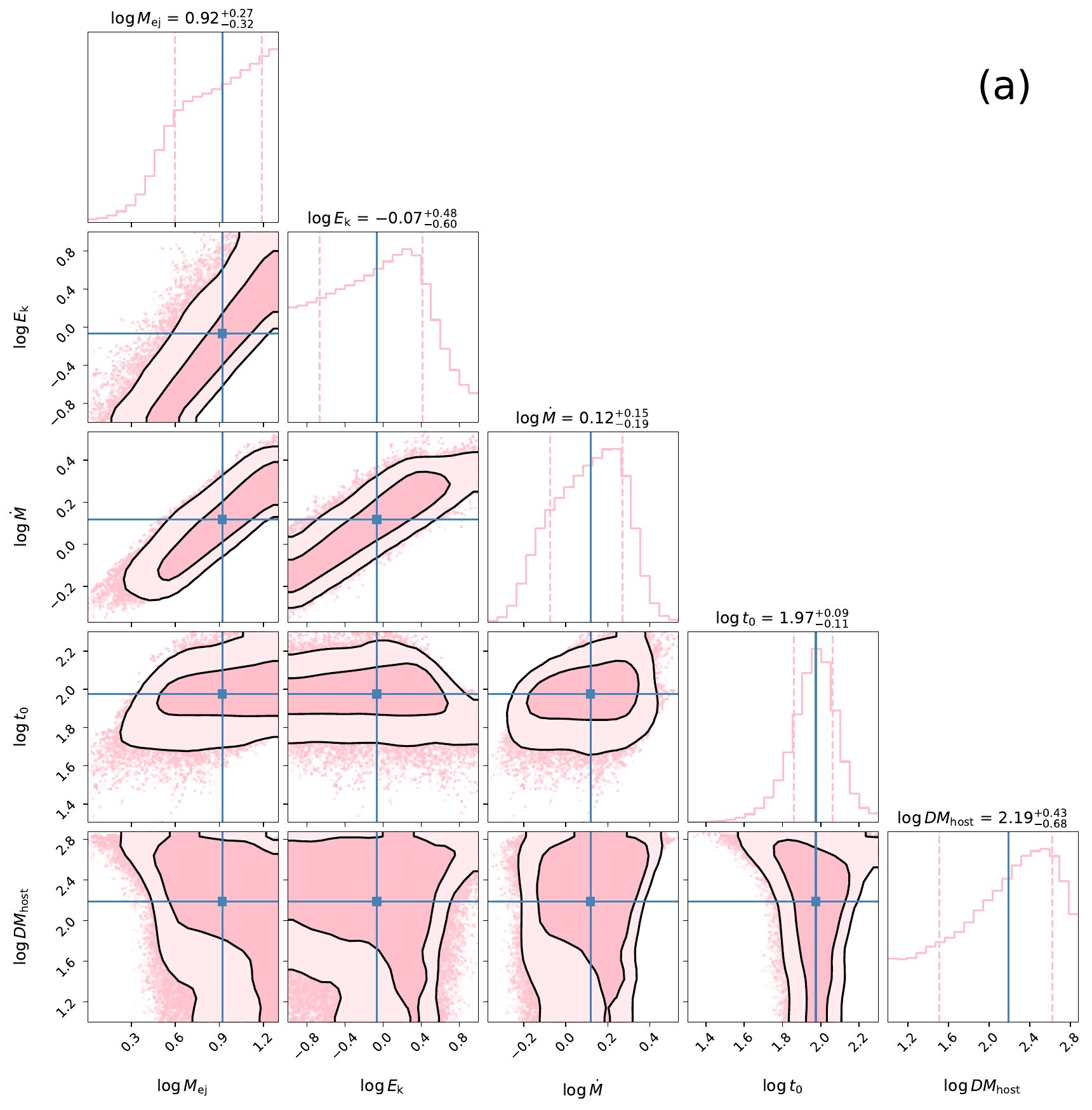}\hfill
    \includegraphics[width=0.33\linewidth]{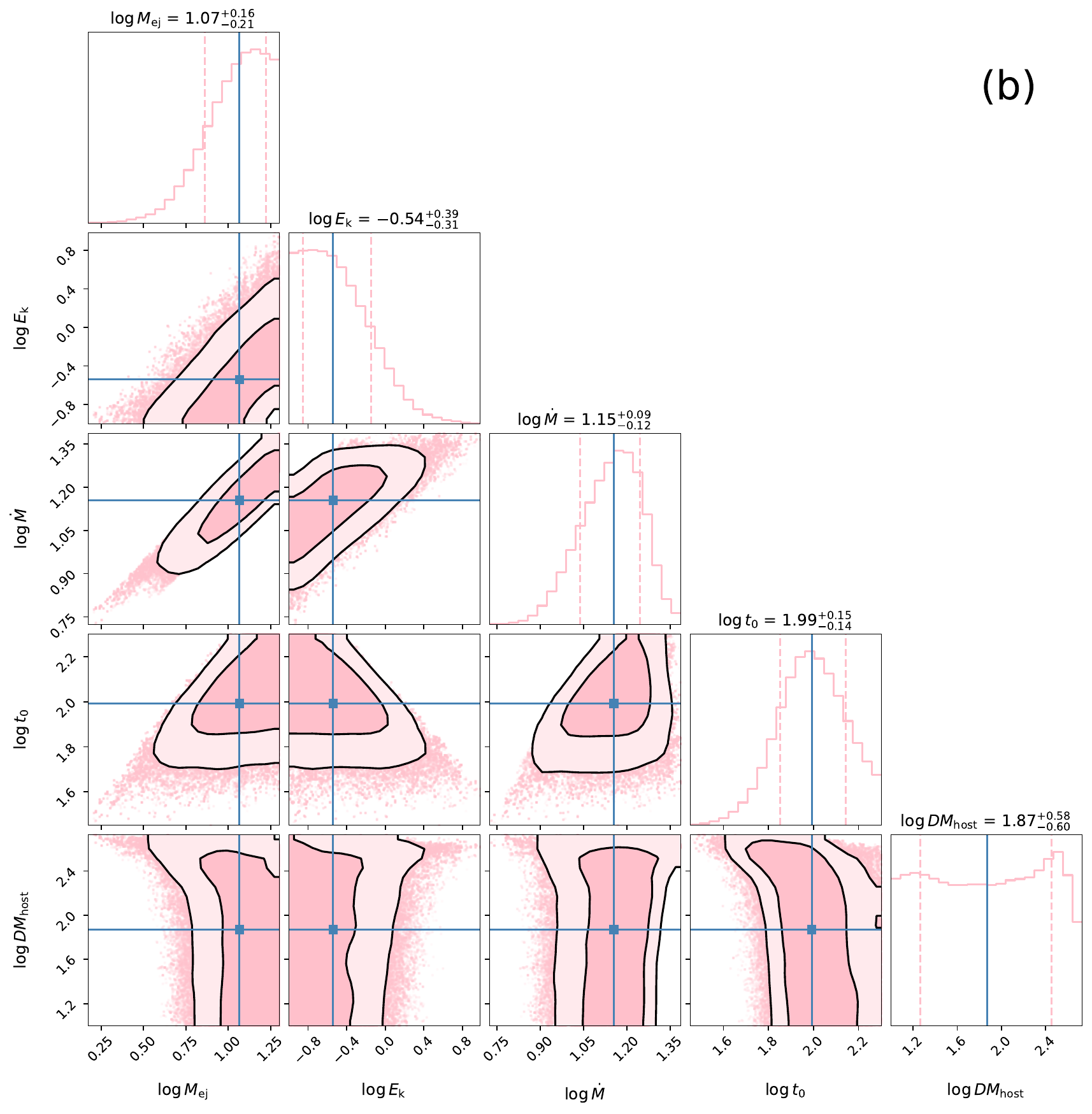}\hfill
    \includegraphics[width=0.33\linewidth]{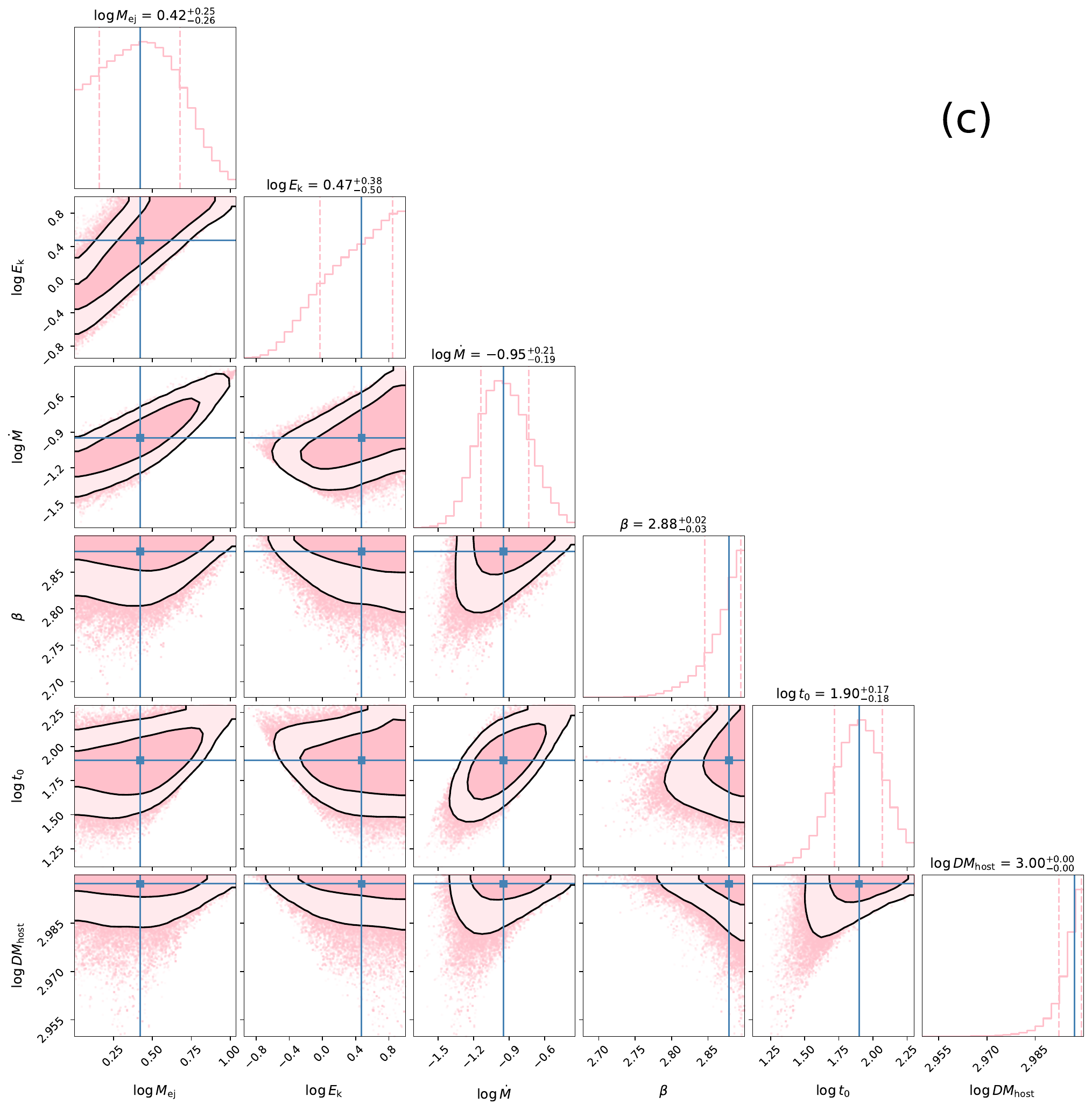}\hfill
    \includegraphics[width=0.33\linewidth]{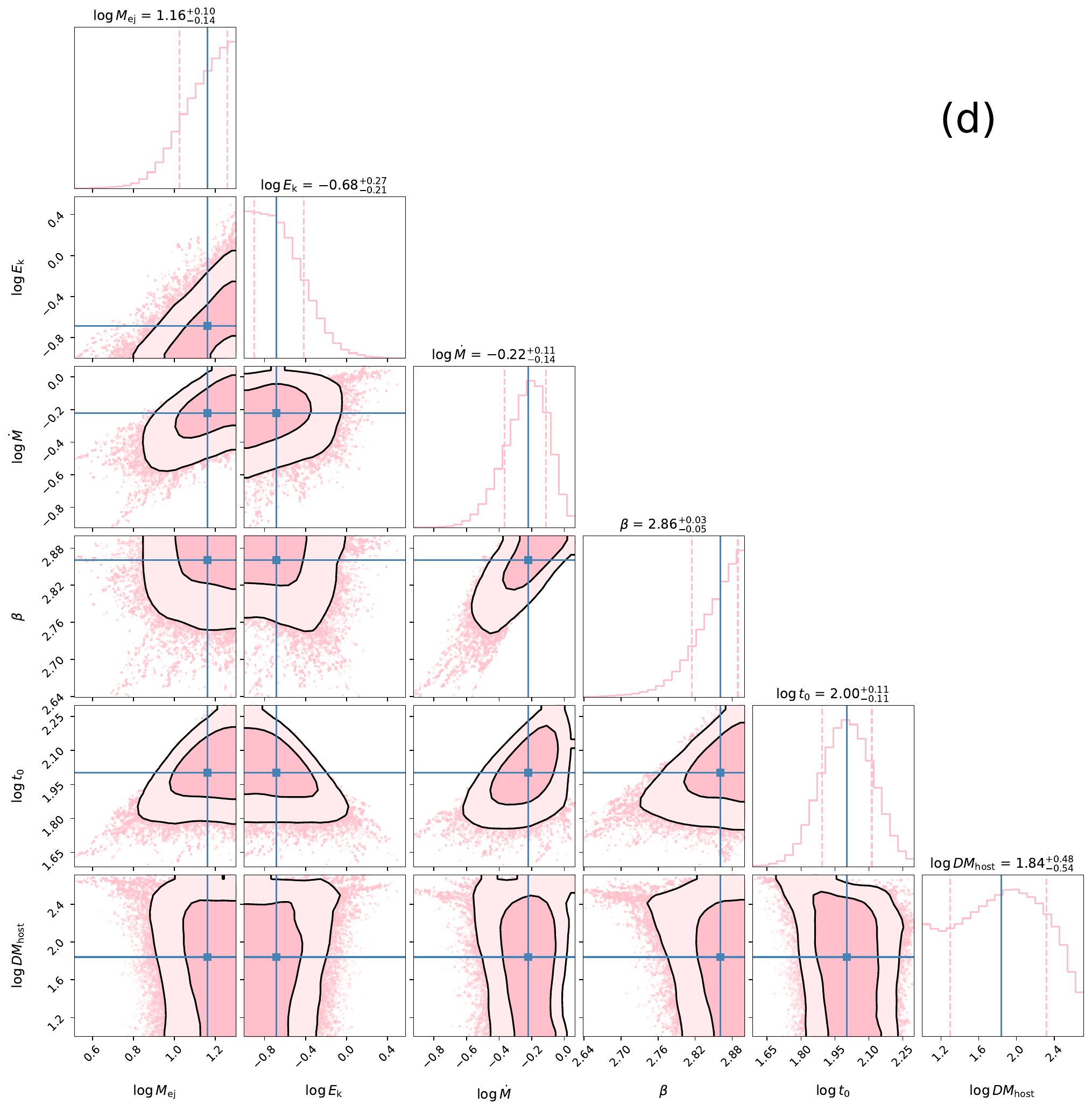}\hfill
    \includegraphics[width=0.33\linewidth]{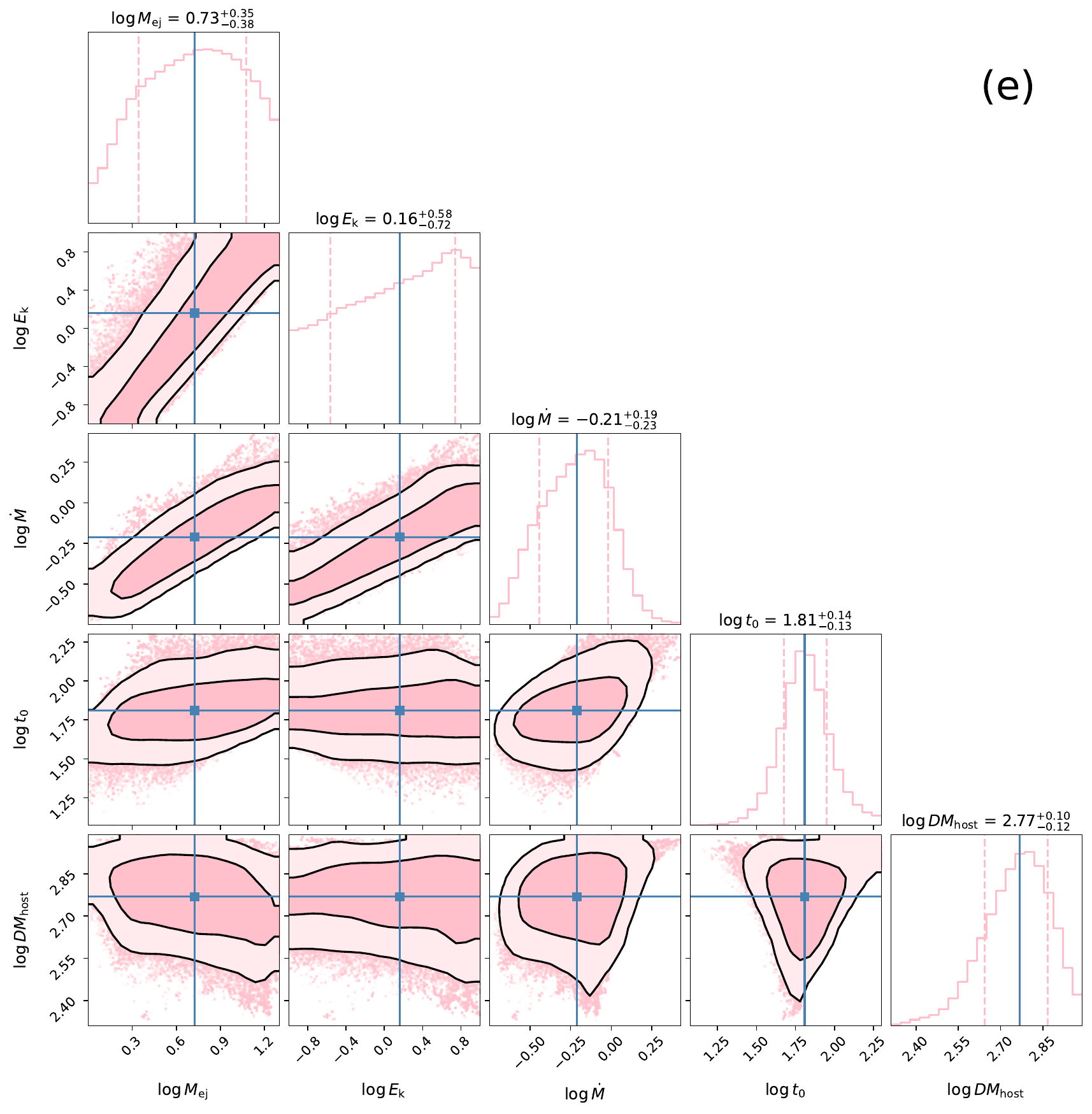}\hfill
    \includegraphics[width=0.33\linewidth]{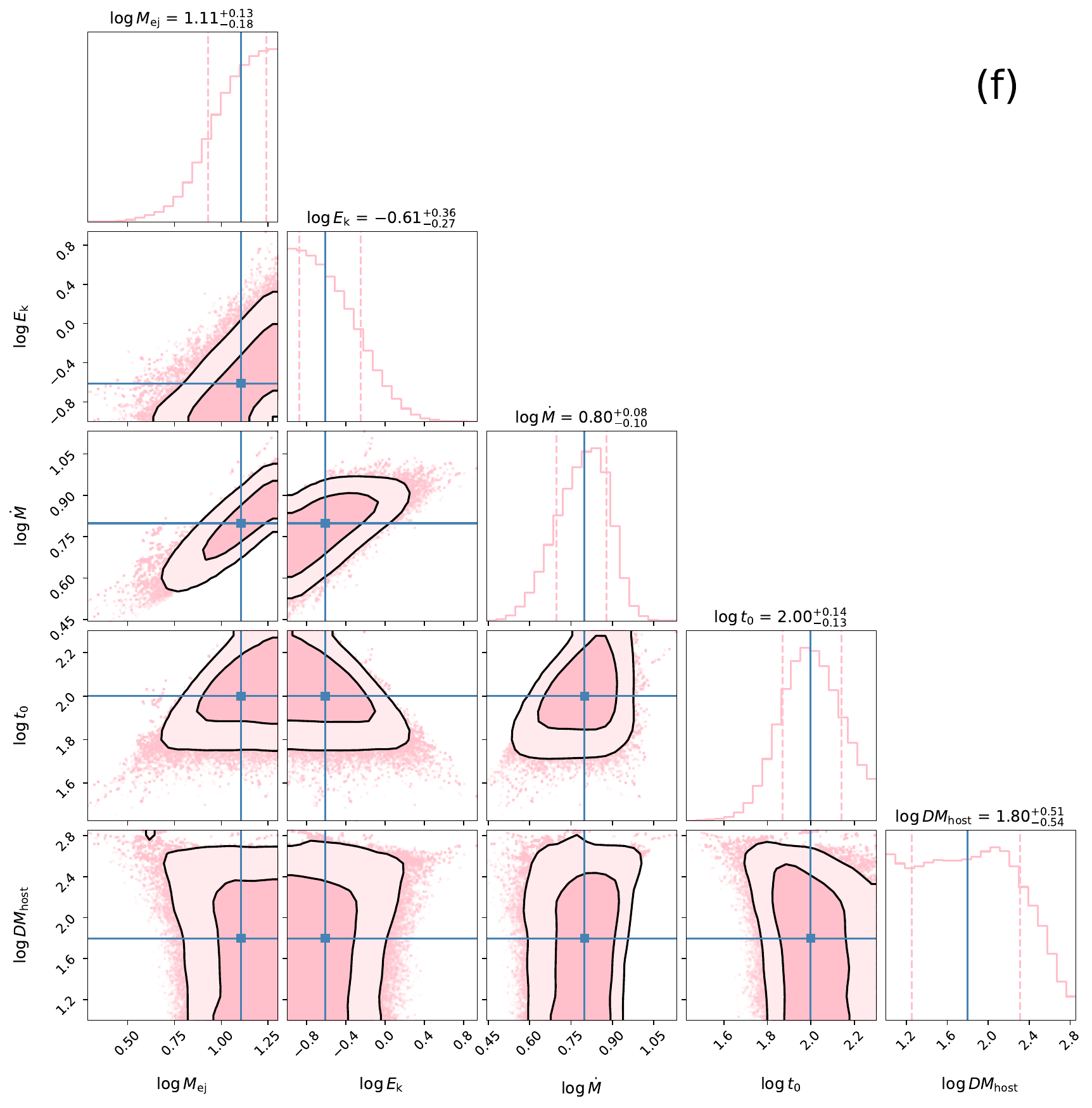}\hfill
    \caption{
  Corner plots of the posterior distributions for all free parameters.
  Panels (a)--(f) correspond to cases A2, A4, B2, B4, E2, and E4, respectively.
  The pink shaded regions mark the $2\sigma$ confidence contours.
  The blue crosses indicate the median values.
  The pink vertical dashed lines in the 1D histograms denote the 16th and 84th percentiles of the samples.
}
    \label{figcorner}
\end{figure}

FRB 190520B is an intriguing repeating source, characterized by its large DMs, extreme and time-variable RM with two sign reversals, and its association with a PRS, all of which point to a complex and evolving local environment.
In this section, we present the results obtained from the joint analysis of the DM and scattering timescale data of FRB 190520B.
Combining the five prescriptions for temporal broadening with ejecta profiles of $n=2,4,6,$ and $8$ yields 20 cases in total.
Among them, we found that only six cases provide acceptable fits to both the DM and scattering timescale data, namely cases A2, A4, B2, B4, E2, and E4.
Notably, all the six cases belong to the FE solution with $n=2$ or $4$, whereas none of the SSDW solution with $n=6$ or $8$ is retained.
This trend suggests that the current data favor shallower ejecta-density profiles.
However steeper profiles with $n>8$ cannot be ruled out completely.
The best-fit parameters and median values with 68\% credible intervals for the six cases, together with the inferred transition times $t_{\rm tran}$, are listed in Table~\ref{tablefit}. 
The posterior distributions of the free parameters are shown in Figure~\ref{figcorner}.

The derived parameter values span a broad range.
The ejecta masses are $M_{\rm ej}=8.7$--$19.2~M_\odot$, while the kinetic energies range from $1\times10^{50}~{\rm erg}$ to $7.3\times10^{51}~{\rm erg}$.
The inferred wind mass-loss rates are $\dot{M}_{\rm w}=0.3$--$14.8\times10^{-5}~M_\odot~{\rm yr^{-1}}$, and the source ages are $t_0=79.8$--$169.8~{\rm yr}$.
Thus, all cases imply that FRB 190520B is embedded in a young, approximately century-old SNR environment.
The DM contributions from host galaxy differ substantially among the cases.
Cases A2, A4, B4, and E4 require extremely small host contributions, $DM_{\rm host}=10.7$--$28.7~{\rm pc~cm^{-3}}$, whereas cases B2 and E2 require larger values of $998.0$ and $407.6~{\rm pc~cm^{-3}}$, respectively.
Cases B2 and B4 give the same best-fit turbulence spectral index, $\beta=2.9$.

\begin{figure}
    \centering
    \includegraphics[width=0.33\linewidth]{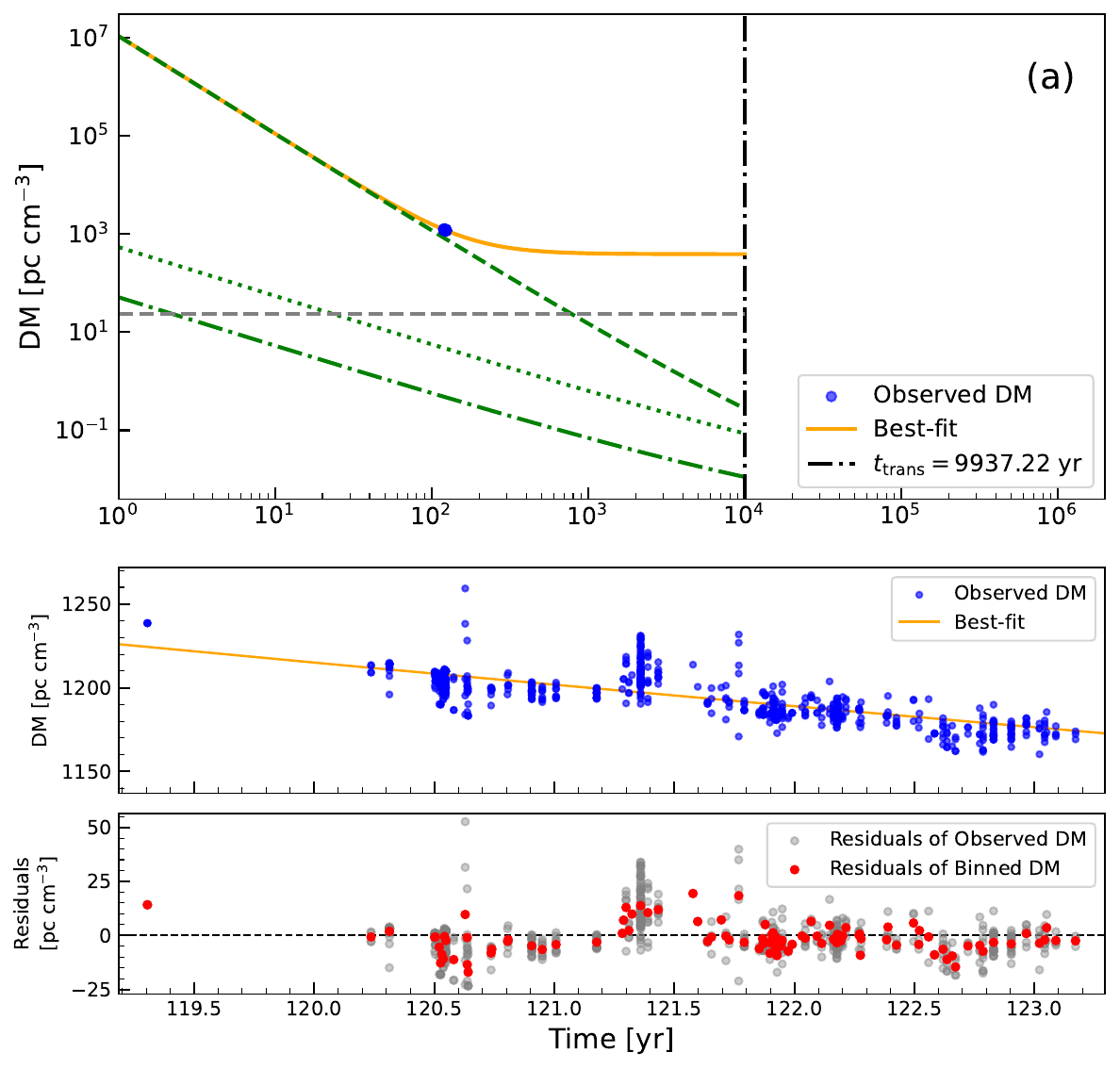}\hfill
    \includegraphics[width=0.33\linewidth]{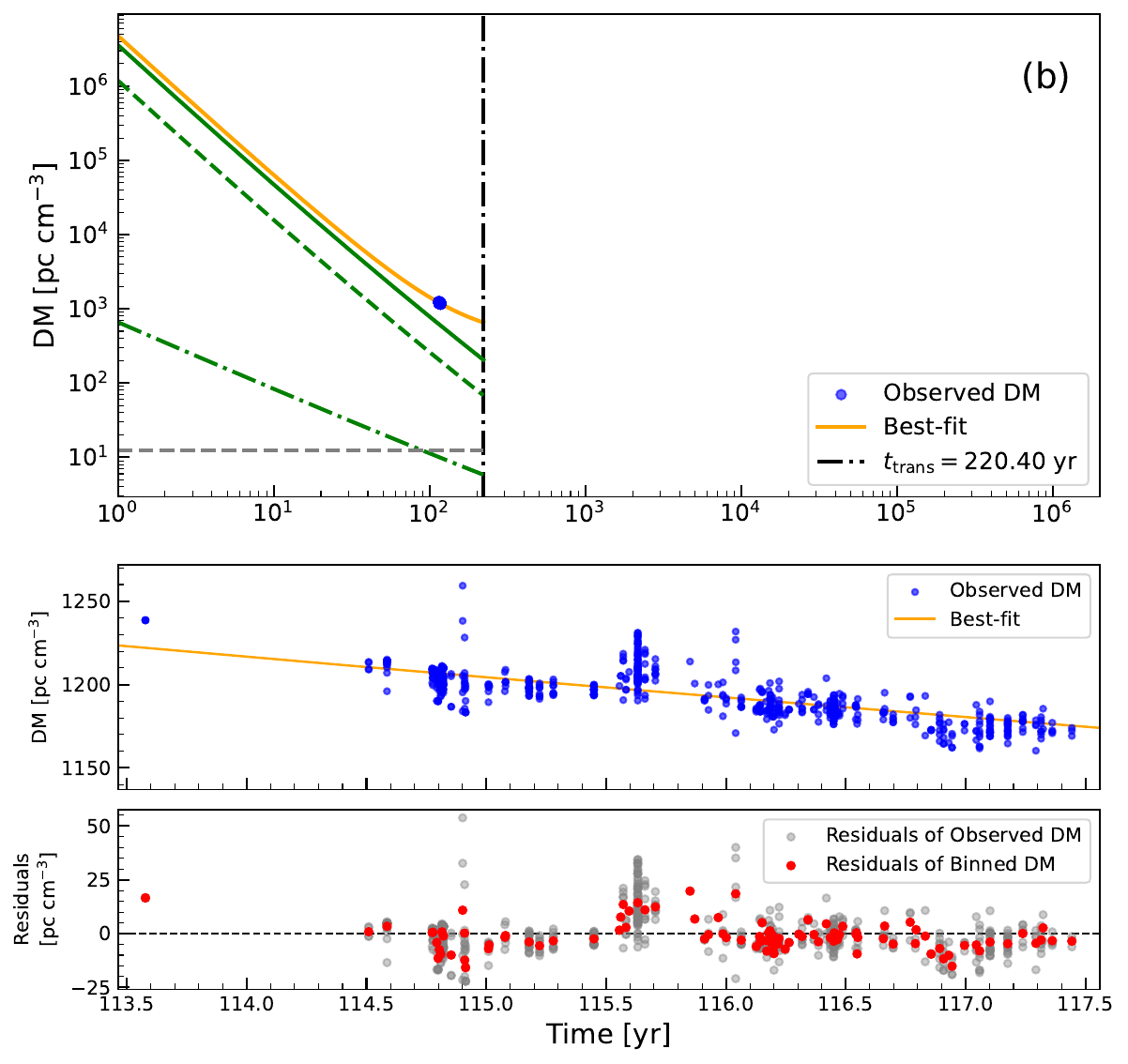}\hfill
    \includegraphics[width=0.33\linewidth]{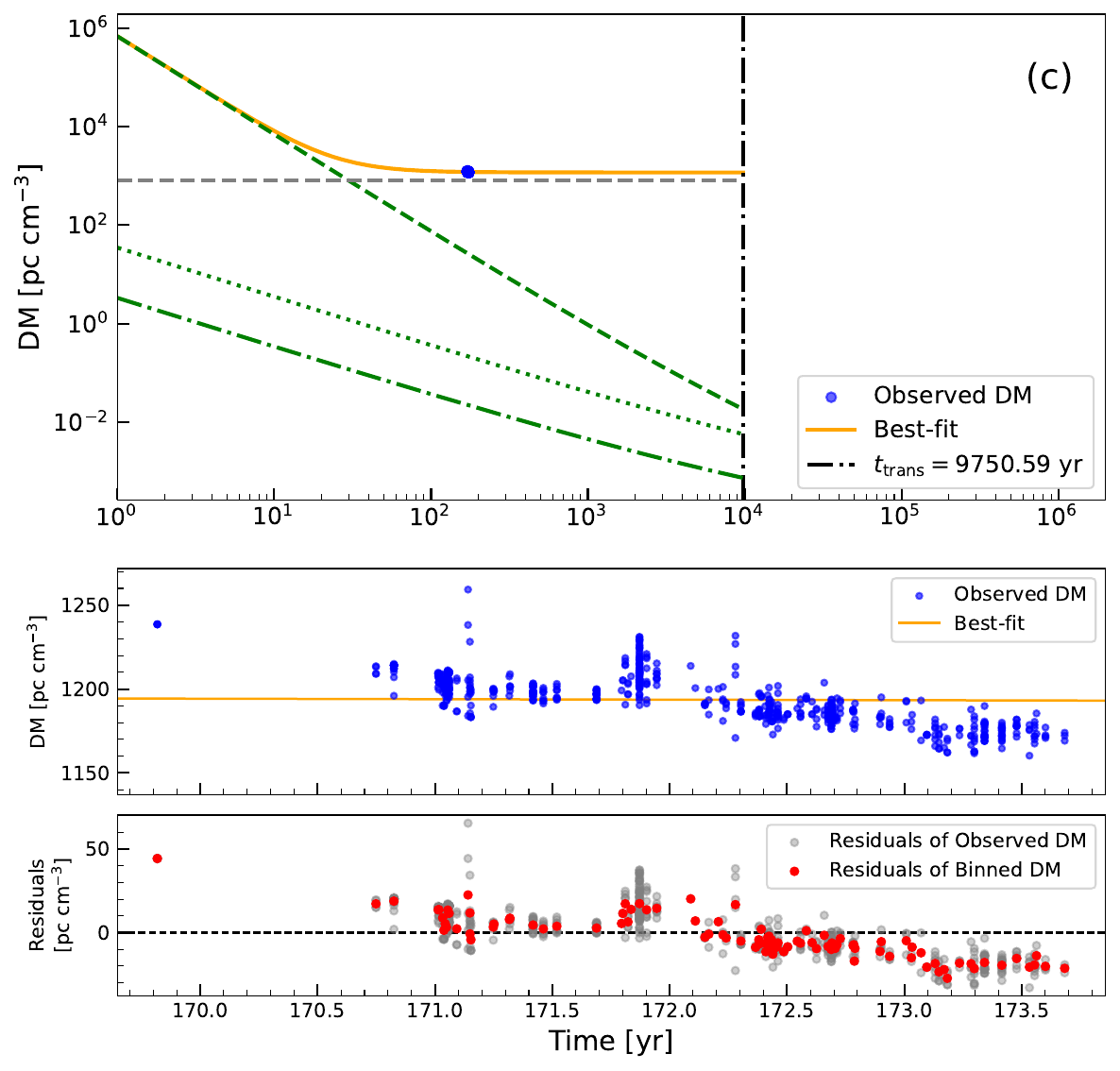}\hfill
    \includegraphics[width=0.33\linewidth]{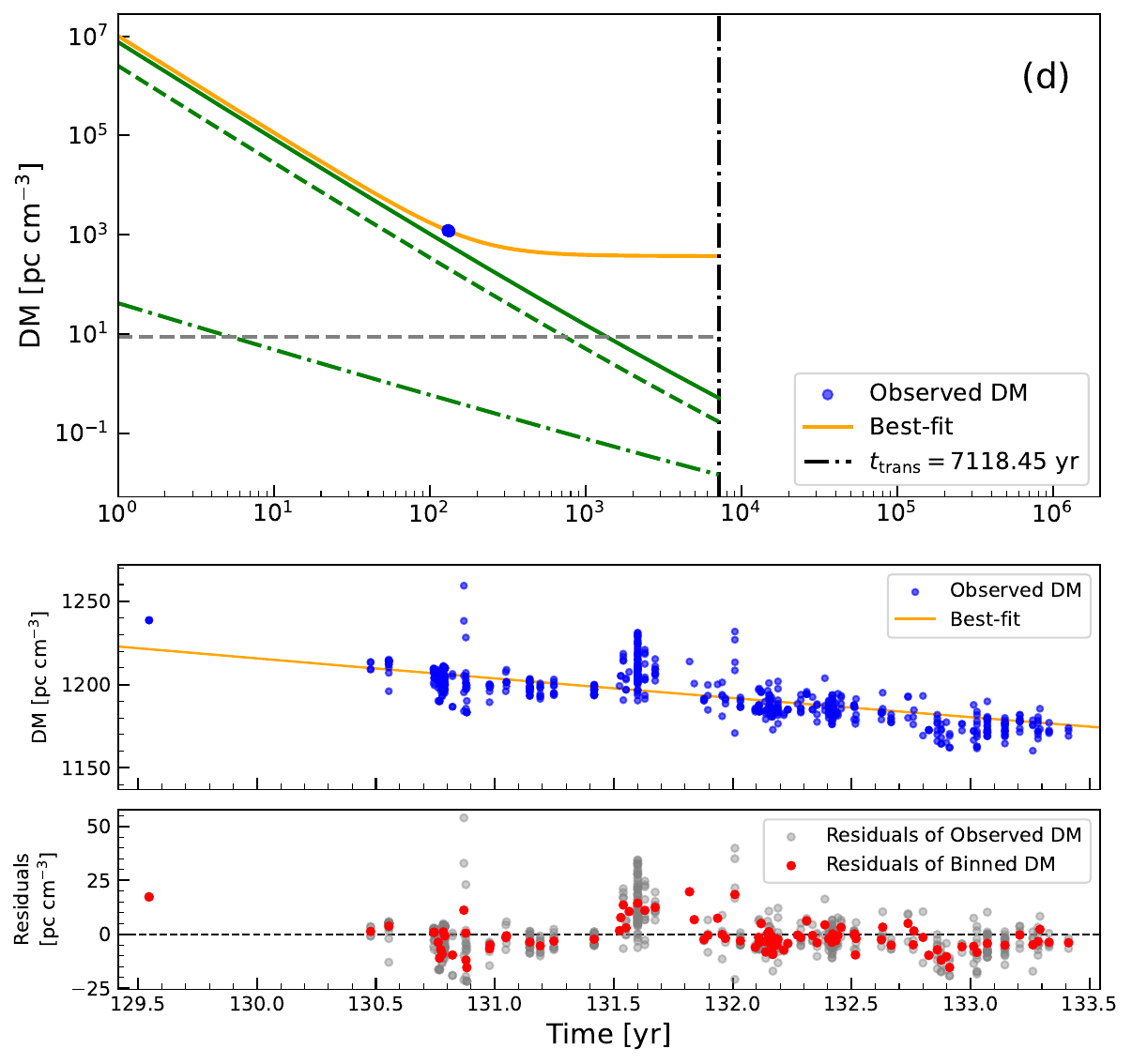}\hfill
    \includegraphics[width=0.33\linewidth]{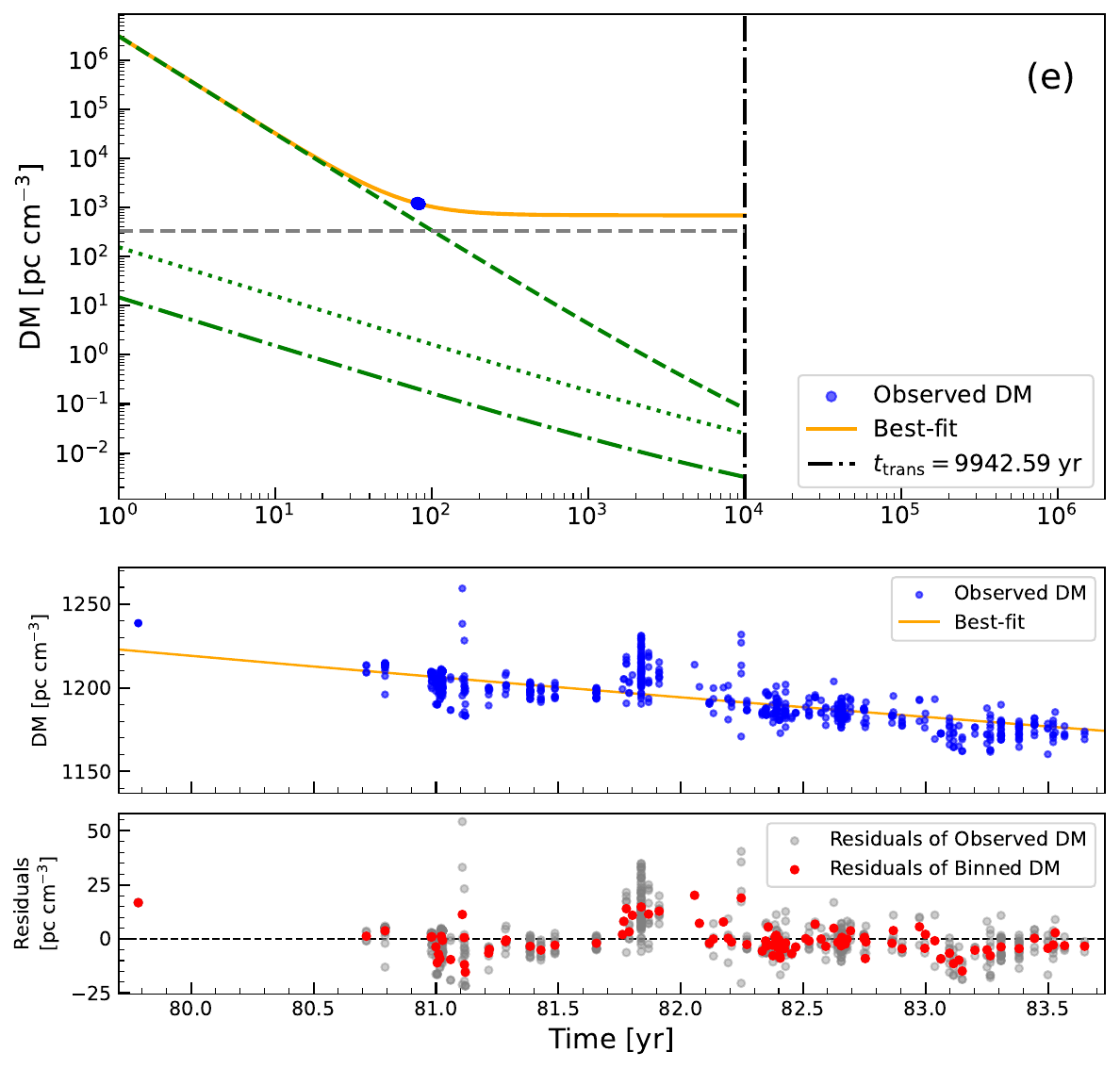}\hfill
    \includegraphics[width=0.33\linewidth]{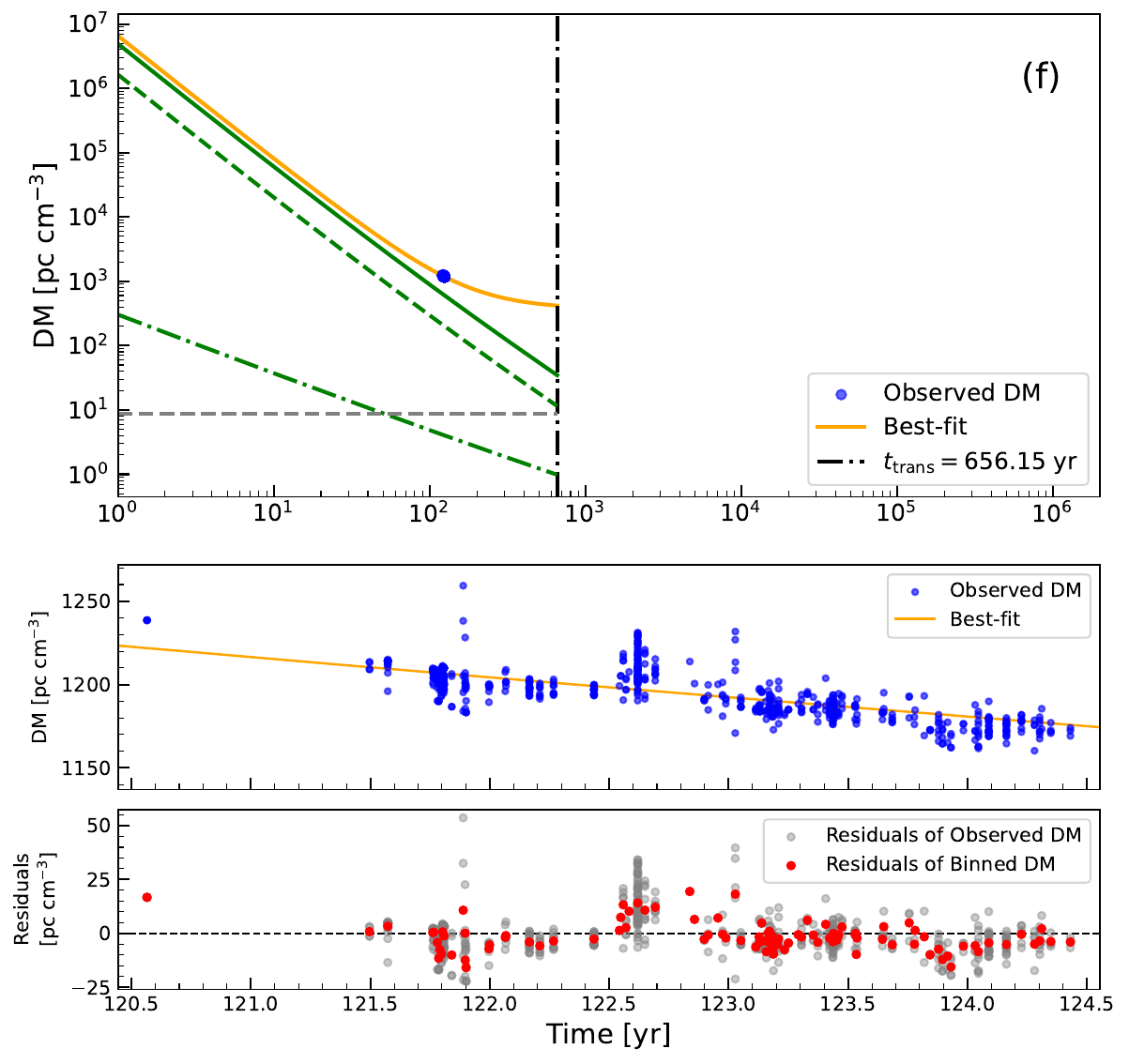}\hfill
    \caption{
Evolution of the DM and the corresponding residuals. Panels (a), (b), (c), (d), (e), and (f) correspond to cases A2, A4, B2, B4,
E2, and E4, respectively. 
In the upper panels, the orange lines show the best-fit DM evolution, and the blue dots are the observed DM. 
The horizontal gray dashed lines are $DM_{\rm host}$, while the green solid, dashed, dotted, and dashdotted lines represent the DM contributions from the unshocked core, unshocked ejecta, shocked ejecta, and shocked ambient medium, respectively. 
The vertical black dash-dotted lines represent the transition times. 
The middle panels show zoomed-in views of the observed DM together with the best-fit curves.
In the lower panels, the gray and red dots are the residuals of observed DM and binned DM, respectively. 
The horizontal black dashed lines indicate the zero residual.
    }
    \label{figDM}
\end{figure}

The corresponding fits to the observed DMs and their residuals are shown in Figure \ref{figDM}.
One can see that, during the burst epoch, the DM evolution becomes nearly stable for B2, indicating that the observed DMs are primarily dominated by the contribution from the host galaxy.
By contrast, other cases of A2, A4, B4, E2 and E4 still exhibit an obvious decreasing trend in DM at their burst epoch.
\cite{WangFY2025} reported a DM decrease of approximately $11~{\rm pc~cm^{-3}~yr^{-1}}$ with a significance above $10\sigma$, while \cite{NiuCH2026} obtained a decrease rate of $12.4\pm0.3~{\rm pc~cm^{-3}~yr^{-1}}$.
The DM slopes predicted by cases A2, A4, B2, B4, E2, and E4 are $-13.2$, $-12.3$, $-0.3$, $-12.0$, $-12.3$, and $-12.1~{\rm pc~cm^{-3}~yr^{-1}}$, respectively.
Except for B2, the predicted slopes are broadly consistent with the observed DM decrease reported by \cite{NiuCH2026}.

To illustrate the physical origin of the DM evolution, the upper panels of Figure~\ref{figDM} display the contributions from different SNR regions with distinct green line styles.
As shown in Section \ref{SNR}, the presence of the ejecta core and the shocked ejecta is subjective to the value of ejecta power-index $n$.
For the $n=2$ cases, A2, B2, and E2, the ejecta core is absent, as expected for $n<3$.
For the $n=4$ cases, A4, B4, and E4, the shocked ejecta contribution is absent because the corresponding electron density vanishes for this density index.
The DM components show similar evolutionary behavior among the $n=2$ cases, and also among the $n=4$ cases.
In A2, B2, and E2, the unshocked ejecta dominates the local DM at early times, whereas the contributions from the shocked ejecta and the shocked ambient medium remain negligible throughout the evolution.
At later times, the host-galaxy contribution becomes dominant in B2 and E2.
By contrast, in A2, where $DM_{\rm host}$ is only several tens of ${\rm pc~cm^{-3}}$, the intervening IGM and the Milky Way make important contributions to the total observed DM budget.
In A4, B4, and E4, both the unshocked core and the unshocked ejecta make significant contributions to the local DM, while the shocked ambient medium remains not important.
In addition, the host-galaxy DM contributions in these three cases are all of order $10~{\rm pc~cm^{-3}}$.
It is worth noting that, even for a low ionization fraction of $\eta\sim0.03$, the two unshocked regions still provide major contributions to the local DM in all six cases and dominate the early-time local DM.
This result is consistent with the simulations of \cite{ZhangZ2026}.

Several previous studies have also investigated the DM evolution of an FRB embedded in an SNR, albeit with more simplified treatments of the density structure.
\cite{piro2016} and \cite{piro2018} attributed the SNR DM solely to the shocked ejecta, assuming a uniform density within this region, and obtained a scaling relation $DM_{\rm SNR}\propto t^{-3/2}$ for the ED phase.
\cite{YangYP2017} and \cite{NiuCH2026} did not distinguish between different SNR components and adopted a uniform density throughout the entire remnant, finding $DM_{\rm SNR}\propto t^{-2}$ for the same phase.
In contrast, our model resolves the SNR into four distinct regions---unshocked core, unshocked ejecta, shocked ejecta, and shocked ambient medium---each governed by a distinct density profile.
The resulting time-dependent local DM during the ED phase, in our work, follows $DM_{\rm local}\propto t^{\gamma}$, with fitted power-law indices $\gamma = -1.9, -1.8, -1.9, -1.9, -1.9, -1.8$ for cases A2, A4, B2, B4, E2, and E4, respectively.
All six retained cases thus yield $\gamma$ between $-1.8$ and $-1.9$.
This deviation from the earlier scaling laws arises because the unshocked and shocked regions contribute with different temporal indices in our model: for $n<5$, an unshocked region evolves as $\gamma=-2$, while a shocked region follows $\gamma=-1$.
The net index of the total local DM is determined by the relative weights of the unshocked and shocked contributions.
Since the local DM is dominated by the unshocked regions, as discussed above, it is natural for the net $\gamma$ to deviate slightly from $-2$.

The vertical dash-dotted black lines in Figure \ref{figDM} indicate
the transition times from the ED phase to the ST phase, which are
derived from the best-fit values of $M_{\rm ej}$, $E_{\rm k}$, and
$\dot{M}_{\rm w}$, together with the fixed wind velocity $v_{\rm
w}=100~{\rm km~s^{-1}}$. The inferred transition times for cases
A2, B2, B4, and E2 are on the order of $10^{3}$--$10^{4}$ yr,
whereas those for cases A4 and E4 are only several hundred years.
Observationally, the transition from the ED phase to the ST phase
is generally expected to occur on timescales of hundreds to
thousands of years (\citealt{Jacco2020}). A relevant example is
the Crab Nebula, which is still broadly in a pre-Sedov,
ejecta-dominated evolutionary state nearly one millennium after SN
1054 (\citealt{Trimble1968}). Therefore, the transition times
inferred for all the cases are not excluded by the expected
evolutionary timescale of young SNRs. Nevertheless, although all
the cases correspond to young systems with $t_0\lesssim170~{\rm
yr}$, they imply very different evolutionary stages relative to
the ED--ST transition. In particular, A4 has already reached about
half of its inferred ED lifetime, and E4 has completed about one
fifth of its inferred ED lifetime. By contrast, A2, B2, B4, and E2
are still at less than about $2\%$ of their inferred ED lifetimes,
indicating that these systems are observed at a much earlier stage
of the ED phase.

The middle and lower panels of Figure \ref{figDM} present zoomed-in views of the observed DM data and show the residuals between the observations and the best-fit model predictions, respectively.
To make the residual distributions clearer, the bursts observed on the same day are binned together and averaged. 
For cases A2, A4, B4, E2, and E4,
the residuals of the binned DMs remain relatively close to zero.
By contrast, B2 exhibits the clearest systematic drift from
positive residuals at earlier epochs to negative residuals at
later epochs. This pattern is consistent with its nearly flat DM
evolution at burst epoch and indicates that B2 does not reproduce
the observed secular DM decrease as well as the other accepted
cases.

\begin{figure}
    \centering
    \includegraphics[width=0.33\linewidth]{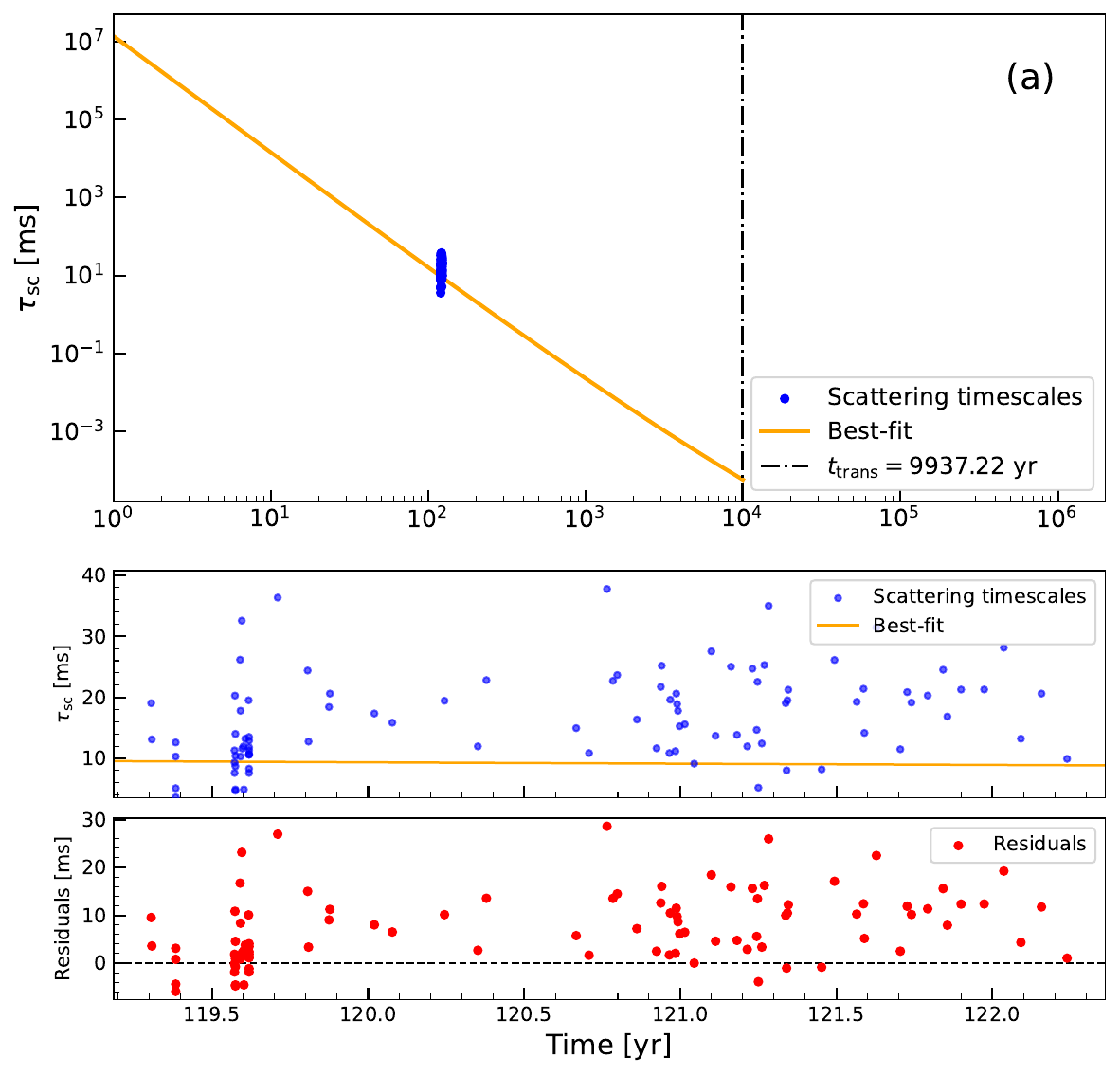}\hfill
    \includegraphics[width=0.33\linewidth]{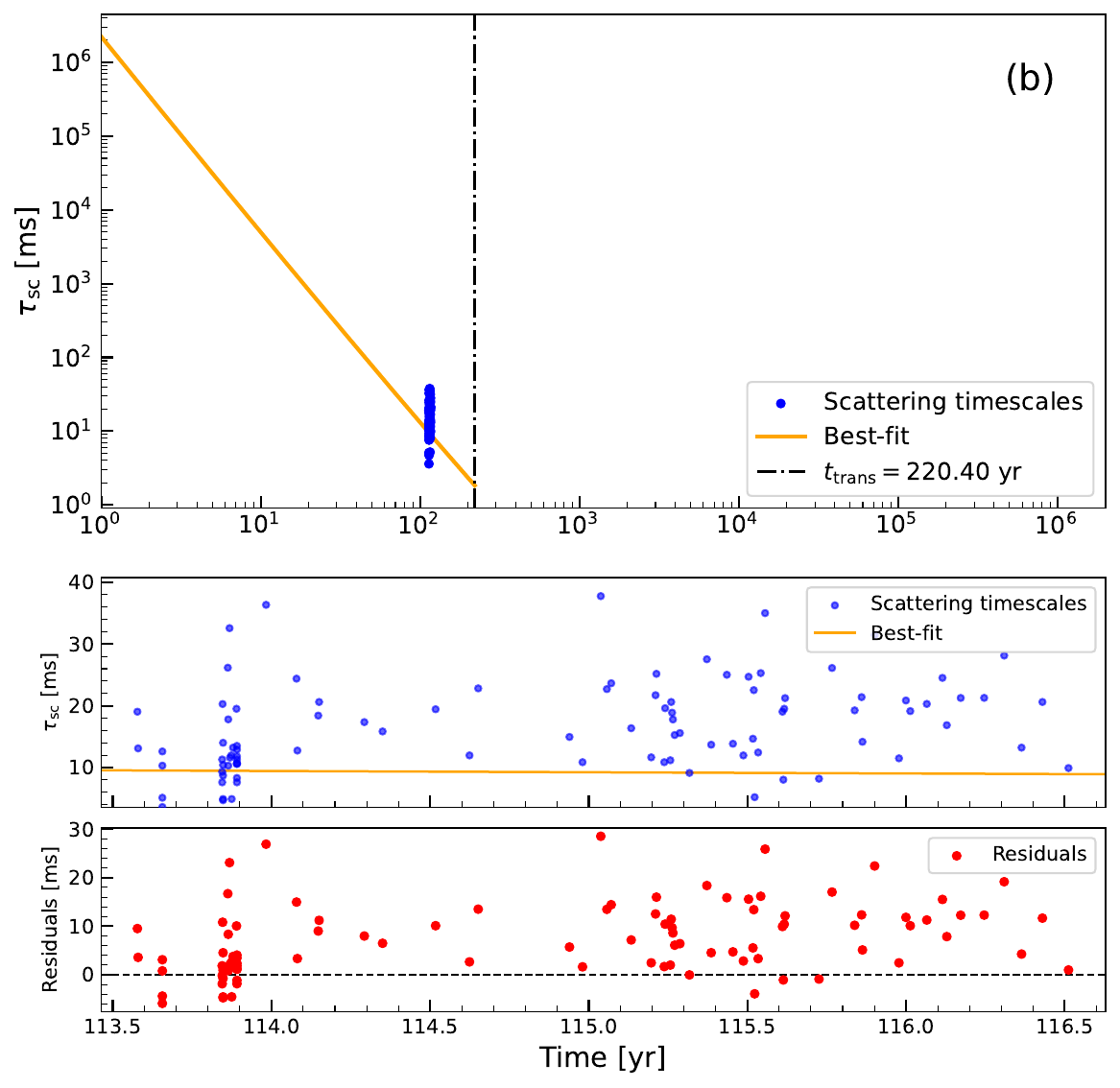}\hfill
    \includegraphics[width=0.33\linewidth]{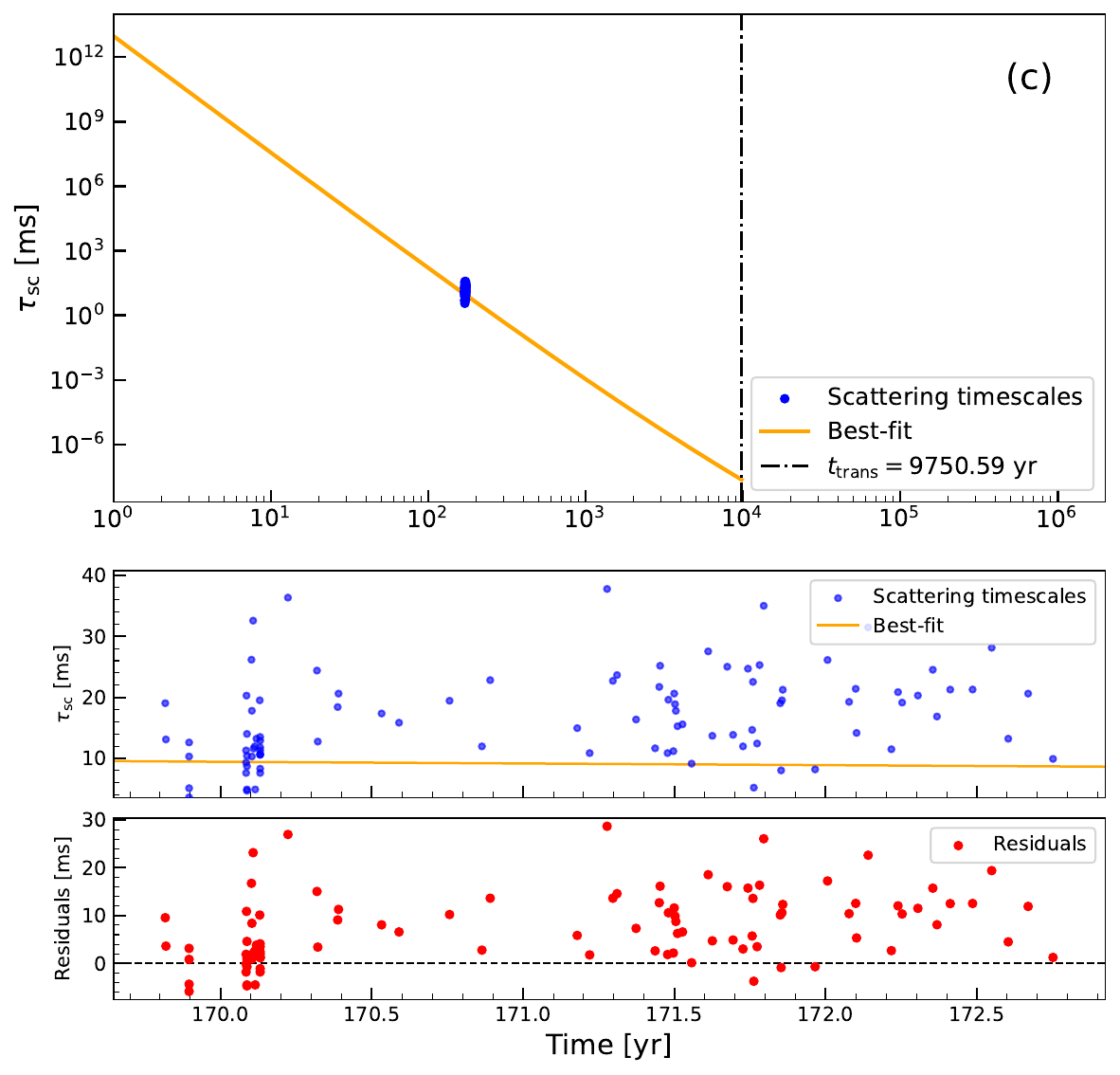}\hfill
    \includegraphics[width=0.33\linewidth]{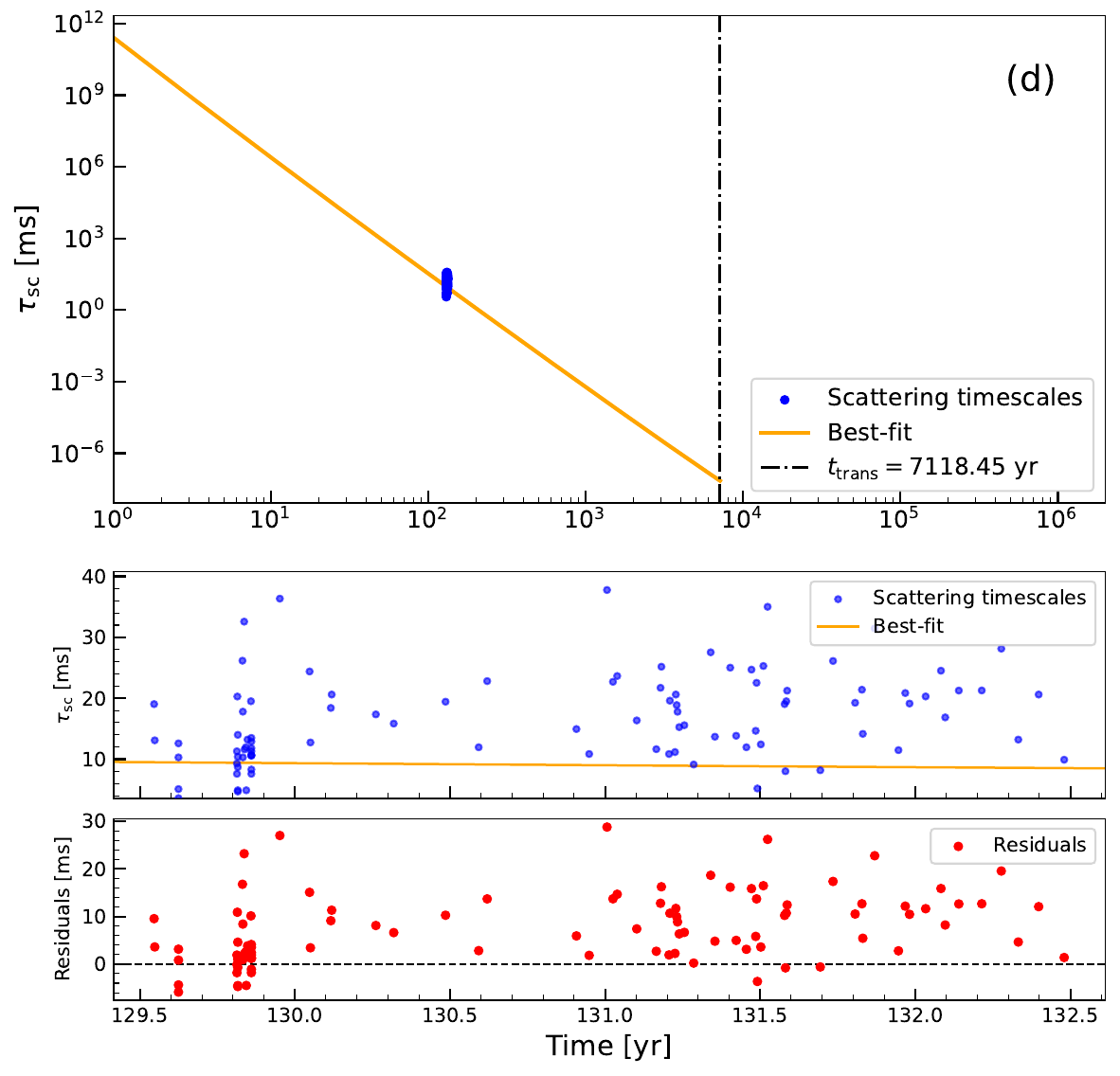}\hfill
    \includegraphics[width=0.33\linewidth]{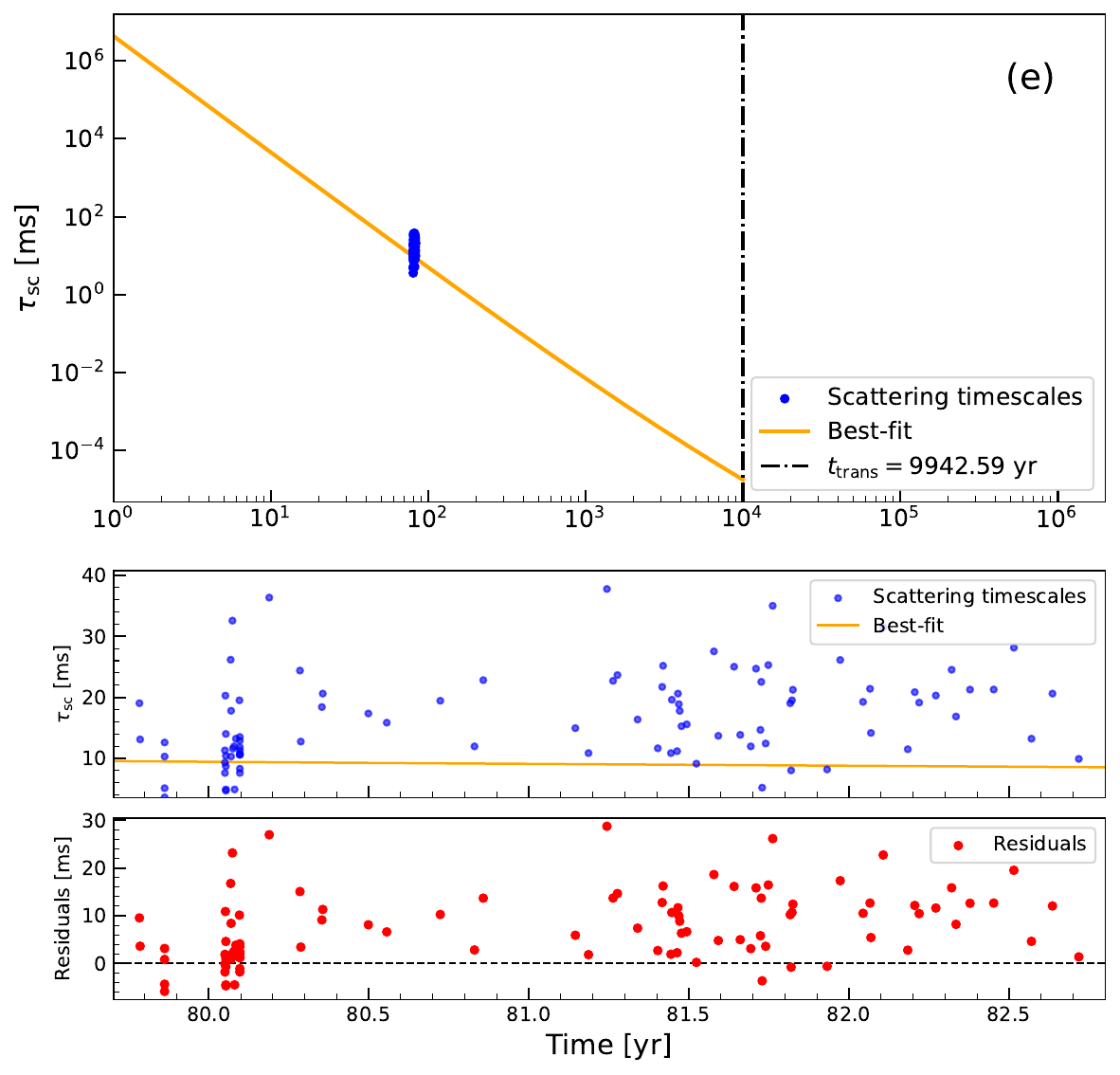}\hfill
    \includegraphics[width=0.33\linewidth]{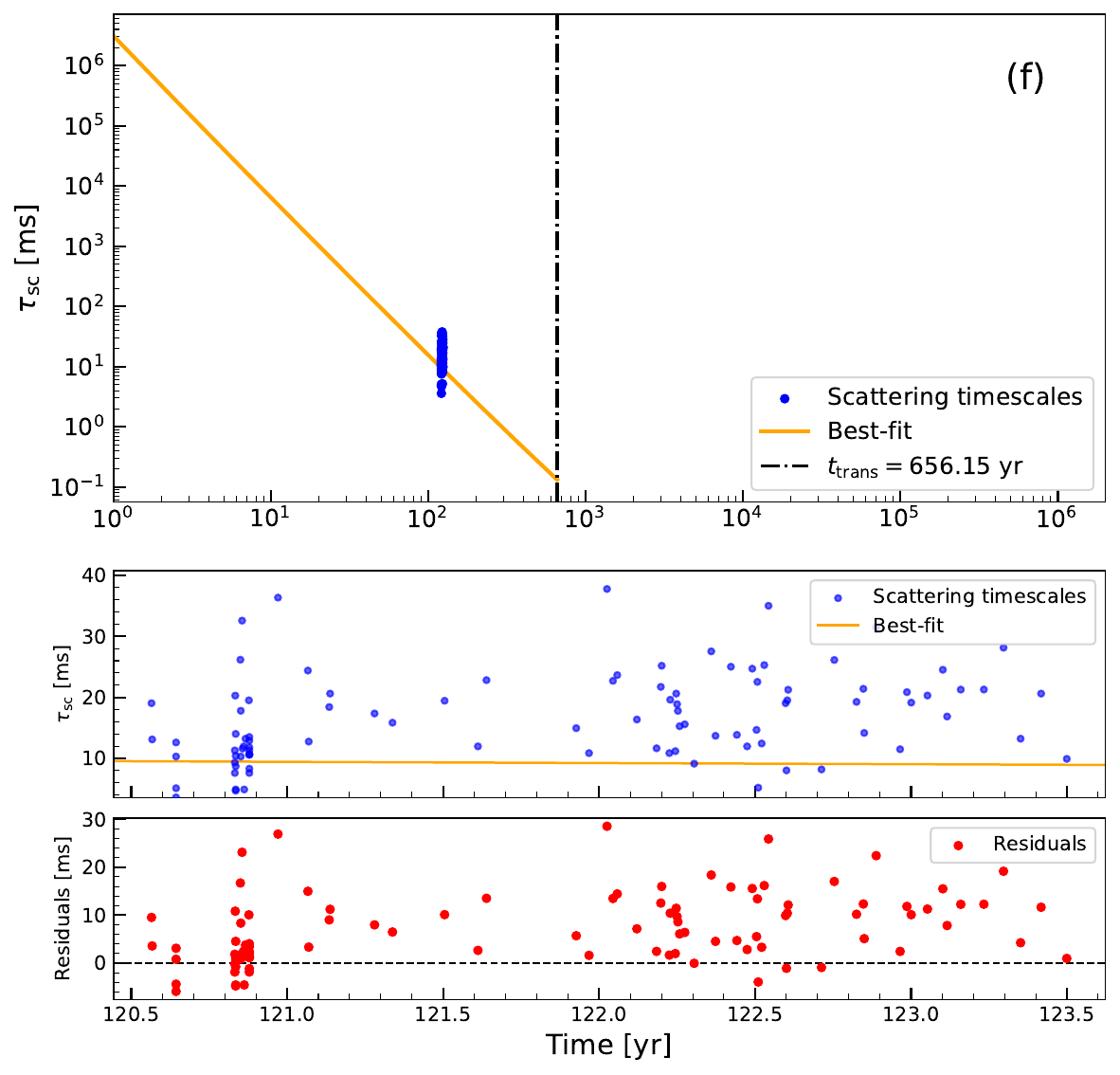}\hfill
    \caption{
Evolution of the scattering timescale and the corresponding residuals. 
Panels (a), (b), (c), (d), (e), and (f) correspond to cases A2, A4, B2, B4, E2, and E4, respectively. 
In the upper panels, the orange lines show the best-fit evolution of the scattering timescale, and the blue dots are the observed scattering timescales. 
The vertical black dash-dotted lines mark the transition times. 
The middle panels show zoomed-in views of the observed scattering timescales together with the best-fit curves. 
During the short burst epoch, the best-fit curves decrease slowly and therefore appear nearly flat.
In the lower panels, the red dots are the residuals of scattering timescales, and the horizontal black dashed lines are the zero residual. }https://www.overleaf.com/project/6a759f7d2eb731a434fcde0c/download/zip
    \label{figscatter}
\end{figure}

In Figure \ref{figscatter}, we present the best-fits to the observed scattering timescales, together with zoomed-in views of the observational data and the corresponding residuals.
It is worth noting that the evolutions of
scattering timescales for all cases are significantly similar but
with different initial conditions. In particular, B2 and B4
predict much larger scattering timescales at early times than the
other cases. This indicates that the retained cases can fit the
current data over the observed epoch while still implying very
different earlier scattering histories.
As can be seen from the middle panels of (a)--(f), the observed scattering timescale shows a nearly flat trend during the burst epoch, indicating that the scattering contribution does not evolve significantly over the burst epoch.
The lower panels of (a)$-$(f) display the
residuals between the observed scattering timescales and the
theoretical predictions. The residuals display a clear systematic
upward offset relative to the zero-residual line and are
remarkably similar among all six cases. Moreover, the positive
residuals are generally more pronounced at later epochs than at
earlier ones. This suggests that the observed scattering
timescales decrease more slowly with time than predicted by the best-fit curves. Therefore, although the single SNR model
reproduces the overall scattering timescale, the drifts of
residuals may indicate that the environment of FRB 190520B is more
complicated than a single expanding SNR. 
Alternatively, the SNR itself may have more turbulent and clumpy small-scale structures.

\begin{figure}
    \centering
    \includegraphics[width=1.0\linewidth]{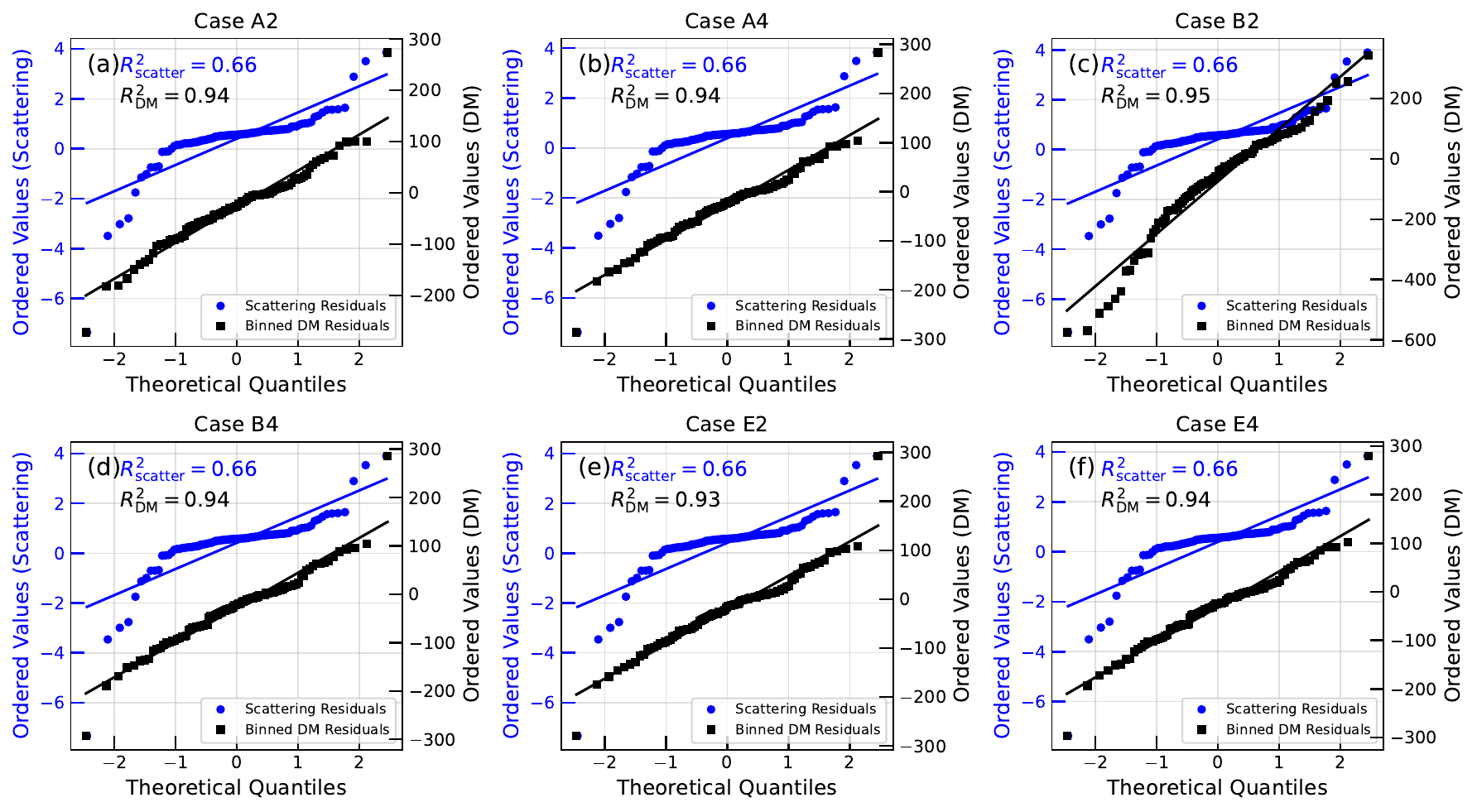}
    \caption{
The Q$-$Q plots of the residuals of DM and scattering timescale.
Panels (a), (b), (c), (d), (e), and (f) correspond to cases A2,
A4, B2, B4, E2, and E4, respectively. The blue points represent
the residuals of the scattering timescales and are referenced to
the left y-axis, while the black points represent the residuals of
the binned DMs and are referenced to the right y-axis. The blue
and black lines are the corresponding linear fits. $R_{\rm
scatter}^2$ and $R^2_{\rm DM}$ are the determined coefficients of residuals of the 
scattering timescales and the observed DM, respectively.
    }
    \label{figresidual}
\end{figure}

To quantify the goodness of fit and compare the retained cases, we
evaluate the Bayesian Information Criterion (BIC) and perform a
quantile--quantile (Q--Q) analysis of the residuals. Figure
\ref{figresidual} presents the Q--Q plots for cases A2, A4, B2, B4, E2,
and E4 in Panels (a)--(f), respectively. These plots compare the
empirical quantiles of the normalized residuals with the
theoretical quantiles of a norm distributions.
Based on these Q--Q plots and the associated coefficients of determination, we characterize the residuals as follows.
For the scattering residuals, all six cases give the same
coefficient of determination, $R_{\rm scatter}^2=0.66$. The
scattering residuals deviate systematically from a normal
distribution, showing that the adopted SNR model does not fully
describe the stochastic or structured variations in the scattering
data. For the binned DM residuals, the Q--Q behavior is closer to
normal, with $R_{\rm DM}^2=0.93$--$0.95$. Case B2 gives the
largest value, $R_{\rm DM}^2=0.95$, but this does not imply the
best DM fit, because its residual shifts and BIC value are much
larger than those of the statistically preferred cases.

The BIC and $\Delta$BIC values for scattering timescales and
observed DMs are summarized in Table \ref{tablestat}. For the
scattering-timescale data, A2, A4, and E4 give the minimum value,
${\rm BIC}_{\rm scatter}=191$. Case E2 is nearly indistinguishable
from them, with $\Delta{\rm BIC}_{\rm scatter}=1$, while B2 and B4
are mildly disfavored, with $\Delta{\rm BIC}_{\rm scatter}=5$ and
$6$, respectively. Therefore, the scattering data alone do not
provide a decisive preference among A2, A4, E2, and E4. The DM
data provide a much stronger statistical discrimination because of
their small observational uncertainties. The minimum value is
obtained for E2, with ${\rm BIC}_{\rm DM}=1569492$. For cases A4,
B4, A2, and E4, we have $\Delta{\rm BIC}_{\rm DM}=68760$, $78996$,
$102588$, and $133151$, respectively. Case B2 is strongly
disfavored, with $\Delta{\rm BIC}_{\rm DM}=4919321$, reflecting
its inability to reproduce the observed secular DM decrease. Since
the DM BIC dominates the statistical comparison, E2 is the
statistically favored case among the six solutions.

\section{Discussions}
\label{discussion}

\subsection{Self-consistency of the scattering theory}
\label{justify}

The formulae for temporal broadening adopted in Section \ref{scatter} rely on three standard assumptions that are usually satisfied in ISM-like environments \citep{Rickett1977}:
(i) the deviations of the refractive index are much smaller than unity everywhere;
(ii) the inner scale $l_0$ is much larger than the wavelength of the electromagnetic wave, $\lambda$; and
(iii) the outer scale $L$ is much smaller than the thickness of the scattering medium, $\Delta R$.
In this subsection, we examine whether these assumptions are satisfied in the shocked ejecta and the shocked ambient medium of the SNR, which are the two regions responsible for temporal broadening in our work.
This provides a self-consistency check for the parameters favored by the fits.

\begin{figure}
    \centering
    \includegraphics[width=0.5\linewidth]{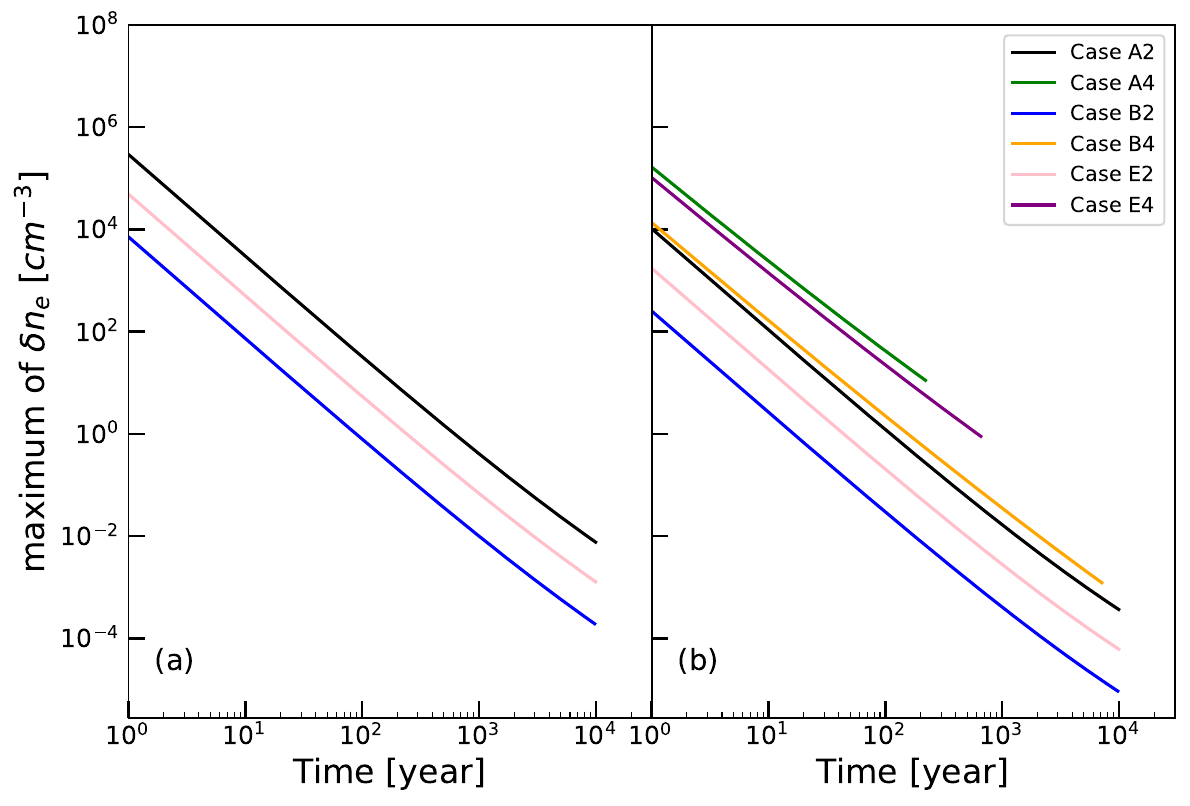}
    \caption{
Evolution of the maximum density fluctuation along radial
direction, $\delta n_{\rm e}$, in the shocked ejecta and shocked
ambient medium, shown in Panels (a) and (b), respectively. The
black, green, blue, orange, pink, and purple curves correspond to
cases A2, A4, B2, B4, E2, and E4, respectively. }
    \label{figdeltane}
\end{figure}

The deviation of the refractive index induced by electron-density
fluctuations is $\delta n_{\rm r}=-r_{\rm e}\lambda^2\delta n_{\rm
e}/2\pi$. For observing frequency of $\nu=1.25$ GHz, corresponding
to a wavelength of $\lambda=24~{\rm cm}$, the condition $\delta
n_{\rm r}\ll1$ is equivalent to $\delta n_{\rm
e}\ll3.9\times10^{10}~{\rm cm^{-3}}$. Here we use the radial
maximum of $\delta n_{\rm e}$ at each time as a conservative
diagnostic. Figure \ref{figdeltane} shows the evolution of this maximum
value up to the transition time for cases A2, A4, B2, B4, E2, and
E4.
Panels (a) and (b) correspond to the shocked ejecta and the shocked ambient medium, respectively. 
Only cases
with $n=2$ appear in the shocked ejecta panel, because the
electron density of the shocked ejecta vanishes for $n=4$. In both
regions, the maximum values of $\delta n_{\rm e}$ remain below
$10^6~{\rm cm^{-3}}$, which is more than four orders of magnitude
smaller than the threshold implied by $\delta n_{\rm r}\ll1$.
Therefore, the small-refractive-index-perturbation condition is
safely satisfied in all the six cases.

\begin{figure}
    \centering
    \includegraphics[width=0.5\linewidth]{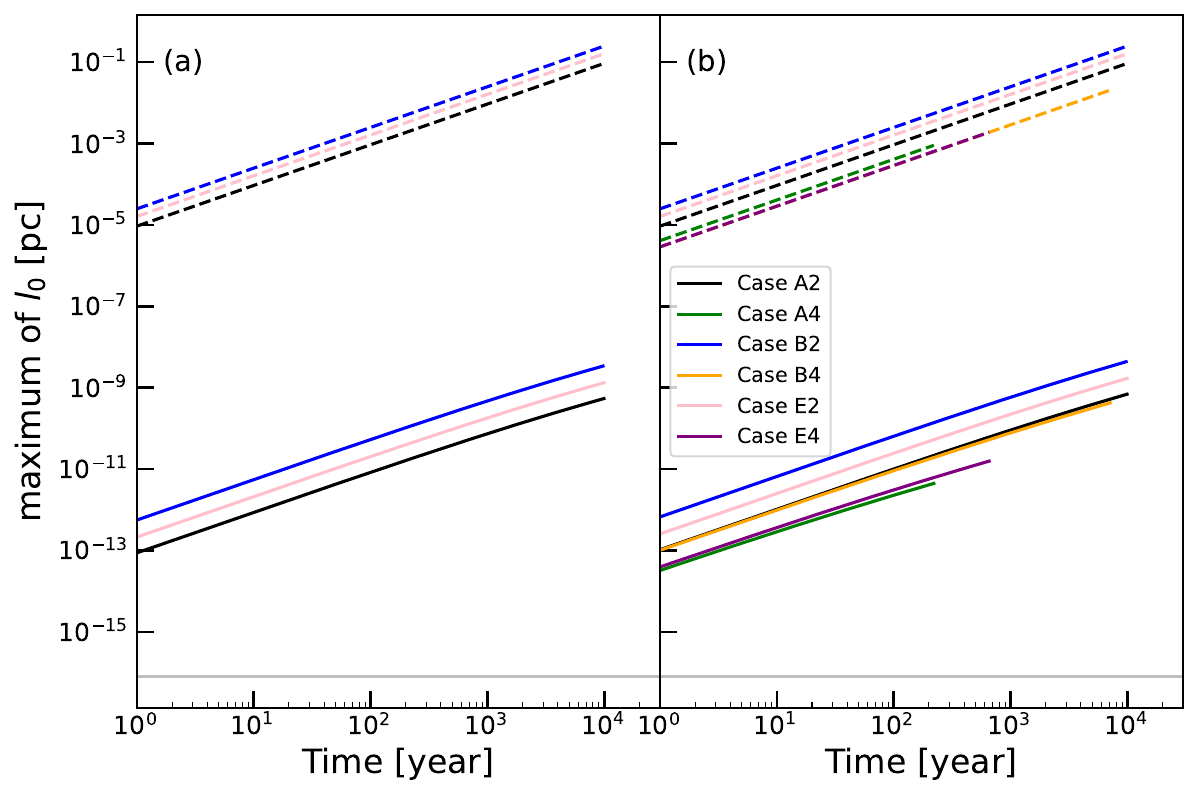}
    \caption{
Evolution of the maximum inner scale $l_0$ along the radial
direction, in the shocked ejecta and shocked ambient medium, shown
in Panels (a) and (b), respectively. The black, green, blue,
orange, pink, and purple curves correspond to cases A2, A4, B2,
B4, E2, and E4, respectively. The solid curves show the evolution
of the maximum inner scale $l_0$ along the radial direction, while
the dashed curves show the evolution of $L/10$, where $L$ is the
outer scale. The horizontal gray lines indicate $10\lambda$.
    }
    \label{figl0}
\end{figure}

The second assumption requires $l_0\gg\lambda$. For
$\lambda=24~{\rm cm}$, this corresponds to
$l_0\gg7.8\times10^{-18}$ pc. In addition, for an irregularity
spectrum in the inertial range, one also requires $l_0\ll L$
(\citealt{Xu2016}). The required scale hierarchy is therefore
$\lambda \ll l_0 \ll L$. Figure \ref{figl0} shows the time
evolution of the radial maximum of $l_0$ in the shocked ejecta and
the shocked ambient medium, corresponding to Panels (a) and (b),
respectively.
The evolution of $L/10$ and the horizontal line representing $10\lambda$ are also shown in Figure \ref{figl0}.
Thus, the separations between $l_0$,
$10\lambda$, and $L/10$ are visualized directly. Because $l_0$ is
estimated as described in Section \ref{scatter} and is related to
$n_{\rm e}$, only cases with $n=2$ have non-zero values of $l_0$
in the shocked ejecta. Figure \ref{figl0} shows that, throughout
the ED phase, the maximum value of $l_0$ stays below $0.1L$ and
above $10\lambda$ in all cases. This indicates that the condition
$\lambda\ll l_0\ll L$ is satisfied with a clear margin. The
treatment of the inner scale $l_0$ remains one of the main sources
of uncertainty in the scattering calculation. In the ISM, $l_0$ is
usually constrained by observations. For example,
\cite{Armstrong1995} used observational data from nearby pulsars
within $\sim1$ kpc to derive an upper limit on the ISM inner scale
of $\sim10^{-8}$ pc, while \cite{Gupta1993} inferred a range of
$l_0\sim10^{-9}$--$10^{-7}$ pc from pulsar observations.

\begin{figure}
    \centering
    \includegraphics[width=0.5\linewidth]{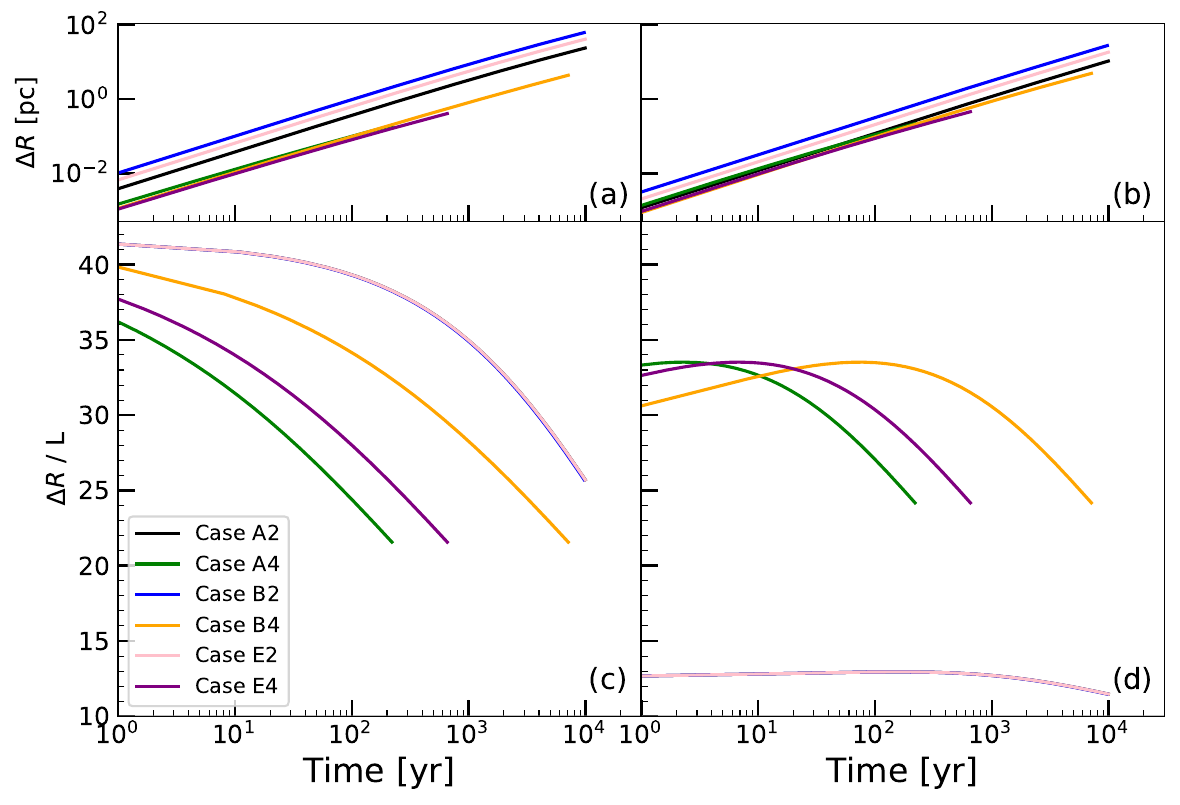}
    \caption{
Evolution of the plasma thickness in the shocked ejecta and
shocked ambient medium, shown in Panels (a) and (b), respectively.
Panels (c) and (d) show the corresponding evolution of the ratio
between the plasma thickness and the outer scale, $\Delta R/L$, in
the shocked ejecta and shocked ambient medium. The black, green,
blue, orange, pink, and purple curves correspond to cases A2, A4,
B2, B4, E2, and E4, respectively. }
    \label{figlthick}
\end{figure}

Figures \ref{figlthick}(a) and (b) show the evolution of the thickness $\Delta R$ in the shocked ejecta and the shocked ambient medium, respectively, while Figures \ref{figlthick}(c) and (d) show the corresponding evolution of $\Delta R/L$ in these two regions.
It can be seen that $\Delta R/L$ is always larger than 10 in all retained cases, which means that the outer scale remains sufficiently smaller than the thickness of the scattering layer throughout the relevant evolutionary stage.
A noteworthy feature is that, although $\Delta R$ itself increases monotonically with time, the ratio $\Delta R/L$ generally decreases.
This indicates that the outer scale grows faster than the shell thickness during the ED phase.
This trend is physically reasonable, because in the ST phase the outer scale is usually expected to become comparable to the thickness of the shocked regions (\citealt{Chevalier1982,Dwarkadas2000}).
Therefore, the three assumptions underlying the scattering formulae are all satisfied with substantial margins.
Overall, these checks show that the six retained cases lie in a regime where the adopted scattering theory is self-consistent.

\subsection{Free-free absorption}
\label{ff}

Another important issue is whether the surrounding SNR is transparent to the radio bursts from FRB 190520B.
Radio waves may experience free-free absorption in SNR.
The absorption coefficient is (\citealt{Rybicki1979})
\begin{equation}\label{func44}
    \alpha_{\nu}^{\rm ff}=0.018T^{-3/2}Z^{2}n_{\rm e}n_{\rm i}\nu^{-2}\bar{g}_{\rm ff},
\end{equation}
where $T$ is the plasma temperature, $Z$ is the ionic charge number, $n_{\rm i}$ is the ion density, $\nu$ is the frequency of radio emission, and $\bar{g}_{\rm ff}$ is the Gaunt factor.
In our calculation, we approximate the free-free optical depth as
\begin{equation}\label{func45}
    \tau_{\rm ff}\simeq \alpha_{\nu}^{\rm ff}\Delta R.
\end{equation}
We adopt $Z\sim1$, $n_{\rm i}\sim n_{\rm e}$, and $\bar{g}_{\rm ff}\sim1$.
For FRB 190520B, the frequency of bursts is set to $\nu=1.25$ GHz.
The temperatures in the two unshocked regions are set to $10^4$ K (\citealt{Metzger2017}; \citealt{ZhaoZY2021}).
For the shocked regions, we use the downstream proton temperature, $T_{\rm p}$, as an approximation for the plasma temperature, where
\begin{equation}\label{func46}
    k_{\rm B}T_{\rm p}=\frac{3}{16}m_{\rm p}V_{\rm s}^{2},
\end{equation}
where $k_{\rm B}$ is the Boltzmann constant.
The reverse- and forward-shock velocities are obtained by differentiating Equations (\ref{func9}) and (\ref{func11}) with respect to time.

\begin{figure}
    \centering
    \includegraphics[width=0.5\linewidth]{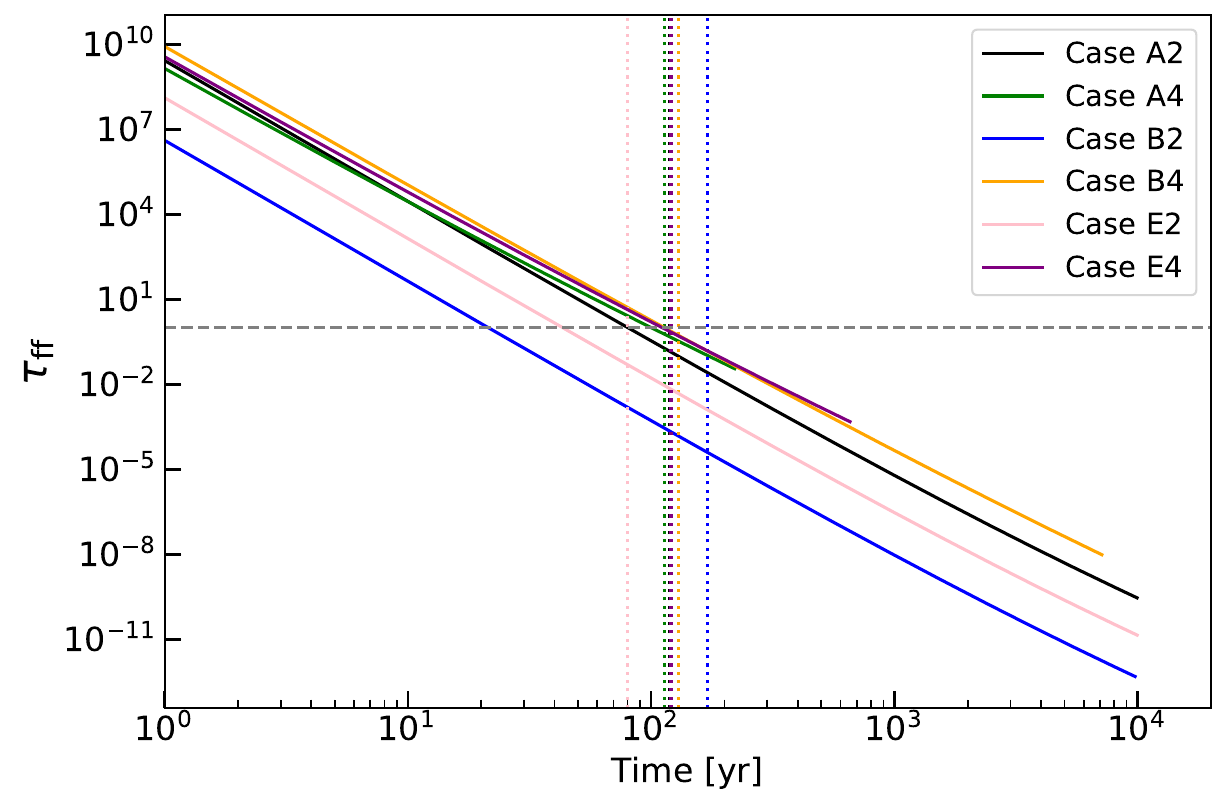}
    \caption{
Evolution of the free-free optical depth. The black, green, blue,
orange, pink, and purple curves correspond to cases A2, A4, B2,
B4, E2, and E4, respectively. The solid curves show the evolution
of the optical depth, while the vertical dotted lines mark the
inferred source ages. The horizontal gray dashed line indicates
$\tau_{\rm ff}=1$. }
    \label{figtauff}
\end{figure}

Figure \ref{figtauff} shows the time evolution of $\tau_{\rm ff}$ for all six cases.
It can be seen that the optical depth decreases rapidly with time and drops below unity from several decades to about one century for all cases.
Among the six cases, B4 remains optically thick for the longest time, becoming transparent at $t\simeq114$ yr.
Cases A4 and E4, with transparency times of $t\simeq100$ and $111$ yr, respectively, also remain opaque longer than A2, B2, and E2, whose transparency times are $t\simeq80$, $21$, and $43$ yr, respectively.
Nevertheless, in all six cases the SNR becomes transparent to 1.25 GHz bursts before the inferred source age of FRB 190520B.
This means that the surrounding remnant is unlikely to prevent the escape of the observed GHz radio bursts.

This result can also be understood together with Figure \ref{figlthick}.
Although the shell thickness $\Delta R$ increases monotonically with time, $\tau_{\rm ff}$ decreases by many orders of magnitude.
Therefore, it can be concluded that the time evolution of the free-free opacity is controlled primarily by the rapid decline of the density, rather than by the increase in the propagation length.
Finally, Equation (\ref{func44}) shows that $\tau_{\rm ff}\propto\nu^{-2}$ for fixed plasma properties.
The transparency condition is therefore frequency dependent: even if the local SNR is already transparent at GHz, it can remain opaque for a longer period at lower radio frequencies.
Hence, within the SNR scenarios, a young local environment can still be compatible with the detection of GHz bursts from FRB 190520B, while displaying a suppression of lower frequency emission.
Together with the evolutions of DMs and scattering timescales, we conclude that the six cases point out to a physically plausible evolutionary stage of the SNR: it is still young enough to provide appreciable local DM and scattering timescales, but already evolved enough to become transparent for GHz radio emissions.

\subsection{Physical implications}
\label{imply}

It is worth noting that the present framework reproduces the secular DM evolution more successfully than the detailed evolution of the scattering timescale. 
As shown in Section~\ref{result}, the DM residuals are closer to a normal distribution, indicating that the local DM evolution of FRB~190520B can be described well by a single expanding SNR. 
The scattering residuals, however, show a systematic positive offset, especially at later epochs.
Notably, during the burst epoch, the scattering timescale shows a nearly flat trend in all six cases.
These two facts together suggest the presence of an extra source that contributes an approximately constant scattering timescale.
One straightforward possibility is that an additional scattering component is present along the line of sight. 
Indeed, \citet{Ocker2023} reported scattering variations of FRB~190520B on timescales from minutes to days, which are not clearly correlated with the observed DMs and were attributed to a dynamic and inhomogeneous circumsource medium.
Alternatively, the residual drift may come from more complicated internal structures within the SNR itself---such as density clumps or asymmetric shocks---that are not captured by the self-similar approximation adopted here.
Therefore, while the SNR model successfully captures the secular DM evolution, the scattering data likely require an extra component associated with the central engine, or a more detailed treatment of the SNR internal structure.

Beyond this residual offset of scattering timescale, there is a contrast existing in observations: the observed DM exhibits a clear secular decreasing trend, whereas the scattering timescale does not show a similarly monotonic evolution and displays much larger intrinsic scatter.
This contrasting behavior, however, has a natural physical origin within our model. 
As discussed above, the local DM is dominated by the unshocked core and unshocked ejecta, which are characterized by smooth, large-scale density distributions that evolve steadily with the SNR expansion. 
The secular DM decrease is therefore driven by the gradual dilution of the unshocked components as the SNR expands. 
In contrast, the scattering is primarily produced by the shocked regions---the shocked ejecta and the shocked ambient medium. 
The scattering timescale is sensitive to small-scale turbulent structures in these regions, which have shorter dynamical timescales and are
subject to stochastic fluctuations.
These naturally account for the larger scatter and the lack of a clear monotonic trend in the scattering data.

The inferred source ages, $t_0=79.8$--$169.8~{\rm yr}$, indicate
that the central engine is very young. This age range is
compatible with the scenario in which an active newborn magnetar
is embedded in an expanding SNR. In our work, $DM_{\rm host}$
represents a time-independent host contribution and provides
information on the host environment of FRB 190520B. Cases A2, A4,
B4, and E4 predict extremely small values of $DM_{\rm host}$, of
the order of several tens of ${\rm pc~cm^{-3}}$. In these cases,
the observed secular DM evolution is mainly driven by the
expanding local environment, especially the young SNR. Such low
host contributions, besides, indicate either the source of FRB
190520B is located an offset region of the host galaxy, or it is
embedded in a bubble cavity. Since FRB 190520B is located at a
star-forming region in its host, if such low $DM_{\rm host}$ is
real, the former scenario is more possible. Case B2 is the
opposite one, requiring the largest host contribution, $DM_{\rm
host}\simeq10^3~{\rm pc~cm^{-3}}$. As a result, the predicted DM
at burst epoch becomes almost flat, with a slope of only
$-0.3~{\rm pc~cm^{-3}~yr^{-1}}$. This is in clear tension with the
observed secular DM decrease of FRB 190520B. Case E2 provides an
intermediate scenario. It has $DM_{\rm host}\simeq400~{\rm
pc~cm^{-3}}$, implying that both the host-galaxy component and the
local SNR component make important contributions to the total
observed DM. Physically, this case corresponds to a young SNR
embedded in a relatively high-density host environment, such as a
dense star-forming region, or an H{\sc ii}-region. This case is
then broadly consistent with the fact that FRB 190520B is located
in an active star-forming environment and is associated with a
compact PRS (\citealt{Niu2022}). Future monitoring of the DM
evolution can help break this degeneracy between the host and the
local contribution: if the observed DM approaches a high floor, a
large $DM_{\rm host}$ component would be favored; if it continues
to decline steadily, the low-$DM_{\rm host}$, local-dominated
solutions would become more plausible.

A potential concern is that $t_0$ and $DM_{\rm host}$ may be strongly degenerate, which would compromise the SNR parameter constraints.
To assess this, we repeated the MCMC sampling three times with different random seeds for each of the six retained cases. 
Across these runs, the Spearman correlation coefficients between $t_0$ and $DM_{\rm host}$ remain at $|\rho| \lesssim 0.3$ for A2, A4, B4, E2, and E4, whereas B2 consistently yields $\rho \simeq 0.5$.
The origin of this pattern is as follows. 
The observed DM decreases steeply at a rate of $\sim 12~{\rm pc~cm^{-3}~yr^{-1}}$ (\citealt{WangFY2025}; \citealt{NiuCH2026}). 
In all cases except B2, the local DM contribution is substantial and its slope is well matched to this observed decrease during the burst epoch, which makes observational data sensitive to $DM_{\rm local}$ and then tightly constrains the time evolution of $DM_{\rm local}$ and hence $t_0$. 
Because $DM_{\rm host}$ is constant in time and contributes only to the absolute DM level, the two parameters are nearly orthogonal in the likelihood. 
The scattering timescale provides an additional, independent handle on $DM_{\rm local}$, further breaking the degeneracy between $t_0$ and $DM_{\rm host}$.
In B2, by contrast, $DM_{\rm host}\simeq998~{\rm pc~cm^{-3}}$ overwhelms the local contribution and the predicted local DM slope is nearly flat ($-0.3~{\rm pc~cm^{-3}~yr^{-1}}$). 
As a result, $DM_{\rm local}$ is insensitive to $t_0$ in this case.
Therefore, neither $t_0$ nor $DM_{\rm host}$ is sharply constrained, and the two parameters drift together in the posterior, producing the strongest degeneracy among all six cases.
The weak $t_0$--$DM_{\rm host}$ degeneracy in all statistically favored cases therefore does not significantly affect the reliability of the SNR parameter constraints.

All the six cases require relatively large ejecta masses, $M_{\rm ej}=8.7$--$19.2~M_\odot$.
However, the inferred kinetic energies and wind mass-loss rates span broad ranges, with $E_{\rm k}=0.1$--$7.3\times10^{51}~{\rm erg}$ and $\dot{M}_{\rm w}=0.3$--$14.8\times10^{-5}~M_\odot~{\rm yr^{-1}}$ for the adopted wind velocity $v_{\rm w}=100~{\rm km~s^{-1}}$.
Therefore, the present fits do not uniquely determine either the explosion energy or the progenitor mass-loss preference.
Thus, while a massive-star progenitor is favored, the current fitting results alone are insufficient to identify a unique supernova subtype.
The conservative picture is that FRB 190520B is associated with a young, massive-star core-collapse SNR expanding into a moderately dense progenitor wind.

It is also noteworthy that the two Kolmogorov-spectrum prescriptions fail to reproduce the observations.
This fact, however, does not indicate the non-existence of Kolmogorov-spectrum turbulence in SNRs.
Observational evidence for Kolmogorov-like turbulence has been reported in many SNRs.
For example, \citet{Shimoda2018} found a Kolmogorov-like magnetic energy spectrum in Tycho's SNR.
The failure of the Kolmogorov-spectrum prescriptions here is more likely related to their specific implementation in the scattering model.
In our calculation, the outer scale $L$ is estimated from a linear fit to the simulation results of \citet{Dwarkadas2000}.
With this prescription, $L$ grows too rapidly for the theoretical scattering timescales to fit the observational data.
If the outer scale instead followed a time-dependent relation with a much smaller growth rate, while still remaining much larger than the inner scale, the Kolmogorov-spectrum prescriptions may reproduce the observations.

Overall, combining the statistical results with the physical interpretation, E2 is the most favored case in the present framework.
It gives the minimum ${\rm BIC}_{\rm DM}$, predicts a DM slope that lies within the range given by \cite{NiuCH2026}, and remains competitive in the scattering fit.
Physically, its fitted SNR parameters, $M_{\rm ej}$, $E_{\rm k}$, and $\dot{M}_{\rm w}$, are all reasonable.
In particular, the partition between the local and host DM contributions in this case is more reasonable than in the other cases.
Although its transition time is somewhat long, it is still acceptable.

\section{Summary}
\label{Summary}

In this work, we test the scenario that FRB 190520B originates
from a central engine formed in the core-collapse of a massive
star and embedded in a SNR expanding into a wind environment. The
SNR evolution is described by the self-similar solution, with an
ejecta-envelope density index of $n=2, 4, 6,$ or $8$. We combine
the self-similar SNR model with five prescriptions for temporal
broadening induced by plasma density irregularities, and use the
observed DM and scattering-timescale data of FRB 190520B to
constrain the properties of its local environment. Our main
conclusions are summarized as below:

\begin{enumerate}[label=(\roman*)]
    \item
    Among the 20 cases explored in this work, only six cases are retained
    by the fitting procedure, namely A2, A4, B2, B4, E2, and E4.
    All retained cases correspond to the FE solution with $n=2$ or $4$,
    while none of the SSDW cases with $n=6$ or $8$ can reproduce the observational data.
    The two Kolmogorov-type turbulence spectra are also inconsistent with the data,
    which may result from the adopted implementation of an overly large outer scale $L$.

    \item
    All retained cases imply a very young SNR for FRB 190520B, with source ages of $t_0=79.8$--$169.8~{\rm yr}$.
    The inferred ejecta masses are relatively large, $M_{\rm ej}=8.7$--$19.2~M_\odot$, favoring a massive-star core-collapse origin.
    However, the inferred kinetic energies and wind mass-loss rates span broad ranges, with $E_{\rm k}=0.1$--$7.3\times10^{51}~{\rm erg}$ and $\dot{M}_{\rm w}=0.3$--$14.8\times10^{-5}~M_\odot~{\rm yr^{-1}}$ for the adopted wind velocity $v_{\rm w}=100~{\rm km~s^{-1}}$.
    Hence, the current modeling is insufficient to identify a unique supernova subtype.

    \item
    The fitted values of $DM_{\rm host}$ imply different physical pictures for the host and its local contributions.
    Cases A2, A4, B4, and E4 require a relatively small $DM_{\rm host}$ value, indicating that the observed DM evolution is dominated mainly by the local SNR contribution.
    B2 represents the opposite extreme, with $DM_{\rm host}\simeq10^3~{\rm pc~cm^{-3}}$, and is strongly disfavored because it predicts an almost flat DM evolution during the burst epoch.
    E2 provides an intermediate and more plausible scenario, in which both the host-galaxy component and the local SNR component are important contributors to the total observed DM.

    \item
    The secular DM evolution is reproduced more successfully than the detailed evolution of the scattering timescale.
    Except for B2, the retained cases predict DM slopes broadly consistent with the observed long-term DM decrease.
    By contrast, the scattering residuals show a systematic upward drift.
    This suggests that an additional scattering component or more turbulent structures in the SNR may be present.

    \item
    We also examined the self-consistency of the scattering theory.
    For the retained cases, the assumptions underlying the temporal-broadening
    formalism are satisfied within the relevant parameter ranges.
    In addition, the optical depth of free-free absorption at 1.25 GHz drops below unity
    before the inferred source ages in all retained cases.
    Therefore, the local SNR can remain young enough to provide appreciable DM
    and scattering, while already being transparent to the observed GHz bursts from the central engine.

    \item
    Combining the statistical results with the physical interpretation, E2 is the most favored case in the SNR framework.

\end{enumerate}

\section*{Acknowledgements}

We would like to thank the anonymous referee for useful comments and suggestions that led to an overall improvement of this study. 
We are grateful to Zhen-Yi Zhao for his helpful suggestions. 
This study is supported by the National Natural Science Foundation of China (Grant Nos. 12233002, 12273113), and by the National Key R\&D Program
of China (2021YFA0718500). 
YFH acknowledges the support from the Xinjiang Tianchi Program.
Jin-Jun Geng acknowledges support from the Youth Innovation Promotion Association (2023331)

\bibliography{sample631}{}

@article{Lorimer2007,
       author = {{Lorimer}, D.~R. and {Bailes}, M. and {McLaughlin}, M.~A. and {Narkevic}, D.~J. and {Crawford}, F.},
        title = "{A Bright Millisecond Radio Burst of Extragalactic Origin}",
      journal = {Science},
         year = 2007,
        month = nov,
       volume = {318},
       number = {5851},
        pages = {777},
          doi = {10.1126/science.1147532},
archivePrefix = {arXiv},
       eprint = {0709.4301},
 primaryClass = {astro-ph},
       adsurl = {https://ui.adsabs.harvard.edu/abs/2007Sci...318..777L}
}

@ARTICLE{Zhang2023,
       author = {{Zhang}, Bing},
        title = "{The physics of fast radio bursts}",
      journal = {Reviews of Modern Physics},
         year = 2023,
        month = jul,
       volume = {95},
       number = {3},
          eid = {035005},
        pages = {035005},
          doi = {10.1103/RevModPhys.95.035005},
archivePrefix = {arXiv},
       eprint = {2212.03972},
 primaryClass = {astro-ph.HE},
       adsurl = {https://ui.adsabs.harvard.edu/abs/2023RvMP...95c5005Z}
}

@ARTICLE{CordesLazio2003,
       author = {{Cordes}, J.~M. and {Lazio}, T.~J.~W.},
        title = "{NE2001.I. A New Model for the Galactic Distribution of Free Electrons and its Fluctuations}",
      journal = {arXiv e-prints},
         year = 2002,
        month = jul,
          eid = {astro-ph/0207156},
        pages = {astro-ph/0207156},
          doi = {10.48550/arXiv.astro-ph/0207156},
archivePrefix = {arXiv},
       eprint = {astro-ph/0207156},
 primaryClass = {astro-ph},
       adsurl = {https://ui.adsabs.harvard.edu/abs/2002astro.ph..7156C}
}

@article{Thornton2013,
       author = {{Thornton}, D. and {Stappers}, B. and {Bailes}, M. and {Barsdell}, B. and {Bates}, S. and {Bhat}, N.~D.~R. and {Burgay}, M. and {Burke-Spolaor}, S. and {Champion}, D.~J. and {Coster}, P. and {D'Amico}, N. and {Jameson}, A. and {Johnston}, S. and {Keith}, M. and {Kramer}, M. and {Levin}, L. and {Milia}, S. and {Ng}, C. and {Possenti}, A. and {van Straten}, W.},
        title = "{A Population of Fast Radio Bursts at Cosmological Distances}",
      journal = {Science},
         year = 2013,
        month = jul,
       volume = {341},
       number = {6141},
        pages = {53-56},
          doi = {10.1126/science.1236789},
archivePrefix = {arXiv},
       eprint = {1307.1628},
 primaryClass = {astro-ph.HE},
       adsurl = {https://ui.adsabs.harvard.edu/abs/2013Sci...341...53T}
}

@article{Beloborodov2020,
       author = {{Beloborodov}, Andrei M.},
        title = "{Blast Waves from Magnetar Flares and Fast Radio Bursts}",
      journal = {\apj},
         year = 2020,
        month = jun,
       volume = {896},
       number = {2},
          eid = {142},
        pages = {142},
          doi = {10.3847/1538-4357/ab83eb},
archivePrefix = {arXiv},
       eprint = {1908.07743},
 primaryClass = {astro-ph.HE},
       adsurl = {https://ui.adsabs.harvard.edu/abs/2020ApJ...896..142B}
}

@article{Spitler2016,
       author = {{Spitler}, L.~G. and {Scholz}, P. and {Hessels}, J.~W.~T. and {Bogdanov}, S. and {Brazier}, A. and {Camilo}, F. and {Chatterjee}, S. and {Cordes}, J.~M. and {Crawford}, F. and {Deneva}, J. and {Ferdman}, R.~D. and {Freire}, P.~C.~C. and {Kaspi}, V.~M. and {Lazarus}, P. and {Lynch}, R. and {Madsen}, E.~C. and {McLaughlin}, M.~A. and {Patel}, C. and {Ransom}, S.~M. and {Seymour}, A. and {Stairs}, I.~H. and {Stappers}, B.~W. and {van Leeuwen}, J. and {Zhu}, W.~W.},
        title = "{A repeating fast radio burst}",
      journal = {\nat},
         year = 2016,
        month = mar,
       volume = {531},
       number = {7593},
        pages = {202-205},
          doi = {10.1038/nature17168},
archivePrefix = {arXiv},
       eprint = {1603.00581},
 primaryClass = {astro-ph.HE},
       adsurl = {https://ui.adsabs.harvard.edu/abs/2016Natur.531..202S}
}

@article{Spitler2014,
       author = {{Spitler}, L.~G. and {Cordes}, J.~M. and {Hessels}, J.~W.~T. and {Lorimer}, D.~R. and {McLaughlin}, M.~A. and {Chatterjee}, S. and {Crawford}, F. and {Deneva}, J.~S. and {Kaspi}, V.~M. and {Wharton}, R.~S. and {Allen}, B. and {Bogdanov}, S. and {Brazier}, A. and {Camilo}, F. and {Freire}, P.~C.~C. and {Jenet}, F.~A. and {Karako-Argaman}, C. and {Knispel}, B. and {Lazarus}, P. and {Lee}, K.~J. and {van Leeuwen}, J. and {Lynch}, R. and {Ransom}, S.~M. and {Scholz}, P. and {Siemens}, X. and {Stairs}, I.~H. and {Stovall}, K. and {Swiggum}, J.~K. and {Venkataraman}, A. and {Zhu}, W.~W. and {Aulbert}, C. and {Fehrmann}, H.},
        title = "{Fast Radio Burst Discovered in the Arecibo Pulsar ALFA Survey}",
      journal = {\apj},
         year = 2014,
        month = aug,
       volume = {790},
       number = {2},
          eid = {101},
        pages = {101},
          doi = {10.1088/0004-637X/790/2/101},
archivePrefix = {arXiv},
       eprint = {1404.2934},
 primaryClass = {astro-ph.HE},
       adsurl = {https://ui.adsabs.harvard.edu/abs/2014ApJ...790..101S}
}

@article{Chatterjee2017,
       author = {{Chatterjee}, S. and {Law}, C.~J. and {Wharton}, R.~S. and {Burke-Spolaor}, S. and {Hessels}, J.~W.~T. and {Bower}, G.~C. and {Cordes}, J.~M. and {Tendulkar}, S.~P. and {Bassa}, C.~G. and {Demorest}, P. and {Butler}, B.~J. and {Seymour}, A. and {Scholz}, P. and {Abruzzo}, M.~W. and {Bogdanov}, S. and {Kaspi}, V.~M. and {Keimpema}, A. and {Lazio}, T.~J.~W. and {Marcote}, B. and {McLaughlin}, M.~A. and {Paragi}, Z. and {Ransom}, S.~M. and {Rupen}, M. and {Spitler}, L.~G. and {van Langevelde}, H.~J.},
        title = "{A direct localization of a fast radio burst and its host}",
      journal = {\nat},
         year = 2017,
        month = jan,
       volume = {541},
       number = {7635},
        pages = {58-61},
          doi = {10.1038/nature20797},
archivePrefix = {arXiv},
       eprint = {1701.01098},
 primaryClass = {astro-ph.HE},
       adsurl = {https://ui.adsabs.harvard.edu/abs/2017Natur.541...58C}
}

@article{Tendulkar2017,
       author = {{Tendulkar}, S.~P. and {Bassa}, C.~G. and {Cordes}, J.~M. and {Bower}, G.~C. and {Law}, C.~J. and {Chatterjee}, S. and {Adams}, E.~A.~K. and {Bogdanov}, S. and {Burke-Spolaor}, S. and {Butler}, B.~J. and {Demorest}, P. and {Hessels}, J.~W.~T. and {Kaspi}, V.~M. and {Lazio}, T.~J.~W. and {Maddox}, N. and {Marcote}, B. and {McLaughlin}, M.~A. and {Paragi}, Z. and {Ransom}, S.~M. and {Scholz}, P. and {Seymour}, A. and {Spitler}, L.~G. and {van Langevelde}, H.~J. and {Wharton}, R.~S.},
        title = "{The Host Galaxy and Redshift of the Repeating Fast Radio Burst FRB 121102}",
      journal = {\apjl},
         year = 2017,
        month = jan,
       volume = {834},
       number = {2},
          eid = {L7},
        pages = {L7},
          doi = {10.3847/2041-8213/834/2/L7},
archivePrefix = {arXiv},
       eprint = {1701.01100},
 primaryClass = {astro-ph.HE},
       adsurl = {https://ui.adsabs.harvard.edu/abs/2017ApJ...834L...7T}
}

@article{Li2021,
       author = {{Li}, D. and {Wang}, P. and {Zhu}, W.~W. and {Zhang}, B. and {Zhang}, X.~X. and {Duan}, R. and {Zhang}, Y.~K. and {Feng}, Y. and {Tang}, N.~Y. and {Chatterjee}, S. and {Cordes}, J.~M. and {Cruces}, M. and {Dai}, S. and {Gajjar}, V. and {Hobbs}, G. and {Jin}, C. and {Kramer}, M. and {Lorimer}, D.~R. and {Miao}, C.~C. and {Niu}, C.~H. and {Niu}, J.~R. and {Pan}, Z.~C. and {Qian}, L. and {Spitler}, L. and {Werthimer}, D. and {Zhang}, G.~Q. and {Wang}, F.~Y. and {Xie}, X.~Y. and {Yue}, Y.~L. and {Zhang}, L. and {Zhi}, Q.~J. and {Zhu}, Y.},
        title = "{A bimodal burst energy distribution of a repeating fast radio burst source}",
      journal = {\nat},
         year = 2021,
        month = oct,
       volume = {598},
       number = {7880},
        pages = {267-271},
          doi = {10.1038/s41586-021-03878-5},
archivePrefix = {arXiv},
       eprint = {2107.08205},
 primaryClass = {astro-ph.HE},
       adsurl = {https://ui.adsabs.harvard.edu/abs/2021Natur.598..267L}
}

@article{Rickett1977,
author = {Rickett, Barney J.},
title = {Interstellar Scattering and Scintillation of Radio Waves},
journal = {Annual Review of Astronomy and Astrophysics},
volume = {15},
number = {1},
pages = {479-504},
year = {1977},
doi = {10.1146/annurev.aa.15.090177.002403},
URL = {https://doi.org/10.1146/annurev.aa.15.090177.002403},
eprint = {https://doi.org/10.1146/annurev.aa.15.090177.002403}
}

@ARTICLE{Rickett1990,
       author = {{Rickett}, B.~J.},
        title = "{Radio propagation through the turbulent interstellar plasma.}",
      journal = {\araa},
         year = 1990,
        month = jan,
       volume = {28},
        pages = {561-605},
          doi = {10.1146/annurev.aa.28.090190.003021},
       adsurl = {https://ui.adsabs.harvard.edu/abs/1990ARA&A..28..561R}
}

@article{Xu2016,
       author = {{Xu}, Siyao and {Zhang}, Bing},
        title = "{On the Origin of the Scatter Broadening of Fast Radio Burst Pulses and Astrophysical Implications}",
      journal = {\apj},
         year = 2016,
        month = dec,
       volume = {832},
       number = {2},
          eid = {199},
        pages = {199},
          doi = {10.3847/0004-637X/832/2/199},
archivePrefix = {arXiv},
       eprint = {1608.03930},
 primaryClass = {astro-ph.HE},
       adsurl = {https://ui.adsabs.harvard.edu/abs/2016ApJ...832..199X}
}

@article{Luan2014,
       author = {{Luan}, Jing and {Goldreich}, Peter},
        title = "{Physical Constraints on Fast Radio Bursts}",
      journal = {\apjl},
         year = 2014,
        month = apr,
       volume = {785},
       number = {2},
          eid = {L26},
        pages = {L26},
          doi = {10.1088/2041-8205/785/2/L26},
archivePrefix = {arXiv},
       eprint = {1401.1795},
 primaryClass = {astro-ph.HE},
       adsurl = {https://ui.adsabs.harvard.edu/abs/2014ApJ...785L..26L}
}

@ARTICLE{Armstrong1995,
       author = {{Armstrong}, J.~W. and {Rickett}, B.~J. and {Spangler}, S.~R.},
        title = "{Electron Density Power Spectrum in the Local Interstellar Medium}",
      journal = {\apj},
         year = 1995,
        month = apr,
       volume = {443},
        pages = {209},
          doi = {10.1086/175515},
       adsurl = {https://ui.adsabs.harvard.edu/abs/1995ApJ...443..209A}
}

@ARTICLE{Lee1975I,
       author = {{Lee}, L.~C. and {Jokipii}, J.~R.},
        title = "{Strong scintillations in astrophysics. I. The Markov approximation, its validity and application to angular broadening.}",
      journal = {\apj},
         year = 1975,
        month = mar,
       volume = {196},
        pages = {695-707},
          doi = {10.1086/153458},
       adsurl = {https://ui.adsabs.harvard.edu/abs/1975ApJ...196..695L}
}

@ARTICLE{Lee1975II,
       author = {{Lee}, L.~C. and {Jokipii}, J.~R.},
        title = "{Strong scintillations in astrophysics. II. A theory of temporal broadening of pulses.}",
      journal = {\apj},
         year = 1975,
        month = oct,
       volume = {201},
        pages = {532-543},
          doi = {10.1086/153916},
       adsurl = {https://ui.adsabs.harvard.edu/abs/1975ApJ...201..532L}
}

@ARTICLE{Yang2022,
       author = {{Yang}, Yuan-Pei and {Lu}, Wenbin and {Feng}, Yi and {Zhang}, Bing and {Li}, Di},
        title = "{Temporal Scattering, Depolarization, and Persistent Radio Emission from Magnetized Inhomogeneous Environments near Repeating Fast Radio Burst Sources}",
      journal = {\apjl},
         year = 2022,
        month = apr,
       volume = {928},
       number = {2},
          eid = {L16},
        pages = {L16},
          doi = {10.3847/2041-8213/ac5f46},
archivePrefix = {arXiv},
       eprint = {2202.09602},
 primaryClass = {astro-ph.HE},
       adsurl = {https://ui.adsabs.harvard.edu/abs/2022ApJ...928L..16Y}
}

@ARTICLE{Niu2022,
       author = {{Niu}, C. -H. and {Aggarwal}, K. and {Li}, D. and {Zhang}, X. and {Chatterjee}, S. and {Tsai}, C. -W. and {Yu}, W. and {Law}, C.~J. and {Burke-Spolaor}, S. and {Cordes}, J.~M. and {Zhang}, Y. -K. and {Ocker}, S.~K. and {Yao}, J. -M. and {Wang}, P. and {Feng}, Y. and {Niino}, Y. and {Bochenek}, C. and {Cruces}, M. and {Connor}, L. and {Jiang}, J. -A. and {Dai}, S. and {Luo}, R. and {Li}, G. -D. and {Miao}, C. -C. and {Niu}, J. -R. and {Anna-Thomas}, R. and {Sydnor}, J. and {Stern}, D. and {Wang}, W. -Y. and {Yuan}, M. and {Yue}, Y. -L. and {Zhou}, D. -J. and {Yan}, Z. and {Zhu}, W. -W. and {Zhang}, B.},
        title = "{A repeating fast radio burst associated with a persistent radio source}",
      journal = {\nat},
         year = 2022,
        month = jun,
       volume = {606},
       number = {7916},
        pages = {873-877},
          doi = {10.1038/s41586-022-04755-5},
archivePrefix = {arXiv},
       eprint = {2110.07418},
 primaryClass = {astro-ph.HE},
       adsurl = {https://ui.adsabs.harvard.edu/abs/2022Natur.606..873N}
}

@ARTICLE{Katz2016,
       author = {{Katz}, J.~I.},
        title = "{Inferences from the Distributions of Fast Radio Burst Pulse Widths, Dispersion Measures, and Fluences}",
      journal = {\apj},
         year = 2016,
        month = feb,
       volume = {818},
       number = {1},
          eid = {19},
        pages = {19},
          doi = {10.3847/0004-637X/818/1/19},
archivePrefix = {arXiv},
       eprint = {1505.06220},
 primaryClass = {astro-ph.HE},
       adsurl = {https://ui.adsabs.harvard.edu/abs/2016ApJ...818...19K}
}

@ARTICLE{CHIMEcatalog2,
       author = {{CHIME/FRB Collaboration} and {Abbott}, Thomas and {Andersen}, Bridget C. and {Andrew}, Shion and {Bandura}, Kevin and {Bhardwaj}, Mohit and {Bhusare}, Yash and {Brar}, Charanjot and {Cassanelli}, Tomas and {Chatterjee}, Shami and {Cliche}, Jean-Francois and {Cook}, Amanda M. and {Curtin}, Alice and {Dobbs}, Matt and {Dong}, Fengqiu Adam and {Eadie}, Gwendolyn and {Eftekhari}, Tarraneh and {Fonseca}, Emmanuel and {Gaensler}, B.~M. and {Good}, Deborah and {Halpern}, Mark and {Hessels}, Jason W.~T. and {Ibik}, Adaeze and {Jain}, Naman and {Joseph}, Ronniy C. and {Kader}, Zarif and {Kaspi}, Victoria M. and {Khan}, Afrokk and {Kharel}, Bikash and {Kumar}, Ajay and {Landecker}, T.~L. and {Lang}, Dustin and {Lanman}, Adam E. and {L'Argent}, Magnus and {Lazda}, Mattias and {Leung}, Calvin and {Li}, Dong Zi and {Lintott}, Chris J. and {Main}, Robert and {Masui}, Kiyoshi W. and {Mate}, Sujay and {McGregor}, Kyle and {McKinven}, Ryan and {Mena-Parra}, Juan and {Meyers}, Bradley W. and {Michilli}, Daniele and {Ng}, Cherry and {Ng}, Mason and {Nimmo}, Kenzie and {Noble}, Gavin and {Pandhi}, Ayush and {Patil}, Swarali S. and {Pearlman}, Aaron B. and {Pen}, Ue-Li and {Pleunis}, Ziggy and {Prochaska}, J. Xavier and {Rafiei-Ravandi}, Masoud and {Ransom}, Scott and {Renard}, Andre and {Sammons}, Mawson W. and {Sand}, Ketan R. and {Scholz}, Paul and {Shah}, Vishwangi and {Shin}, Kaitlyn and {Siegel}, Seth R. and {Sirota}, Sloane and {Smith}, Kendrick and {Stairs}, Ingrid and {Stenning}, David C. and {Tendulkar}, Shriharsh P. and {Vanderlinde}, Keith and {Walmsley}, Mike and {Wang}, Haochen and {Wulf}, Dallas},
        title = "{The Second CHIME/FRB Catalog of Fast Radio Bursts}",
      journal = {\apjs},
         year = 2026,
        month = mar,
       volume = {283},
       number = {1},
          eid = {34},
        pages = {34},
          doi = {10.3847/1538-4365/ae3828},
archivePrefix = {arXiv},
       eprint = {2601.09399},
 primaryClass = {astro-ph.HE},
       adsurl = {https://ui.adsabs.harvard.edu/abs/2026ApJS..283...34C}
}

@ARTICLE{Ricci2021,
       author = {{Ricci}, Roberto and {Piro}, Luigi and {Panessa}, Francesca and {O'Connor}, Brendan and {Lotti}, Simone and {Bruni}, Gabriele and {Zhang}, Bing},
        title = "{Detection of a persistent radio source at the location of FRB20201124A with VLA}",
      journal = {The Astronomer's Telegram},
         year = 2021,
        month = apr,
       volume = {14549},
        pages = {1},
       adsurl = {https://ui.adsabs.harvard.edu/abs/2021ATel14549....1R}
}

@ARTICLE{GengHuang2015,
       author = {{Geng}, J.~J. and {Huang}, Y.~F.},
        title = "{Fast Radio Bursts: Collisions between Neutron Stars and Asteroids/Comets}",
      journal = {\apj},
         year = 2015,
        month = aug,
       volume = {809},
       number = {1},
          eid = {24},
        pages = {24},
          doi = {10.1088/0004-637X/809/1/24},
archivePrefix = {arXiv},
       eprint = {1502.05171},
 primaryClass = {astro-ph.HE},
       adsurl = {https://ui.adsabs.harvard.edu/abs/2015ApJ...809...24G}
}

@ARTICLE{DaiZG2016,
       author = {{Dai}, Z.~G. and {Wang}, J.~S. and {Wu}, X.~F. and {Huang}, Y.~F.},
        title = "{Repeating Fast Radio Bursts from Highly Magnetized Pulsars Traveling through Asteroid Belts}",
      journal = {\apj},
         year = 2016,
        month = sep,
       volume = {829},
       number = {1},
          eid = {27},
        pages = {27},
          doi = {10.3847/0004-637X/829/1/27},
archivePrefix = {arXiv},
       eprint = {1603.08207},
 primaryClass = {astro-ph.HE},
       adsurl = {https://ui.adsabs.harvard.edu/abs/2016ApJ...829...27D}
}

@ARTICLE{ZhangB2017,
       author = {{Zhang}, Bing},
        title = "{A {\textquotedblleft}Cosmic Comb{\textquotedblright} Model of Fast Radio Bursts}",
      journal = {\apjl},
         year = 2017,
        month = feb,
       volume = {836},
       number = {2},
          eid = {L32},
        pages = {L32},
          doi = {10.3847/2041-8213/aa5ded},
archivePrefix = {arXiv},
       eprint = {1701.04094},
 primaryClass = {astro-ph.HE},
       adsurl = {https://ui.adsabs.harvard.edu/abs/2017ApJ...836L..32Z}
}

@ARTICLE{Lyubarsky2020,
       author = {{Lyubarsky}, Yuri},
        title = "{Fast Radio Bursts from Reconnection in a Magnetar Magnetosphere}",
      journal = {\apj},
         year = 2020,
        month = jul,
       volume = {897},
       number = {1},
          eid = {1},
        pages = {1},
          doi = {10.3847/1538-4357/ab97b5},
archivePrefix = {arXiv},
       eprint = {2001.02007},
 primaryClass = {astro-ph.HE},
       adsurl = {https://ui.adsabs.harvard.edu/abs/2020ApJ...897....1L}
}

@ARTICLE{Bochenek2020,
       author = {{Bochenek}, C.~D. and {Ravi}, V. and {Belov}, K.~V. and {Hallinan}, G. and {Kocz}, J. and {Kulkarni}, S.~R. and {McKenna}, D.~L.},
        title = "{A fast radio burst associated with a Galactic magnetar}",
      journal = {\nat},
         year = 2020,
        month = nov,
       volume = {587},
       number = {7832},
        pages = {59-62},
          doi = {10.1038/s41586-020-2872-x},
archivePrefix = {arXiv},
       eprint = {2005.10828},
 primaryClass = {astro-ph.HE},
       adsurl = {https://ui.adsabs.harvard.edu/abs/2020Natur.587...59B}
}

@ARTICLE{CHIME2020magnetar,
       author = {{CHIME/FRB Collaboration} and {Andersen}, B.~C. and {Bandura}, K.~M. and {Bhardwaj}, M. and {Bij}, A. and {Boyce}, M.~M. and {Boyle}, P.~J. and {Brar}, C. and {Cassanelli}, T. and {Chawla}, P. and {Chen}, T. and {Cliche}, J.-F. and {Cook}, A. and {Cubranic}, D. and {Curtin}, A.~P. and {Denman}, N.~T. and {Dobbs}, M. and {Dong}, F.~Q. and {Fandino}, M. and {Fonseca}, E. and {Gaensler}, B.~M. and {Giri}, U. and {Good}, D.~C. and {Halpern}, M. and {Hill}, A.~S. and {Hinshaw}, G.~F. and {H{\"o}fer}, C. and {Josephy}, A. and {Kania}, J.~W. and {Kaspi}, V.~M. and {Landecker}, T.~L. and {Leung}, C. and {Li}, D.~Z. and {Lin}, H.-H. and {Masui}, K.~W. and {McKinven}, R. and {Mena-Parra}, J. and {Merryfield}, M. and {Meyers}, B.~W. and {Michilli}, D. and {Milutinovic}, N. and {Mirhosseini}, A. and {M{\"u}nchmeyer}, M. and {Naidu}, A. and {Newburgh}, L.~B. and {Ng}, C. and {Patel}, C. and {Pen}, U.-L. and {Pinsonneault-Marotte}, T. and {Pleunis}, Z. and {Quine}, B.~M. and {Rafiei-Ravandi}, M. and {Rahman}, M. and {Ransom}, S.~M. and {Renard}, A. and {Sanghavi}, P. and {Scholz}, P. and {Shaw}, J.~R. and {Shin}, K. and {Siegel}, S.~R. and {Singh}, S. and {Smegal}, R.~J. and {Smith}, K.~M. and {Stairs}, I.~H. and {Tan}, C.~M. and {Tendulkar}, S.~P. and {Tretyakov}, I. and {Vanderlinde}, K. and {Wang}, H. and {Wulf}, D. and {Zwaniga}, A.~V.},
        title = "{A bright millisecond-duration radio burst from a Galactic magnetar}",
      journal = {\nat},
         year = 2020,
        month = nov,
       volume = {587},
       number = {7832},
        pages = {54-58},
          doi = {10.1038/s41586-020-2863-y},
archivePrefix = {arXiv},
       eprint = {2005.10324},
 primaryClass = {astro-ph.HE},
       adsurl = {https://ui.adsabs.harvard.edu/abs/2020Natur.587...54C}
}

@ARTICLE{Giacomazzo2013,
       author = {{Giacomazzo}, Bruno and {Perna}, Rosalba},
        title = "{Formation of Stable Magnetars from Binary Neutron Star Mergers}",
      journal = {\apjl},
         year = 2013,
        month = jul,
       volume = {771},
       number = {2},
          eid = {L26},
        pages = {L26},
          doi = {10.1088/2041-8205/771/2/L26},
archivePrefix = {arXiv},
       eprint = {1306.1608},
 primaryClass = {astro-ph.HE},
       adsurl = {https://ui.adsabs.harvard.edu/abs/2013ApJ...771L..26G}
}

@ARTICLE{Yoon2007,
       author = {{Yoon}, S.-C. and {Podsiadlowski}, Ph. and {Rosswog}, S.},
        title = "{Remnant evolution after a carbon-oxygen white dwarf merger}",
      journal = {\mnras},
         year = 2007,
        month = sep,
       volume = {380},
       number = {3},
        pages = {933-948},
          doi = {10.1111/j.1365-2966.2007.12161.x},
archivePrefix = {arXiv},
       eprint = {0704.0297},
 primaryClass = {astro-ph},
       adsurl = {https://ui.adsabs.harvard.edu/abs/2007MNRAS.380..933Y}
}

@ARTICLE{Zhong2020,
       author = {{Zhong}, Shu-Qing and {Dai}, Zi-Gao},
        title = "{Magnetars from Neutron Star-White Dwarf Mergers: Application to Fast Radio Bursts}",
      journal = {\apj},
         year = 2020,
        month = apr,
       volume = {893},
       number = {1},
          eid = {9},
        pages = {9},
          doi = {10.3847/1538-4357/ab7bdf},
archivePrefix = {arXiv},
       eprint = {2002.11975},
 primaryClass = {astro-ph.HE},
       adsurl = {https://ui.adsabs.harvard.edu/abs/2020ApJ...893....9Z}
}

@BOOK{Jacco2020,
       author = {{Vink}, Jacco},
        title = "{Physics and Evolution of Supernova Remnants}",
        publisher={Cham: Springer},
         year = 2020,
          doi = {10.1007/978-3-030-55231-2},
       adsurl = {https://ui.adsabs.harvard.edu/abs/2020pesr.book.....V}
}

@ARTICLE{TM99,
       author = {{Truelove}, J. Kelly and {McKee}, Christopher F.},
        title = "{Evolution of Nonradiative Supernova Remnants}",
      journal = {\apjs},
         year = 1999,
        month = feb,
       volume = {120},
       number = {2},
        pages = {299-326},
          doi = {10.1086/313176},
       adsurl = {https://ui.adsabs.harvard.edu/abs/1999ApJS..120..299T},
        shorthand = {TM99}
}

@ARTICLE{Chevalier1982,
       author = {{Chevalier}, R.~A.},
        title = "{Self-similar solutions for the interaction of stellar ejecta with an external medium.}",
      journal = {\apj},
         year = 1982,
        month = jul,
       volume = {258},
        pages = {790-797},
          doi = {10.1086/160126},
       adsurl = {https://ui.adsabs.harvard.edu/abs/1982ApJ...258..790C}
}

@ARTICLE{Nadezhin1985,
       author = {{Nadezhin}, D.~K.},
        title = "{On the Initial Phase of Interaction Between Expanding Stellar Envelopes and Surrounding Medium}",
      journal = {\apss},
         year = 1985,
        month = may,
       volume = {112},
       number = {2},
        pages = {225-249},
          doi = {10.1007/BF00653506},
       adsurl = {https://ui.adsabs.harvard.edu/abs/1985Ap&SS.112..225N}
}

@ARTICLE{anna2023,
       author = {{Anna-Thomas}, Reshma and {Connor}, Liam and {Dai}, Shi and {Feng}, Yi and {Burke-Spolaor}, Sarah and {Beniamini}, Paz and {Yang}, Yuan-Pei and {Zhang}, Yong-Kun and {Aggarwal}, Kshitij and {Law}, Casey J. and {Li}, Di and {Niu}, Chenhui and {Chatterjee}, Shami and {Cruces}, Marilyn and {Duan}, Ran and {Filipovic}, Miroslav D. and {Hobbs}, George and {Lynch}, Ryan S. and {Miao}, Chenchen and {Niu}, Jiarui and {Ocker}, Stella K. and {Tsai}, Chao-Wei and {Wang}, Pei and {Xue}, Mengyao and {Yao}, Ju-Mei and {Yu}, Wenfei and {Zhang}, Bing and {Zhang}, Lei and {Zhu}, Shiqiang and {Zhu}, Weiwei},
        title = "{Magnetic field reversal in the turbulent environment around a repeating fast radio burst}",
      journal = {Science},
         year = 2023,
        month = may,
       volume = {380},
       number = {6645},
        pages = {599-603},
          doi = {10.1126/science.abo6526},
archivePrefix = {arXiv},
       eprint = {2202.11112},
 primaryClass = {astro-ph.HE},
       adsurl = {https://ui.adsabs.harvard.edu/abs/2023Sci...380..599A}
}

@ARTICLE{Wei2025,
       author = {{Wei}, Jia-Peng and {Huang}, Yong-Feng and {Cui}, Lang and {Liu}, Xiang and {Geng}, Jin-Jun and {Wu}, Xue-Feng},
        title = "{Intrinsic Pulse Widths of FRB 20121102A and Calculation of Broadening from Propagation and Instrumental Effects}",
      journal = {\apj},
         year = 2025,
        month = feb,
       volume = {980},
       number = {1},
          eid = {114},
        pages = {114},
          doi = {10.3847/1538-4357/adace3},
archivePrefix = {arXiv},
       eprint = {2402.02360},
 primaryClass = {astro-ph.HE},
       adsurl = {https://ui.adsabs.harvard.edu/abs/2025ApJ...980..114W}
}

@ARTICLE{ZhaoZY2021,
       author = {{Zhao}, Z.~Y. and {Zhang}, G.~Q. and {Wang}, Y.~Y. and {Tu}, Zuo-Lin and {Wang}, F.~Y.},
        title = "{Dispersion and Rotation Measures from the Ejecta of Compact Binary Mergers: Clue to the Progenitors of Fast Radio Bursts}",
      journal = {\apj},
         year = 2021,
        month = feb,
       volume = {907},
       number = {2},
          eid = {111},
        pages = {111},
          doi = {10.3847/1538-4357/abd321},
archivePrefix = {arXiv},
       eprint = {2010.10702},
 primaryClass = {astro-ph.HE},
       adsurl = {https://ui.adsabs.harvard.edu/abs/2021ApJ...907..111Z}
}

@ARTICLE{Tang2017,
       author = {{Tang}, Xiaping and {Chevalier}, Roger A.},
        title = "{Shock evolution in non-radiative supernova remnants}",
      journal = {\mnras},
         year = 2017,
        month = mar,
       volume = {465},
       number = {4},
        pages = {3793-3802},
          doi = {10.1093/mnras/stw2978},
archivePrefix = {arXiv},
       eprint = {1607.06391},
 primaryClass = {astro-ph.HE},
       adsurl = {https://ui.adsabs.harvard.edu/abs/2017MNRAS.465.3793T}
}

@ARTICLE{Micelotta2016,
       author = {{Micelotta}, Elisabetta R. and {Dwek}, Eli and {Slavin}, Jonathan D.},
        title = "{Dust destruction by the reverse shock in the Cassiopeia A supernova remnant}",
      journal = {\aap},
         year = 2016,
        month = may,
       volume = {590},
          eid = {A65},
        pages = {A65},
          doi = {10.1051/0004-6361/201527350},
archivePrefix = {arXiv},
       eprint = {1602.02754},
 primaryClass = {astro-ph.GA},
       adsurl = {https://ui.adsabs.harvard.edu/abs/2016A&A...590A..65M}
}

@ARTICLE{Hwang2012,
       author = {{Hwang}, Una and {Laming}, J. Martin},
        title = "{A Chandra X-Ray Survey of Ejecta in the Cassiopeia A Supernova Remnant}",
      journal = {\apj},
         year = 2012,
        month = feb,
       volume = {746},
       number = {2},
          eid = {130},
        pages = {130},
          doi = {10.1088/0004-637X/746/2/130},
archivePrefix = {arXiv},
       eprint = {1111.7316},
 primaryClass = {astro-ph.HE},
       adsurl = {https://ui.adsabs.harvard.edu/abs/2012ApJ...746..130H}
}

@ARTICLE{Laming2003,
       author = {{Laming}, J. Martin and {Hwang}, Una},
        title = "{On the Determination of Ejecta Structure and Explosion Asymmetry from the X-Ray Knots of Cassiopeia A}",
      journal = {\apj},
         year = 2003,
        month = nov,
       volume = {597},
       number = {1},
        pages = {347-361},
          doi = {10.1086/378268},
archivePrefix = {arXiv},
       eprint = {astro-ph/0306119},
 primaryClass = {astro-ph},
       adsurl = {https://ui.adsabs.harvard.edu/abs/2003ApJ...597..347L}
}

@INCOLLECTION{Chevalier2017,
       author = {{Chevalier}, Roger A. and {Fransson}, Claes},
        title = "{Thermal and Non-thermal Emission from Circumstellar Interaction}",
        publisher={Cham: Springer},
    booktitle = {Handbook of Supernovae},
         year = 2017,
       editor = {{Alsabti}, Athem W. and {Murdin}, Paul},
        pages = {875},
          doi = {10.1007/978-3-319-21846-5_34},
       adsurl = {https://ui.adsabs.harvard.edu/abs/2017hsn..book..875C}
}

@BOOK{ZB2018,
       author = {{Zhang}, Bing},
        title = "{The Physics of Gamma-Ray Bursts}",
     publisher = {Cambridge University Press},
         year = 2018,
          doi = {10.1017/9781139226530},
       adsurl = {https://ui.adsabs.harvard.edu/abs/2018pgrb.book.....Z}
}

@ARTICLE{Dwarkadas2000,
       author = {{Dwarkadas}, Vikram V.},
        title = "{Interaction of Type IA Supernovae with Their Surroundings: The Exponential Profile in Two Dimensions}",
      journal = {\apj},
         year = 2000,
        month = sep,
       volume = {541},
       number = {1},
        pages = {418-427},
          doi = {10.1086/309406},
archivePrefix = {arXiv},
       eprint = {astro-ph/0008076},
 primaryClass = {astro-ph},
       adsurl = {https://ui.adsabs.harvard.edu/abs/2000ApJ...541..418D}
}

@ARTICLE{Shirkey1978,
       author = {{Shirkey}, R.~C.},
        title = "{The radio dynamical evolution of young supernova remnants.}",
      journal = {\apj},
         year = 1978,
        month = sep,
       volume = {224},
        pages = {477-487},
          doi = {10.1086/156395},
       adsurl = {https://ui.adsabs.harvard.edu/abs/1978ApJ...224..477S}
}

@ARTICLE{Fraschetti2010,
       author = {{Fraschetti}, F. and {Teyssier}, R. and {Ballet}, J. and {Decourchelle}, A.},
        title = "{Simulation of the growth of the 3D Rayleigh-Taylor instability in supernova remnants using an expanding reference frame}",
      journal = {\aap},
         year = 2010,
        month = jun,
       volume = {515},
          eid = {A104},
        pages = {A104},
          doi = {10.1051/0004-6361/200912692},
archivePrefix = {arXiv},
       eprint = {1002.5048},
 primaryClass = {astro-ph.HE},
       adsurl = {https://ui.adsabs.harvard.edu/abs/2010A&A...515A.104F}
}

@ARTICLE{smith2014,
       author = {{Smith}, Nathan},
        title = "{Mass Loss: Its Effect on the Evolution and Fate of High-Mass Stars}",
      journal = {\araa},
         year = 2014,
        month = aug,
       volume = {52},
        pages = {487-528},
          doi = {10.1146/annurev-astro-081913-040025},
archivePrefix = {arXiv},
       eprint = {1402.1237},
 primaryClass = {astro-ph.SR},
       adsurl = {https://ui.adsabs.harvard.edu/abs/2014ARA&A..52..487S}
}

@ARTICLE{Smartt2009,
       author = {{Smartt}, Stephen J.},
        title = "{Progenitors of Core-Collapse Supernovae}",
      journal = {\araa},
         year = 2009,
        month = sep,
       volume = {47},
       number = {1},
        pages = {63-106},
          doi = {10.1146/annurev-astro-082708-101737},
archivePrefix = {arXiv},
       eprint = {0908.0700},
 primaryClass = {astro-ph.SR},
       adsurl = {https://ui.adsabs.harvard.edu/abs/2009ARA&A..47...63S}
}

@ARTICLE{Rahaman2025,
       author = {{Rahaman}, Sk. Minhajur and {Acharya}, Sandeep Kumar and {Beniamini}, Paz and {Granot}, Jonathan},
        title = "{Persistent Radio Sources Associated with Fast Radio Bursts: Implications from Magnetar Progenitors}",
      journal = {\apj},
         year = 2025,
        month = aug,
       volume = {988},
       number = {2},
          eid = {276},
        pages = {276},
          doi = {10.3847/1538-4357/ade70c},
archivePrefix = {arXiv},
       eprint = {2504.01125},
 primaryClass = {astro-ph.HE},
       adsurl = {https://ui.adsabs.harvard.edu/abs/2025ApJ...988..276R}
}

@ARTICLE{Bhattacharya2025,
       author = {{Bhattacharya}, Mukul and {Murase}, Kohta and {Kashiyama}, Kazumi},
        title = "{Quasi-steady emission from repeating fast radio bursts can be explained by magnetar wind nebulae}",
      journal = {\mnras},
         year = 2025,
        month = dec,
          doi = {10.1093/mnras/staf2175},
archivePrefix = {arXiv},
       eprint = {2412.19358},
 primaryClass = {astro-ph.HE},
       adsurl = {https://ui.adsabs.harvard.edu/abs/2025MNRAS.tmp.2057B}
}

@ARTICLE{Trimble1968,
       author = {{Trimble}, Virginia},
        title = "{Motions and Structure of the Filamentary Envelope of the Crab Nebula}",
      journal = {\aj},
         year = 1968,
        month = sep,
       volume = {73},
        pages = {535},
          doi = {10.1086/110658},
       adsurl = {https://ui.adsabs.harvard.edu/abs/1968AJ.....73..535T}
}

@ARTICLE{WangFY2025,
       author = {{Wang}, F.~Y. and {Lan}, H.~T. and {Zhao}, Z.~Y. and {Wu}, Q. and {Feng}, Y. and {Yi}, S.~X. and {Dai}, Z.~G. and {Cheng}, K.~S.},
        title = "{Evidence of young magnetars in massive binary embedded in a supernova remnant as sources of active fast radio bursts}",
      journal = {arXiv e-prints},
         year = 2025,
        month = dec,
          eid = {arXiv:2512.07140},
        pages = {arXiv:2512.07140},
archivePrefix = {arXiv},
       eprint = {2512.07140},
 primaryClass = {astro-ph.HE},
       adsurl = {https://ui.adsabs.harvard.edu/abs/2025arXiv251207140W}
}

@ARTICLE{Gupta1993,
       author = {{Gupta}, Y. and {Rickett}, B.~J. and {Coles}, W.~A.},
        title = "{Refractive Interstellar Scintillation of Pulsar Intensities at 74 MHz}",
      journal = {\apj},
         year = 1993,
        month = jan,
       volume = {403},
        pages = {183},
          doi = {10.1086/172193},
       adsurl = {https://ui.adsabs.harvard.edu/abs/1993ApJ...403..183G}
}

@ARTICLE{Coles1989,
       author = {{Coles}, W.~A. and {Harmon}, J.~K.},
        title = "{Propagation Observations of the Solar Wind near the Sun}",
      journal = {\apj},
         year = 1989,
        month = feb,
       volume = {337},
        pages = {1023},
          doi = {10.1086/167173},
       adsurl = {https://ui.adsabs.harvard.edu/abs/1989ApJ...337.1023C}
}

@BOOK{Rybicki1979,
       author = {{Rybicki}, George B. and {Lightman}, Alan P.},
        title = "{Radiative processes in astrophysics}",
        publisher={New York: Wiley},
         year = 1979,
       adsurl = {https://ui.adsabs.harvard.edu/abs/1979rpa..book.....R}
}

@ARTICLE{Metzger2017,
       author = {{Metzger}, Brian D. and {Berger}, Edo and {Margalit}, Ben},
        title = "{Millisecond Magnetar Birth Connects FRB 121102 to Superluminous Supernovae and Long-duration Gamma-Ray Bursts}",
      journal = {\apj},
         year = 2017,
        month = may,
       volume = {841},
       number = {1},
          eid = {14},
        pages = {14},
          doi = {10.3847/1538-4357/aa633d},
archivePrefix = {arXiv},
       eprint = {1701.02370},
 primaryClass = {astro-ph.HE},
       adsurl = {https://ui.adsabs.harvard.edu/abs/2017ApJ...841...14M}
}

@ARTICLE{NiuCH2026,
       author = {{Niu}, Chen-Hui and {Li}, Di and {Yang}, Yuan-Pei and {Zhu}, Yuhao and {Zhang}, Yongkun and {Zhang}, Jia-Heng and {Du}, Zexin and {Yao}, Jumei and {Zheng}, Xiaoping and {Wang}, Pei and {Feng}, Yi and {Zhang}, Bing and {Zhu}, Weiwei and {Yu}, Wenfei and {Jiang}, Ji-An and {Dai}, Shi and {Tsai}, Chao-Wei and {Chen}, A. Ming and {Hou}, Yijun and {Niu}, Jiarui and {Wang}, Weiyang and {Miao}, Chenchen and {Li}, Xinming and {Zhang}, Junshuo},
        title = "{A persistently active fast radio burst source embedded in an expanding supernova remnant}",
      journal = {Science Bulletin},
         year = 2026,
        month = jan,
       volume = {71},
       number = {1},
        pages = {76-82},
          doi = {10.1016/j.scib.2025.11.023},
archivePrefix = {arXiv},
       eprint = {2512.05448},
 primaryClass = {astro-ph.HE},
       adsurl = {https://ui.adsabs.harvard.edu/abs/2026SciBu..71...76N}
}

@ARTICLE{ZhangZ2026,
       author = {{Zhang}, Zhao Joseph and {Kawashima}, Gaku and {Lee}, Shiu-Hang and {Nagamine}, Kentaro and {Zhang}, Bing and {Fujimaru}, Yusei},
        title = "{Probing the Dispersion and Rotation Measure Contributions from Supernova Remnants in Fast Radio Burst Source Environments with 1D SNR Simulation}",
      journal = {\apj},
         year = 2026,
        month = jul,
       volume = {1005},
       number = {2},
          eid = {234},
        pages = {234},
          doi = {10.3847/1538-4357/ae7b2a},
archivePrefix = {arXiv},
       eprint = {2603.07012},
 primaryClass = {astro-ph.HE},
       adsurl = {https://ui.adsabs.harvard.edu/abs/2026ApJ..1005..234Z}
}

@ARTICLE{Ocker2023,
       author = {{Ocker}, Stella Koch and {Cordes}, James M. and {Chatterjee}, Shami and {Li}, Di and {Niu}, Chen-Hui and {McKee}, James W. and {Law}, Casey J. and {Anna-Thomas}, Reshma},
        title = "{Scattering variability detected from the circumsource medium of FRB 20190520B}",
      journal = {\mnras},
         year = 2023,
        month = feb,
       volume = {519},
       number = {1},
        pages = {821-830},
          doi = {10.1093/mnras/stac3547},
archivePrefix = {arXiv},
       eprint = {2210.01975},
 primaryClass = {astro-ph.HE},
       adsurl = {https://ui.adsabs.harvard.edu/abs/2023MNRAS.519..821O}
}

@ARTICLE{Shimoda2018,
       author = {{Shimoda}, Jiro and {Akahori}, Takuya and {Lazarian}, A. and {Inoue}, Tsuyoshi and {Fujita}, Yutaka},
        title = "{Discovery of Kolmogorov-like magnetic energy spectrum in Tycho's supernova remnant by two-point correlations of synchrotron intensity}",
      journal = {\mnras},
         year = 2018,
        month = oct,
       volume = {480},
       number = {2},
        pages = {2200-2205},
          doi = {10.1093/mnras/sty2034},
archivePrefix = {arXiv},
       eprint = {1807.10089},
 primaryClass = {astro-ph.HE},
       adsurl = {https://ui.adsabs.harvard.edu/abs/2018MNRAS.480.2200S}
}

@ARTICLE{piro2016,
       author = {{Piro}, Anthony L.},
        title = "{The Impact of a Supernova Remnant on Fast Radio Bursts}",
      journal = {\apjl},
         year = 2016,
        month = jun,
       volume = {824},
       number = {2},
          eid = {L32},
        pages = {L32},
          doi = {10.3847/2041-8205/824/2/L32},
archivePrefix = {arXiv},
       eprint = {1604.04909},
 primaryClass = {astro-ph.HE},
       adsurl = {https://ui.adsabs.harvard.edu/abs/2016ApJ...824L..32P}
}

@ARTICLE{YangYP2017,
       author = {{Yang}, Yuan-Pei and {Zhang}, Bing},
        title = "{Dispersion Measure Variation of Repeating Fast Radio Burst Sources}",
      journal = {\apj},
         year = 2017,
        month = sep,
       volume = {847},
       number = {1},
          eid = {22},
        pages = {22},
          doi = {10.3847/1538-4357/aa8721},
archivePrefix = {arXiv},
       eprint = {1707.02923},
 primaryClass = {astro-ph.HE},
       adsurl = {https://ui.adsabs.harvard.edu/abs/2017ApJ...847...22Y}
}

@ARTICLE{piro2018,
       author = {{Piro}, Anthony L. and {Gaensler}, B.~M.},
        title = "{The Dispersion and Rotation Measure of Supernova Remnants and Magnetized Stellar Winds: Application to Fast Radio Bursts}",
      journal = {\apj},
         year = 2018,
        month = jul,
       volume = {861},
       number = {2},
          eid = {150},
        pages = {150},
          doi = {10.3847/1538-4357/aac9bc},
archivePrefix = {arXiv},
       eprint = {1804.01104},
 primaryClass = {astro-ph.HE},
       adsurl = {https://ui.adsabs.harvard.edu/abs/2018ApJ...861..150P}
}

@ARTICLE{Katz2022,
       author = {{Katz}, J.~I.},
        title = "{The environment and constraints on the mass of FRB 190520B}",
      journal = {\mnras},
         year = 2022,
        month = jul,
       volume = {514},
       number = {1},
        pages = {L27-L30},
          doi = {10.1093/mnrasl/slac051},
archivePrefix = {arXiv},
       eprint = {2203.14943},
 primaryClass = {astro-ph.HE},
       adsurl = {https://ui.adsabs.harvard.edu/abs/2022MNRAS.514L..27K}
}
\bibliographystyle{aasjournal}

\begin{table}[]
\centering
\caption{Summary of the 20 cases. The first letter denotes the scattering model (A--E, corresponding to Equations~\eqref{func37a}--\eqref{func39}), and the appended number indicates the ejecta power-law index $n$.}
\label{tablecases}
\begin{tabular}{ccclc}
\hline
Cases & Equations & Spectrum & Retained cases \\
\hline
A2, A4, A6, A8  & \eqref{func37a} & Power-law, $\beta<3$ &  A2, A4 \\
B2, B4, B6, B8  & \eqref{func37b} & Power-law, $\beta<3$ &  B2, B4 \\
C2, C4, C6, C8  & \eqref{func38a} & Power-law, $\beta=11/3$ (Kolmogorov)& -- \\
D2, D4, D6, D8  & \eqref{func38b} & Power-law, $\beta=11/3$ (Kolmogorov)& -- \\
E2, E4, E6, E8  & \eqref{func39}  & Gaussian  & E2, E4 \\
\hline
\end{tabular}
\end{table}

\begin{table}[]
  \centering
  \begin{threeparttable}
  \caption{
     Best-fit and median values with 68\% credible intervals, and transition times for cases A2, A4, B2, B4, E2, and E4.}
  \begin{tabular}{c|c c c c c c |c}
\hline
\hline
Cases & $M_{\rm ej}/M_{\odot}$ & $E_{\rm k}/10^{51}~{\rm erg}$ & $\dot{M}_{\rm w}/10^{-5}~M_{\odot}~{\rm yr}^{-1}$ & $t_0/{\rm yr}$ & $DM_{\rm host}/{\rm pc~cm^{-3}}$ & $\beta$
& $t_{\rm tran}/{\rm yr}$\\
\hline
A2 & $19.2^{a}/8.3^{+7.2\,b}_{-4.3}$ & $2.3/0.85^{+1.72}_{-0.64}$ & $1.9/1.3^{+0.5}_{-0.5}$ & $119.3/93.3^{+21.5}_{-20.9}$ & $28.7/154.9^{+262.0}_{-122.5}$ & $\times$ & $9937.2$
\\
A4 & $13.7/11.7^{+5.2}_{-4.5}$ & $0.3/0.29^{+0.42}_{-0.15}$ & $14.8/14.1^{+3.3}_{-3.4}$ & $113.6/97.7^{+40.3}_{-26.9}$ & $15.3/74.1^{+207.7}_{-55.5}$ & $\times$ & $220.4$ \\
B2 & $8.7/2.6^{+2.0}_{-1.2}$ & $7.3/3.0^{+4.1}_{-2.0}$ & $0.3/0.11^{+0.07}_{-0.04}$ & $169.8/79.4^{+38.1}_{-27.0}$ & $998.0/1000.0^{+0.0}_{-0.0}$ & $2.9/2.88^{+0.02}_{-0.03}$ &
$9750.6$ \\
B4 & $17.5/14.5^{+3.7}_{-4.0}$ & $0.2/0.21^{+0.18}_{-0.08}$ & $0.8/0.60^{+0.17}_{-0.17}$ & $129.5/100.0^{+28.8}_{-22.4}$ & $11.0/69.2^{+139.7}_{-49.2}$ &
$2.9/2.86^{+0.03}_{-0.05}$ & $7118.5$ \\
E2 & $16.7/5.4^{+6.7}_{-3.1}$ & $6.0/1.4^{+4.0}_{-1.2}$ & $1.0/0.62^{+0.34}_{-0.25}$ & $79.8/64.6^{+24.6}_{-16.7}$ & $407.6/588.8^{+152.5}_{-142.2}$ & $\times$ & $9942.6$ \\
E4 & $10.2/12.9^{+4.5}_{-4.4}$ & $0.1/0.25^{+0.32}_{-0.11}$ & $5.2/6.3^{+1.3}_{-1.3}$ & $120.6/100.0^{+38.0}_{-25.9}$ & $10.7/63.1^{+141.1}_{-44.9}$ & $\times$ & $656.2$ \\
\hline
  \end{tabular}
\label{tablefit}
\begin{tablenotes}
    \footnotesize
    \item[a] Best-fit parameters.
    \item[b] Posterior median with the 68\% credible interval.
\end{tablenotes}
\end{threeparttable}
\end{table}

\begin{table}[]
    \centering
    \caption{
BIC and $\Delta$BIC values for the scattering-timescale and DM
fits of cases A2, A4, B2, B4, E2, and E4.}
    \begin{tabular}{c|c c c c}
\hline
\hline
Cases & BIC$_{\rm scatter}$ & $\Delta$BIC$_{\rm scatter}$ & BIC$_{\rm DM}$ & $\Delta$BIC$_{\rm DM}$ \\
\hline
A2 & 191 & 0 & 1672080  & 102588     \\
A4 & 191 & 0 & 1638253  & 68760      \\
B2 & 196 & 5 & 6488813  & 4919321    \\
B4 & 197 & 6 & 1648488  & 78996      \\
E2 & 192 & 1 & 1569492  & 0          \\
E4 & 191 & 0 & 1702643  & 133151     \\
\hline
    \end{tabular}
    \label{tablestat}
\end{table}

\appendix
\section{Derivation of pulse broadening for a Gaussian density fluctuation spectrum}\label{appe}

Following \cite{Rickett1977}, the density fluctuation of a
Gaussian spectrum is
\begin{equation}
    P_{3N}(k)=C_N^2\exp{[-k^2/k_0^2]},
\end{equation}
where $k_0=2\pi/l_0$. Since the normalization of the Gaussian
spectrum is performed over the entire wavenumber space, we obtain
\begin{equation}\label{A2}
    C_N^2=\delta n_{\rm e}^2/k_0^3\pi^{3/2}.
\end{equation}
According to  Equation A7 in \cite{Rickett1990}, the wave structure function can be written as
\begin{equation}
D_{\phi} =4\pi r_{\rm e}^2\lambda^2C_N^2\Delta D\int_0^{2\pi}\int_0^{\infty}[1-\cos{(\vec{k}\cdot\vec{\sigma_{\rm s}})}]\exp{(-k^2/k_0^2)}kdkd\theta,
\end{equation}
where $\theta$ is the angle between $k$ and $\sigma_{\rm s}$.
Performing the integration gives
\begin{equation}
    D_{\phi}=4\pi^2\lambda^2r_{\rm e}^2C_N^2\Delta Dk_0^2[1-\exp{(-k_0^2\sigma_{\rm s}^2/4)}].
\end{equation}
One can see that when $k_0\sigma_{\rm s}\gg1$, the exponential term becomes negligible and the wave structure function saturates at a constant value.
We therefore only consider the scenario of $k_0\sigma_{\rm s}\ll1$.

Combining Equation (\ref{A2}), $D_{\phi}$ can be written as
\begin{equation}
    D_{\phi}=2\pi^{3/2}r_{\rm e}^2\lambda^2l_0^{-1}\sigma_{\rm s}^2\Delta D\delta n_{\rm
    e}^2.
\end{equation}
The diffractive length, $\sigma_{\rm diff}$, is therefore given by
\begin{equation}
    \sigma_{\rm diff} = [2\pi^{3/2}r_{\rm e}^2\lambda^2l_0^{-1}\Delta D\delta n_{\rm e}^2]^{-1/2}.
\end{equation}
Finally, following the assumptions introduced in Section
\ref{scatter} and considering Equations (\ref{func35}) and
(\ref{func36}), the temporal broadening caused by a Gaussian
spectrum is
\begin{equation}
    \tau_{\rm sc}=\frac{r_{\rm e}^2\lambda^4DM^2}{4c\pi^{1/2}(1+z_0)^3}(\frac{\delta n_{\rm e}}{n_{\rm e}})^2f~l_0^{-1}.
\end{equation}

\end{CJK*}
\end{document}